\documentclass[a4paper,twocolumn,11pt,accepted=2026-04-08]{quantumarticle}
\pdfoutput=1

\usepackage{amsmath}
\usepackage[dvipsnames]{xcolor}
\usepackage{hyperref}
\usepackage{tikz}

\usepackage[english]{babel}
\usepackage[utf8]{inputenc}
\usepackage[colorinlistoftodos, color=green!40, prependcaption]{todonotes}
\usepackage{amsthm}
\usepackage{mathtools}
\usepackage{physics}
\usepackage{braket}
\usepackage{listings}
\usepackage{cancel}
\usepackage{graphicx}
\usepackage[left=18.5mm,right=15mm,top=30mm,columnsep=15pt]{geometry} 
\usepackage{adjustbox}
\usepackage{placeins}
\usepackage[T1]{fontenc}
\usepackage{lipsum}
\usepackage[numbers, sort&compress]{natbib}
\usepackage{amssymb}
\usepackage{bm}
\usepackage{xr}
\usepackage[ruled,vlined]{algorithm2e}
\usepackage{placeins}
\usepackage[normalem]{ulem}
\usepackage{float}
\usepackage{dcolumn}
\usepackage{bm}
\usepackage[font=normalsize]{caption}
\usepackage{subcaption}
\usepackage{diagbox}
\usepackage{multirow}
\newcolumntype{?}{!{\vrule width 1pt}}
\usepackage{makecell}

\begin{document}

\title{Clifford Deformed Compass Codes}

\author{Julie A. Campos}
\email{julie.campos@duke.edu}
\affiliation{Duke Quantum Center, Duke University, Durham, NC 27701, USA}
\affiliation{Department of Physics, Duke University, Durham, NC 27708, USA
}
\author{Kenneth R. Brown}
\email{ken.brown@duke.edu}
\affiliation{Duke Quantum Center, Duke University, Durham, NC 27701, USA}
\affiliation{Department of Physics, Duke University, Durham, NC 27708, USA
}
\affiliation{Department of Electrical and Computer Engineering, Duke University, Durham, NC 27708, USA}
\affiliation{Department of Chemistry, Duke University, Durham, NC 27708, USA}


\begin{abstract}
We can design efficient quantum error-correcting (QEC) codes by tailoring them to our choice of quantum architecture. Useful tools for constructing such codes include Clifford deformations and appropriate gauge fixings of compass codes. In this work, we find Clifford deformations that can be applied to elongated compass codes resulting in QEC codes with improved performance under noise models with errors biased towards dephasing commonly seen in quantum computing architectures. These Clifford deformations enhance decoder performance by introducing symmetries, while the stabilizers of compass codes can be selected to obtain more information on high-rate errors. As a result, the codes exhibit thresholds that increase with bias and lower logical error rates under both code capacity and phenomenological noise models. One of the Clifford deformations we explore yields QEC codes with better thresholds and logical error rates than those of the XZZX surface code at moderate biases under code capacity noise.
\end{abstract}

\maketitle

\section{\label{sec:intro} Introduction}
The advancement of quantum computers is limited by noise leading to errors in computation. One way to handle these errors is by using quantum error-correcting (QEC) \cite{calderbank1996good} codes to encode logical qubits in several physical qubits. By doing this, the logical error rate can be suppressed exponentially, enabling us to achieve fault-tolerant computation when the physical error rate is less than a threshold value \cite{knill1998resilient,
 aharonov1997fault}. 

While depolarizing noise is the most common choice of error model when evaluating the performance of a QEC code \cite{bombin2012universal, tomita2014low, bravyi2014efficient, sahay2022decoder}, it is not the most representative of real noise. The depolarizing noise model assumes that any Pauli error can occur with the same probability, but it is often the case that quantum systems exhibit a more structured noise model and in some cases qubits are engineered to experience biased noise. For example, superconducting cat qubits can be designed to have dominant Pauli-$X$ or Pauli-$Z$ errors \cite{grimm2020stabilization,lescanne2020exponential, berdou2023one, bocquet2024quantum}. Additionally, there are methods to engineer bias toward erasure errors in superconducting \cite{chou2024superconducting, kubica2023erasure, teoh2023dual}, neutral atom \cite{cong2022hardware_neutralatoms, wu2022erasure} and trapped ion qubits \cite{kang2023quantum}. In this work, we are primarily motivated by noise models with dominant dephasing errors which have been observed in trapped ion \cite{nigg2014quantum,ballance2016high}, spin \cite{taylor2005fault} and superconducting qubits \cite{aliferis2009fault, rosenblum2018fault}.

A benefit of having biased errors in quantum computing architectures is that we can design QEC codes that extract more information on dominant errors. The resulting codes can achieve high thresholds under biased noise models \cite{ tuckett2018ultrahigh, tuckett2019tailoring, stephens2013high,li20192d, bonilla2021xzzx, Sahay_Neutralatomswithbiasederasure,dua2022clifford, tiurev2023correcting}. Two examples of such codes are the XZZX surface code \cite{bonilla2021xzzx} and elongated compass codes \cite{li20192d, huang2020fault}, which are effective at detecting and correcting errors biased toward dephasing.

Elongated compass codes are a class of 2D compass codes \cite{li20192d} which result from a choice of fixed gauges \cite{paetznick2013universal,bombin2015gauge}. The stabilizers of elongated compass codes are determined by the elongation parameter, which dictates the amount of weight-2 $X$ stabilizers. As the number of these weight-2 $X$ stabilizers increases, the code can detect and correct more $Z$ errors. Compass codes have also been studied under noise models with coherent errors \cite{pato2024logical}. 

The XZZX surface code is equivalent to the surface code \cite{kitaev2003fault,dennis2002topological,fowler2012towards} up to the application of a Hadamard transformation on every other qubit. This simple modification to the surface code stabilizers introduces a symmetry that can provide extra information about the location of errors to the decoder (Figure \ref{fig:XZZX_l2}). This characteristic of the XZZX surface code leads it to have a threshold of 50\% under a noise model with infinite bias towards any Pauli error. 

The XZZX surface code is an example of a Clifford deformation of the surface code \cite{dua2022clifford,tiurev2023correcting}. The Clifford deformation of a stabilizer code refers to the modification of the stabilizers through the application of a set of single-qubit Clifford operators. There have been extensive studies on the application of this procedure to surface codes \cite{tuckett2018ultrahigh, bonilla2021xzzx, tiurev2023correcting, dua2022clifford} and similar procedures have been developed for color codes \cite{san2023cellular, tiurev2024domain} and Floquet codes \cite{setiawan2025tailoring}. Clifford deformed codes have been implemented in experiments. A Pauli deformed Shor code was shown to improve quantum memories in a logical qubit in trapped ions  \cite{debroy2021optimizing}. The XZZX surface code has been implemented experimentally in a superconducting qubit platform \cite{google2023suppressing}. 

In this work, we explore sets of Clifford deformations that add structure to the stabilizers of the elongated compass codes leading to improved thresholds and logical error rates under biased noise models. To preserve the advantage of the elongated compass codes, we consider two sets of Clifford deformations we call the XZZX${\square}$ and the ZXXZ${\square}$ deformations (Figure \ref{fig:Deformations_l6}). These deformations are chosen to preserve the weight-2 $X$ stabilizers of the elongated compass codes while introducing a symmetry that restricts the spread of defects. We present thresholds of these codes under code capacity and phenomenological noise models.

The paper is structured as follows. The noise model is defined in Section \ref{sec:noisemodel}. Compass codes and elongated compass codes are described in more detail in Section \ref{sec:ElongatedCompassCodes}.
We introduce the Clifford deformations we apply to the elongated compass codes in Section \ref{sec:XZZXDefCompasscodes}. 
In Section \ref{sec:decoder}, we give a brief description of the minimum-weight perfect matching (MWPM) algorithm as our decoder. In this section, we also discuss how Clifford deformations affect the decoder graphs. We describe the process we followed to determine thresholds in Section \ref{sec:methods}. Thresholds and logical error rate comparisons are reported and discussed in Section \ref{sec:results}. Concluding remarks are in Section \ref{sec:conclusions}. In the Appendix, we include additional decoder graphs (Appendix \ref{appendix:graphs}) and threshold plots (Appendix \ref{appendix:Thresholds}).)

\begin{figure}[h!] 
\centering
\includegraphics[width=0.9\linewidth]{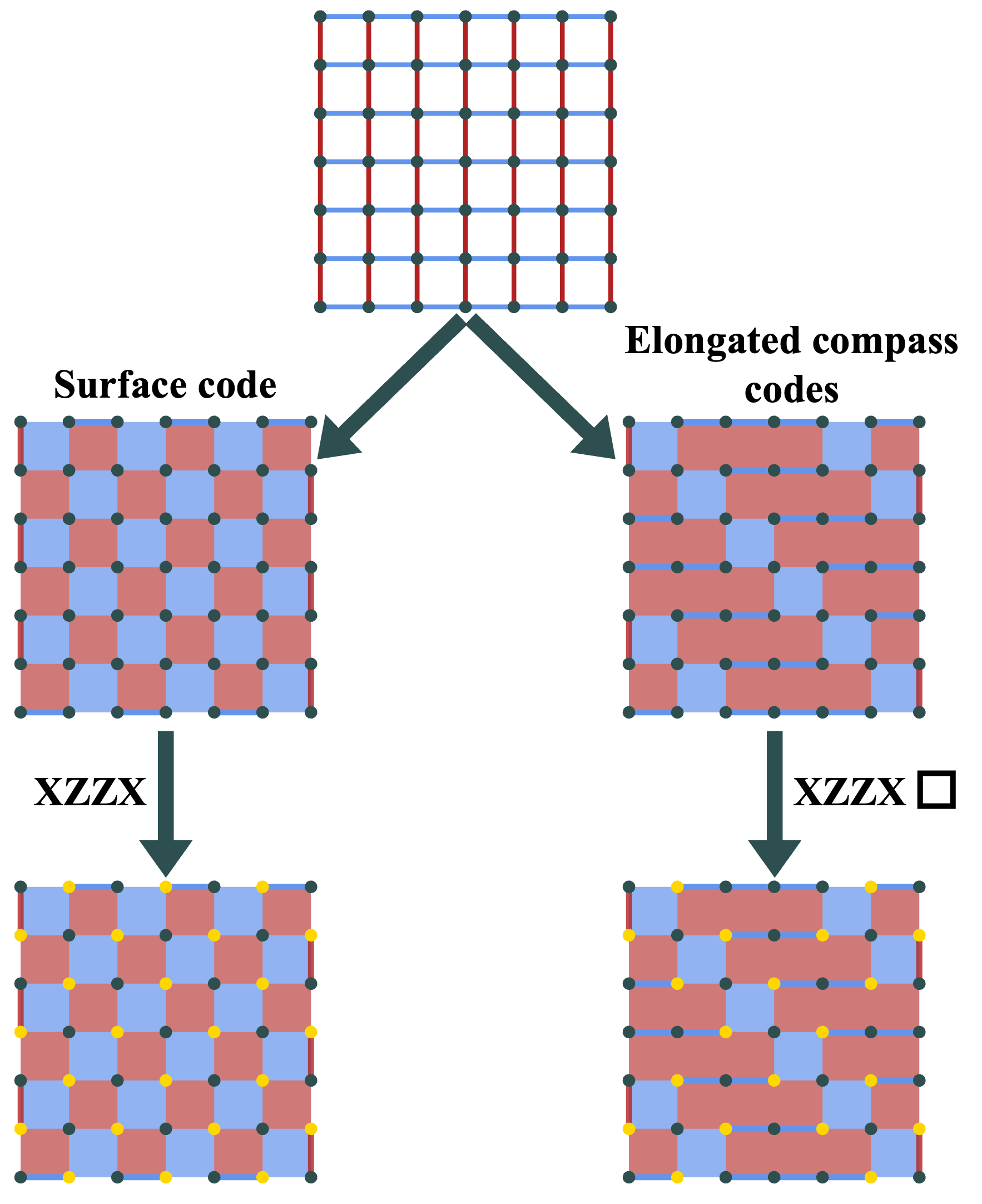}
    \caption{Starting with weight-2 gauge operators ($Z$ in red, $X$ in blue, qubits are gray dots) corresponding to the interaction terms of the quantum compass model Hamiltonian, we construct stabilizer codes through gauge fixing. The surface code and elongated compass codes (here with $\ell = 4$) are examples of such codes. By applying a Clifford deformation on the surface code, we obtain the XZZX surface code which has improved performance under biased noise models. Yellow dots indicate qubits that undergo a Hadamard transformation. The elongated compass codes perform better than the surface code under biased noise models, but can be improved further by applying the XZZX$\square$ Clifford deformation introduced in this work.}
    \label{fig:main_story}
\end{figure}

\section{Codes and Noise Model}

\subsection{\label{sec:noisemodel} Noise Model}

We consider a single-qubit Pauli noise channel where all qubits can experience a Pauli error with probability $p = p_x + p_y + p_z$. Here, $p_x, p_y$ and $p_z$ correspond to the probabilities of $X$, $Y$ and $Z$ errors respectively. These errors occur independently and uniformly across the lattice. The noise channel is expressed in the following way: 

\begin{equation}
\mathcal{E}[\rho] = (1-p)\rho + p_x X\rho X +p_y Y\rho Y + p_z Z\rho Z 
\label{eq:NoiseModel}
\end{equation}

We assume that the errors are biased towards dephasing which, in Pauli representation of the noise channel, is a bias towards Pauli-$Z$ errors. This bias is quantified by $\eta = \frac{p_z}{p_x+p_y}$. For simplicity, we assume $p_x = p_y$. We obtain the depolarizing channel when $\eta =0.5$. In several quantum architectures, $\eta$ can reach values as high as $10^2$ \cite{lescanne2020exponential, grimm2020stabilization}. Motivated by this, we evaluate our codes under biased noise models with $0.5 \leq \eta \leq 100$.

A key concern with the high-weight stabilizers of elongated compass codes is that they require deeper syndrome extraction circuits. As a result, it would be ideal to analyze the performance of the codes under circuit-level noise. However, this requires considerations of gate schedules, parallelization, and bias-preserving gates, which we leave for future work. Instead, we account for the resulting increase in error rates by studying the codes under a weighted phenomenological noise model. In this noise model, we include measurement error rates that scale with the weight of the stabilizers, in addition to the memory errors described above. Furthermore, we normalize the measurement errors so that they match the typical phenomenological noise model on the surface code at standard depolarizing noise. Specifically, the probability of measurement error $p_m$ is $w(p_y+p_z)/4$ where $w\geq 4$ is the weight of the stabilizer. For weight-2 stabilizers, we set $p_m = p_y+p_z$. This measurement noise model extends the noise model presented for the XZZX code \cite{bonilla2021xzzx}.

\subsection{\label{sec:ElongatedCompassCodes} Elongated Compass Codes}

Compass codes are subsystem stabilizer codes \cite{kribs2005operator, poulin2005stabilizer} whose stabilizer group $\mathcal{S}$ \cite{gottesman1997stabilizer} results from a choice of gauge fixes \cite{paetznick2013universal, bombin2015gauge}. The complete gauge group from which we start is generated by the interaction terms of the 2D quantum compass model Hamiltonian \cite{kugel1973crystal, dorier2005quantum, douccot2005protected} on a square lattice where qubits are on the vertices (Figure \ref{fig:main_story}).

One can go between distinct compass codes by fixing a different set of gauges \cite{paetznick2013universal,bombin2015gauge}. Well-known compass codes include the Bacon-Shor code \cite{shor1995scheme, bacon2006operator} and the surface code \cite{kitaev2003fault, dennis2002topological}. Here, we focus on the elongated compass codes which are appropriate for correcting dominant $Z$ errors \cite{li20192d}. Elongated compass codes are classified according to an elongation parameter $\ell$. To illustrate the process of gauge fixing to obtain elongated compass codes, we label the coordinates of the plaquettes on the square lattice $(i,j)$ where the origin is at the top left. Then, we fix the product of the $X$ gauges that are supported by qubits on the plaquettes with $ i-j \equiv 0 \mod\ell$, creating weight-4 $X$ stabilizers. In each row, we fix the product of $Z$ gauges between the $X$ stabilizers we fixed. The resulting $Z$ stabilizers are rectangles of length $\ell -1$ and weight 2$\ell$. Finally, we fix all of the remaining weight-2 $X$ gauges surrounding the $Z$ stabilizer rectangles, ensuring commutativity of all stabilizers. See Figure \ref{fig:main_story} for a depiction of an elongated compass code with $\ell = 4$. An elongated compass code with $\ell=2$ is the rotated surface code (Figure \ref{fig:main_story}). Note that elongated compass codes are Calderbank-Shor-Steane (CSS) codes since their stabilizers are either a product of only Pauli-$X$ or only Pauli-$Z$ \cite{calderbank1996good, steane1996multiple}.

As the elongation parameter grows, the $X$ and $Z$ stabilizers become more asymmetric. The increasing weight of the $Z$ stabilizers makes them less informative about the location of $X$ errors, but by gaining more weight-2 $X$ stabilizers, we obtain more information on the location of $Z$ errors. This leads to a trade-off in the $X$ and $Z$ decoding performance as the bias increases.
A consequence of this trade-off is that there is an optimal bias at which elongated compass codes achieve a maximum threshold \cite{li20192d}. The optimal bias balances the performance of the $X$ and $Z$ decoders for depolarizing noise. The optimal biases ($\eta_{\ell}^{opt}$) found in \cite{li20192d} for elongated compass code with $\ell = 2,3,4,5,6$ are $\eta_{2}^{opt} = 0.5$, $\eta_{3}^{opt} = 1.67$, $\eta_{4}^{opt} = 3.0$, $\eta_{5}^{opt} = 4.26$, and $\eta_{6}^{opt} = 5.89$. The maximum threshold reached at the optimal bias increases with elongation parameter making higher elongations desirable for higher biases. 

\subsection{\label{sec:XZZXDefCompasscodes} Clifford Deformations}
Clifford deformations are modifications of stabilizer codes that can lead to significant improvements in the thresholds of codes under biased noise models \cite{tuckett2018ultrahigh, bonilla2021xzzx, dua2022clifford, tiurev2023correcting, tiurev2024domain}.  
The Clifford deformation of a stabilizer code is the application of an arbitrary set of single-qubit unitary transformations from the Clifford group on the codespace yielding a new stabilizer group. Clifford transformations map Pauli operators to other Pauli operators and thus preserve the commutativity of the stabilizers. Additionally, a Clifford deformation preserves the weight and support qubits of each stabilizer. However, one possible consequence of a Clifford deformation is that the resulting code may be non-CSS since the operators making up the stabilizers are modified. For example, the XZZX surface code is not a CSS code. As a result, we cannot directly decode $X$ and $Z$ syndromes independently, as is standard with CSS codes. We discuss our decoding methods in Section \ref{sec:decoder}.

\begin{figure}[h!]
\centering
\begin{subfigure}{0.45\columnwidth}
\caption{}
\includegraphics[width=\linewidth]{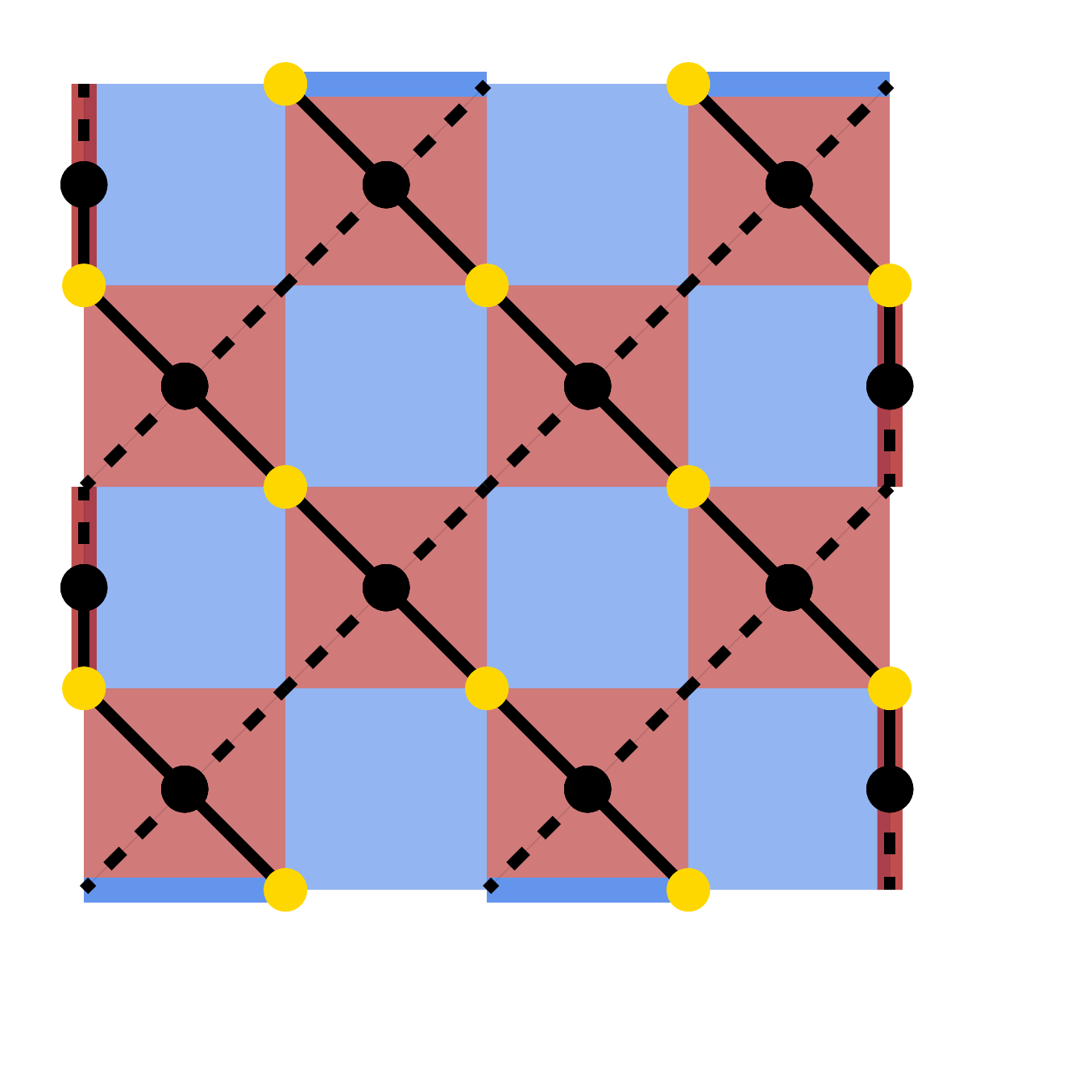}
    \label{fig:l2_XZZX_Zstab}
\end{subfigure}
\hfill
\begin{subfigure}{0.45\columnwidth}
\caption{}
\includegraphics[width=\linewidth]{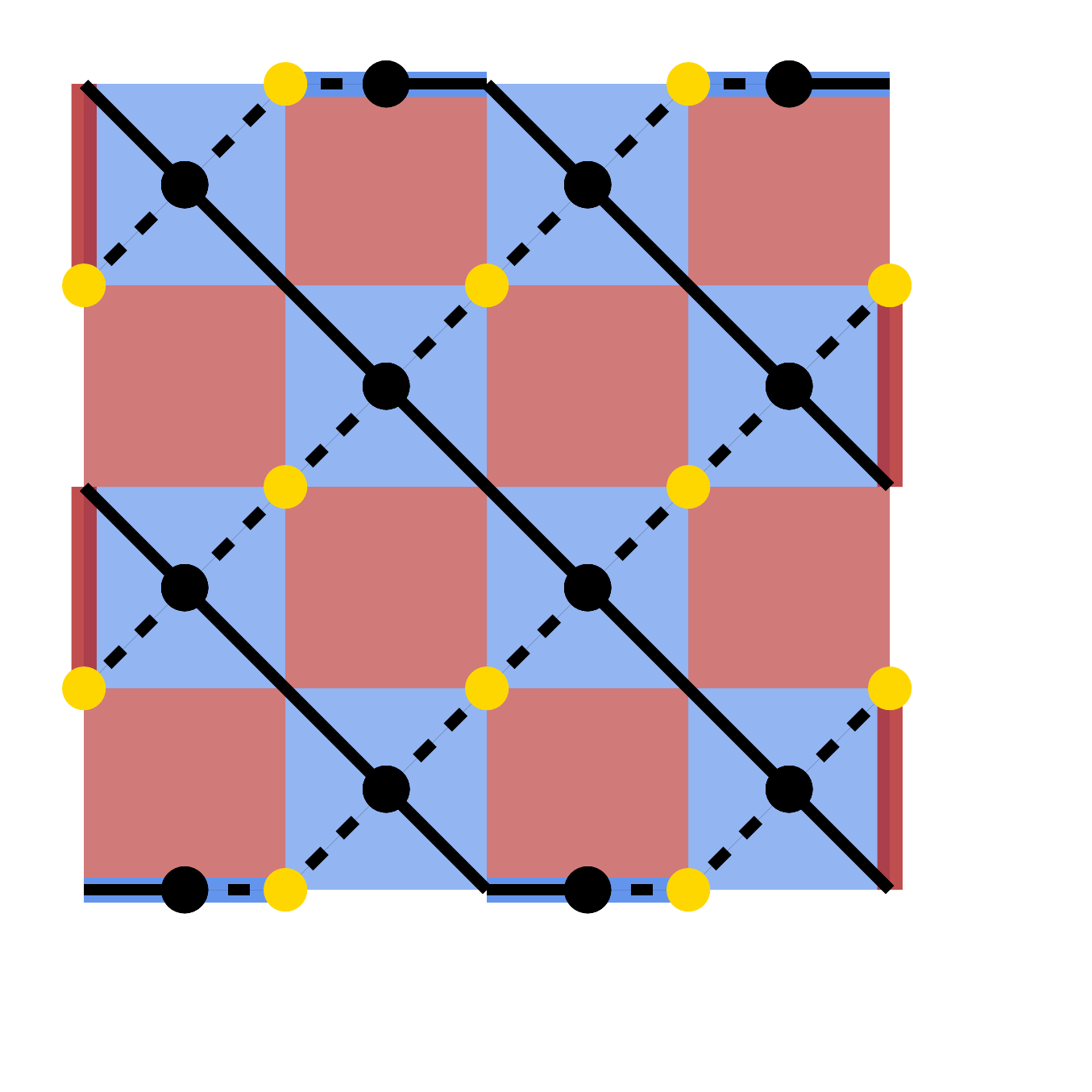}
    \label{fig:l2_XZZX_Xstab}
\end{subfigure}
\hfill
\begin{subfigure}{0.45\columnwidth}
    \caption{}    \includegraphics[width=\linewidth]{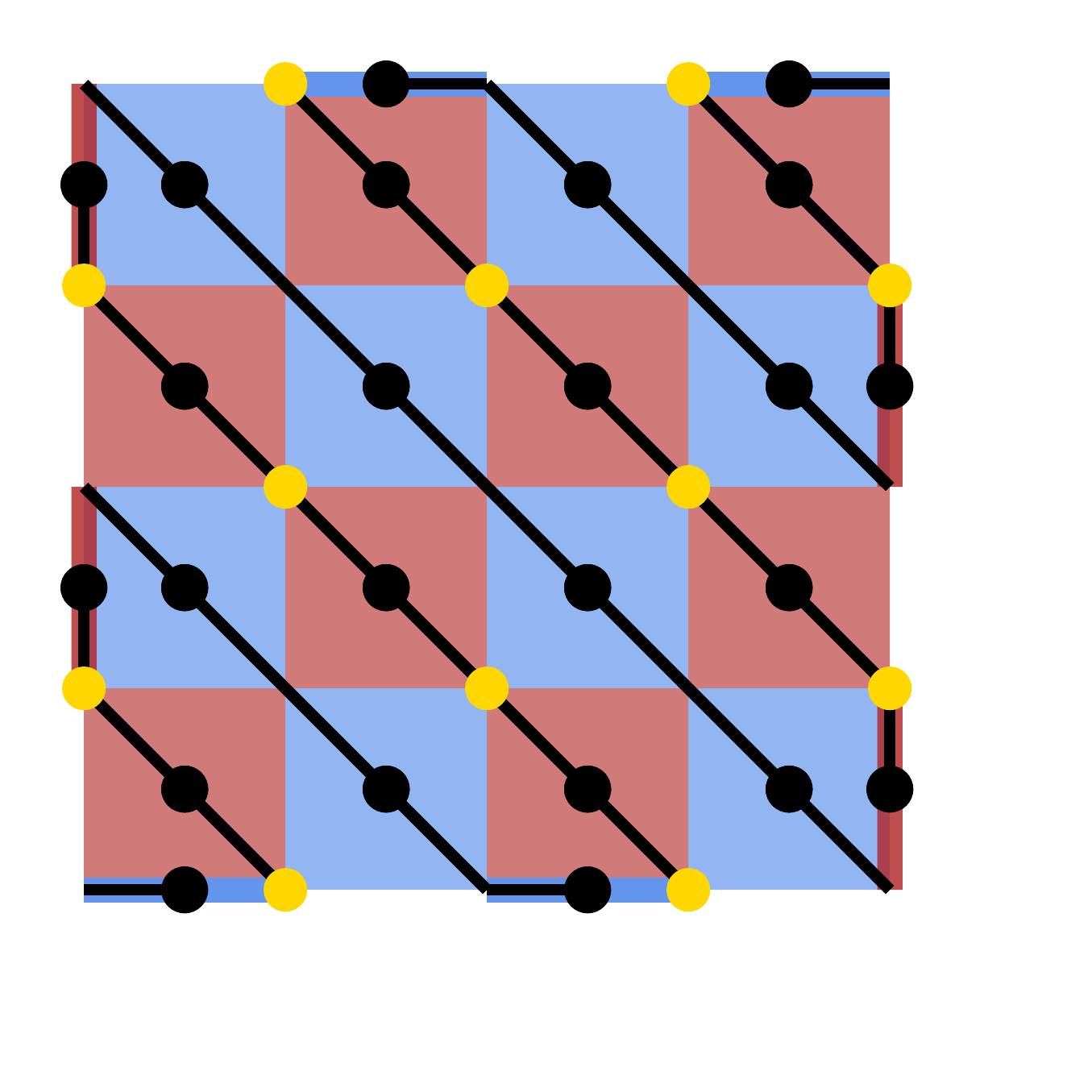}
    \label{fig:combinedlow_l2}
\end{subfigure}
\hfill
\begin{subfigure}{0.45\columnwidth}
    \caption{}
\includegraphics[width=\linewidth]{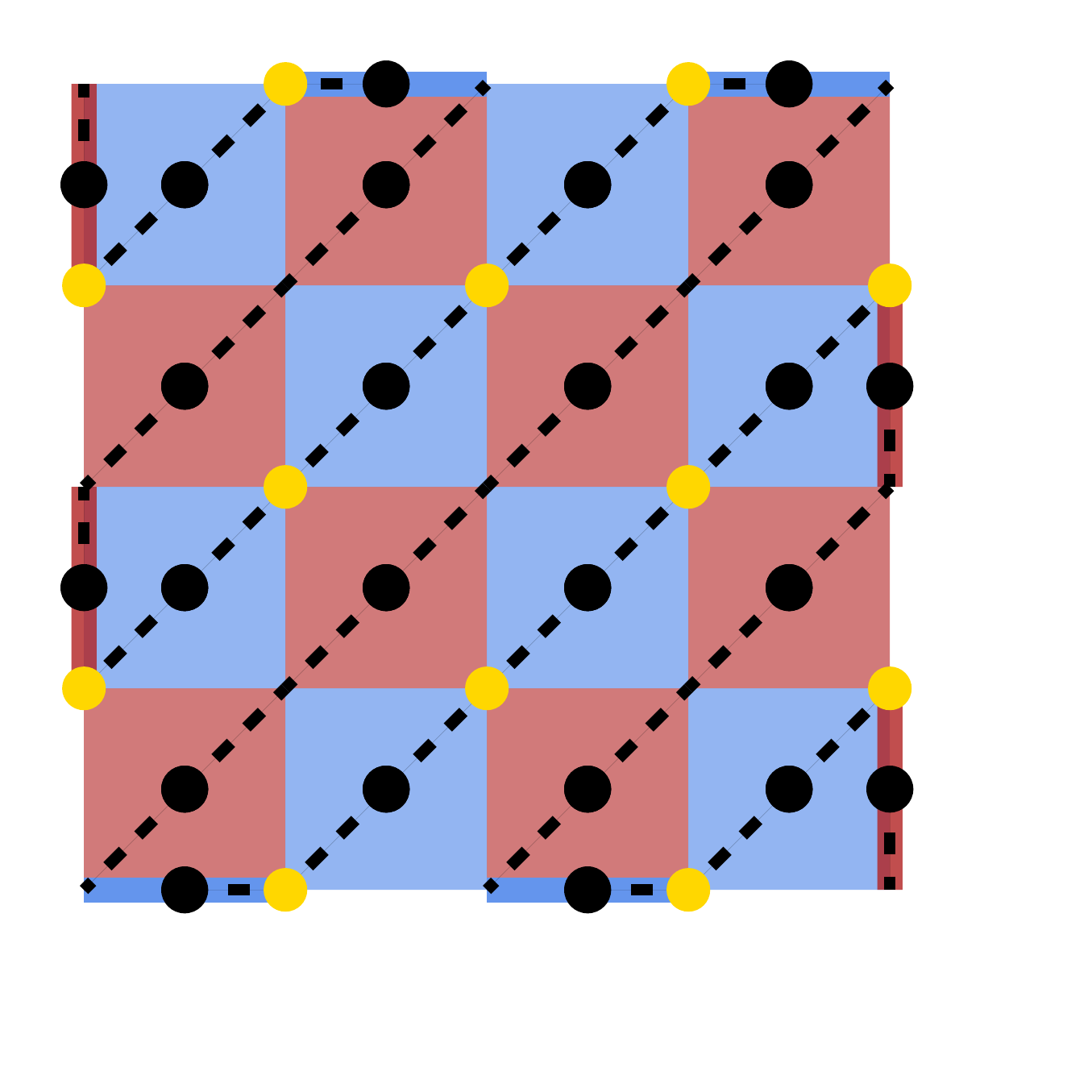}
    \label{fig:combinedhigh_l2}
\end{subfigure}
\caption{\textbf{XZZX surface code} Decoder graphs for \textbf{a}) $Z$ stabilizers (in red) and \textbf{b}) $X$ stabilizers (in blue) of the surface code (elongated compass code with $\ell = 2$). Qubits lie on the vertices of the lattice and those with yellow dots undergo a Hadamard transformation according to the XZZX deformation. Black dots are vertices of the decoder graph and correspond to stabilizers. Decoder graph edges travel across the qubit locations. Solid edges have a low weight (high error probability) and dashed edges have a high weight (low error probability). These graphs are more relevant at low biases since the distinction between the solid and dashed lines is less significant in our decoding method. The combination of the low(high)-weight edges from \textbf{a} and \textbf{b} results in the matching graph shown in \textbf{c}(\textbf{d}). As the bias increases, the decoding graph \textbf{c} dominates the decoding procedure.}
\label{fig:XZZX_l2}
\end{figure}

The XZZX surface code \cite{bonilla2021xzzx} results from a Clifford deformation of the surface code where a Hadamard transformation is applied on every other qubit of the lattice (Figure \ref{fig:XZZX_l2}). All plaquette stabilizers acquire the form XZZX, giving the code its name. This change introduces a symmetry that restricts the propagation of defects to one dimension. Regardless of their location, $X$ or $Z$ errors will produce defects aligned in a particular direction. Furthermore, these directions are perpendicular to each other. This is illustrated in Figures \ref{fig:combinedlow_l2} and \ref{fig:combinedhigh_l2} which depict the directions in which $Z$ and $X$ defects spread respectively. Since defects are restricted to one dimension, a pair of defects aligned in one of these directions will be the endpoints of a string of errors of the same type. This allows us to decode Pauli-$X$ and Pauli-$Z$ errors as disjoint sets of repetition codes under noise models with infinite bias.

The XZZX surface code outperforms the CSS surface code under all biased Pauli noise models and even surpasses the hashing bound for some biases \cite{bonilla2021xzzx}. We can understand the improvement over the CSS surface code by noting that the surface code stabilizers gather more information about $Y$ errors than $X$ and $Z$ errors. In general, this is a consequence of the fact that all surface code stabilizers are sensitive to $Y$ errors, giving us more syndrome bits. As a result, the surface code does well under biased noise models only if the bias is towards $Y$ errors \cite{tuckett2018ultrahigh,tuckett2019tailoring}. In contrast, the additional symmetries of the XZZX surface code provide additional information on $X$ and $Z$ errors to the decoder, making the code efficient in the case of any Pauli bias. 

We could apply the XZZX deformation to elongated compass codes in the same way it was applied to the surface code to get the XZZX surface code. That is, we could apply a Hadamard transformation on every other qubit. However, this is not ideal for elongated compass codes because it will change the weight-2 $X$ stabilizers, removing the advantage of the elongated compass codes. Instead, we consider a similar Clifford deformation that applies a Hadamard transformation to the top right and bottom left qubits supporting weight-4 $X$ stabilizers (Figures \ref{fig:XZZXonSq_l6_low} - \ref{fig:XZZXonSq_l6_high}). After doing this, the weight-4 $X$ stabilizers become of the form XZZX. We refer to the resulting codes as the XZZX$\square$-deformed compass codes.

Another deformation we consider is the ZXXZ$\square$ deformation. This deformation applies Hadamard transformations on the top left and bottom right qubits of the weight-4 $X$ stabilizers (Figures \ref{fig:ZXXZonSq_l6_low} - \ref{fig:ZXXZonSq_l6_high}). The ZXXZ$\square$ deformation only changes the weight-2 $X$ stabilizers in the top and bottom rows of the code while the XZZX$\square$ deformation affects all rows. In the case of $\ell = 2$, the ZXXZ$\square$ deformation is equivalent to the XZZX and XZZX$\square$ deformations since it only switches the directions in which the low and high-weight edges are aligned (Figure \ref{fig:XZZX_l2}). 

\section{\label{sec:decoder} Decoder}
The decoder determines a correction based on the measured syndrome. For the codes we consider, an efficient and sufficiently accurate decoding algorithm is the minimum-weight perfect matching (MWPM) decoder, which we implement using PyMatching \cite{edmonds1965paths, higgott2022pymatching}.

The input of the MWPM algorithm is a weighted graph $\mathcal{G} = (\mathcal{V}, \mathcal{E}, \mathcal{W})$ where $\mathcal{V} = \{v_i\}, \mathcal{E} = \{e_{ij}\} = \{(v_i, v_j)\}$ and $\mathcal{W} = \{w_{ij}\}$ are sets of vertices, edges and weights respectively. The vertices of the graph correspond to stabilizers, the edges correspond to qubits, and the weights of each edge are a logarithmic function of the probability of error of the qubit it corresponds to ($w_{ij} = \log(\frac{1-p_{ij}}{p_{ij}})$). A matching $M$ is a subset of disjoint edges in $E$. A perfect matching is a matching such that $\forall v \in \mathcal{V}, \exists e \in M$ s.t. $v\in e$. Thus, the MWPM of a graph is a perfect matching that minimizes the sum of the weights in the matching. The output is a set of edges that correspond to the most probable set of errors. 

In the case of a CSS code, $X$ and $Z$ syndromes can be decoded independently by running the MPWM algorithm on $X$ and $Z$ decoder graphs. This simplifies decoding since we are dividing the decoding problem into two. The XZZX$\square$ and ZXXZ$\square$- deformed codes we consider are not CSS codes, which means that we cannot decode them in this way directly. However, we can get around this because our decoding problem is equivalent to that of decoding a CSS code under an inhomogeneous noise model.

\begin{figure}[h!]
    \centering
    \begin{subfigure}{0.45\linewidth}
    \caption{}
        \includegraphics[width = \linewidth]{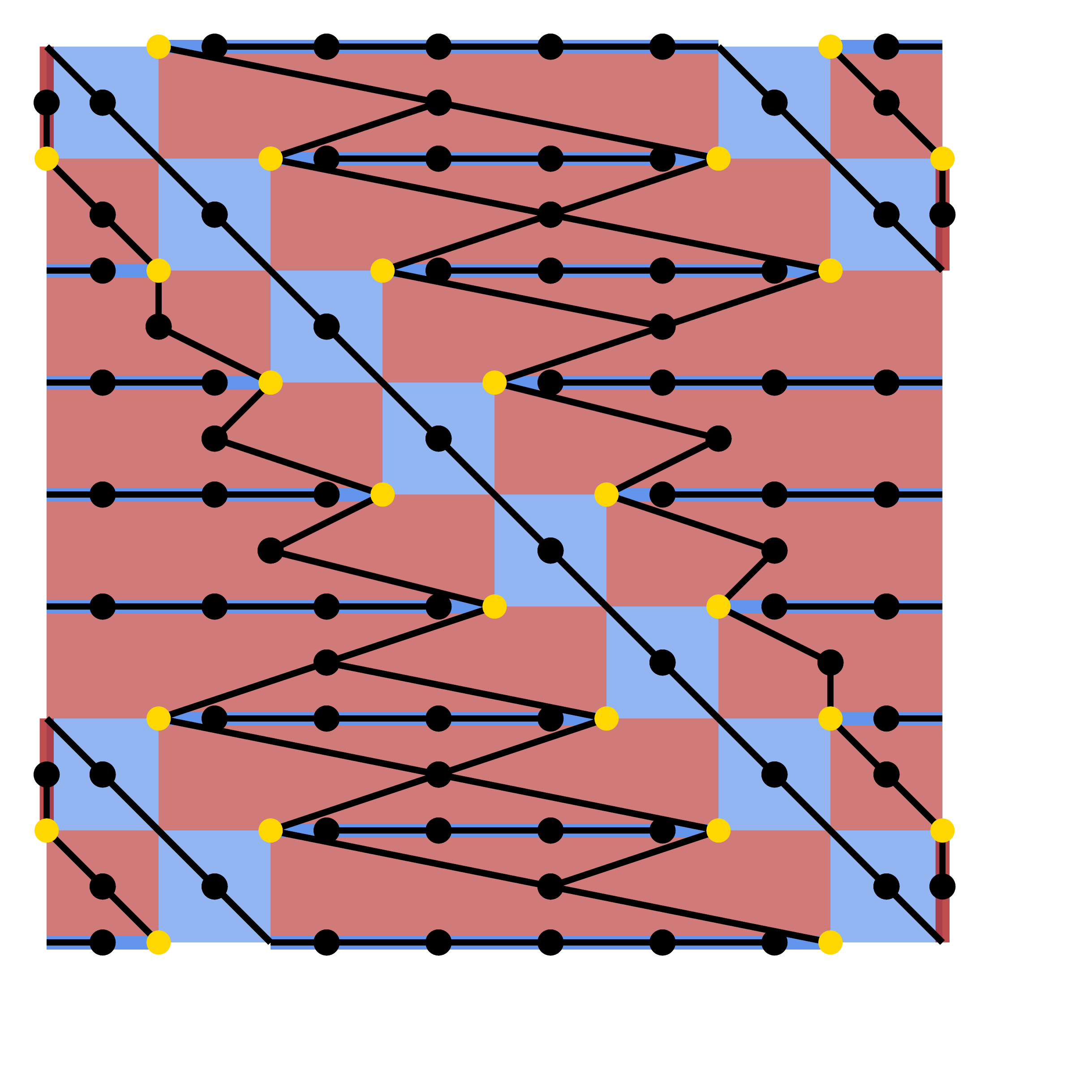}
        \label{fig:XZZXonSq_l6_low}
    \end{subfigure}
    \hfill
    \begin{subfigure}{0.45\linewidth}
    \caption{}
        \includegraphics[width = \linewidth]{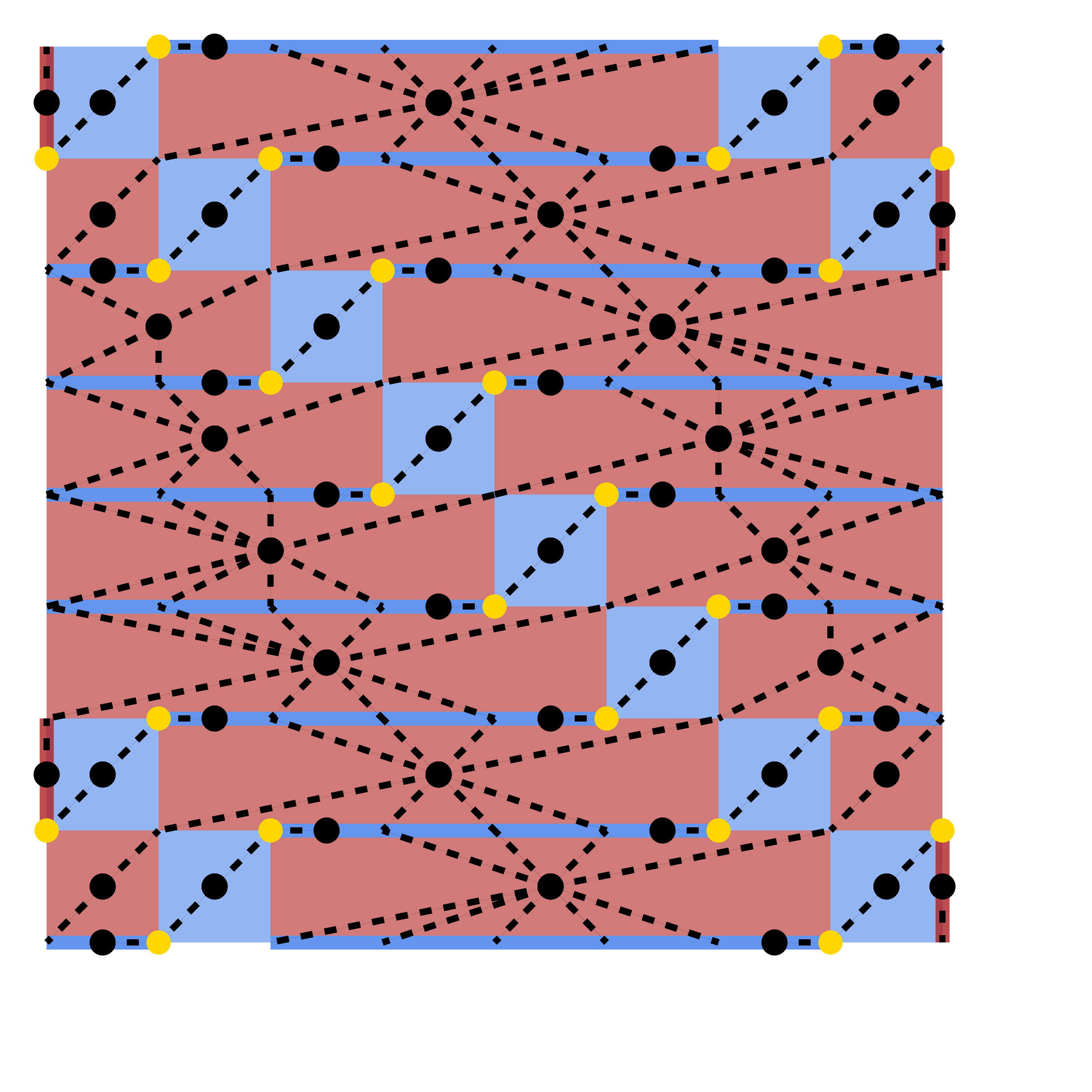}
        \label{fig:XZZXonSq_l6_high}
    \end{subfigure}
        \begin{subfigure}{0.45\linewidth}
    \caption{}
        \includegraphics[width = \linewidth]{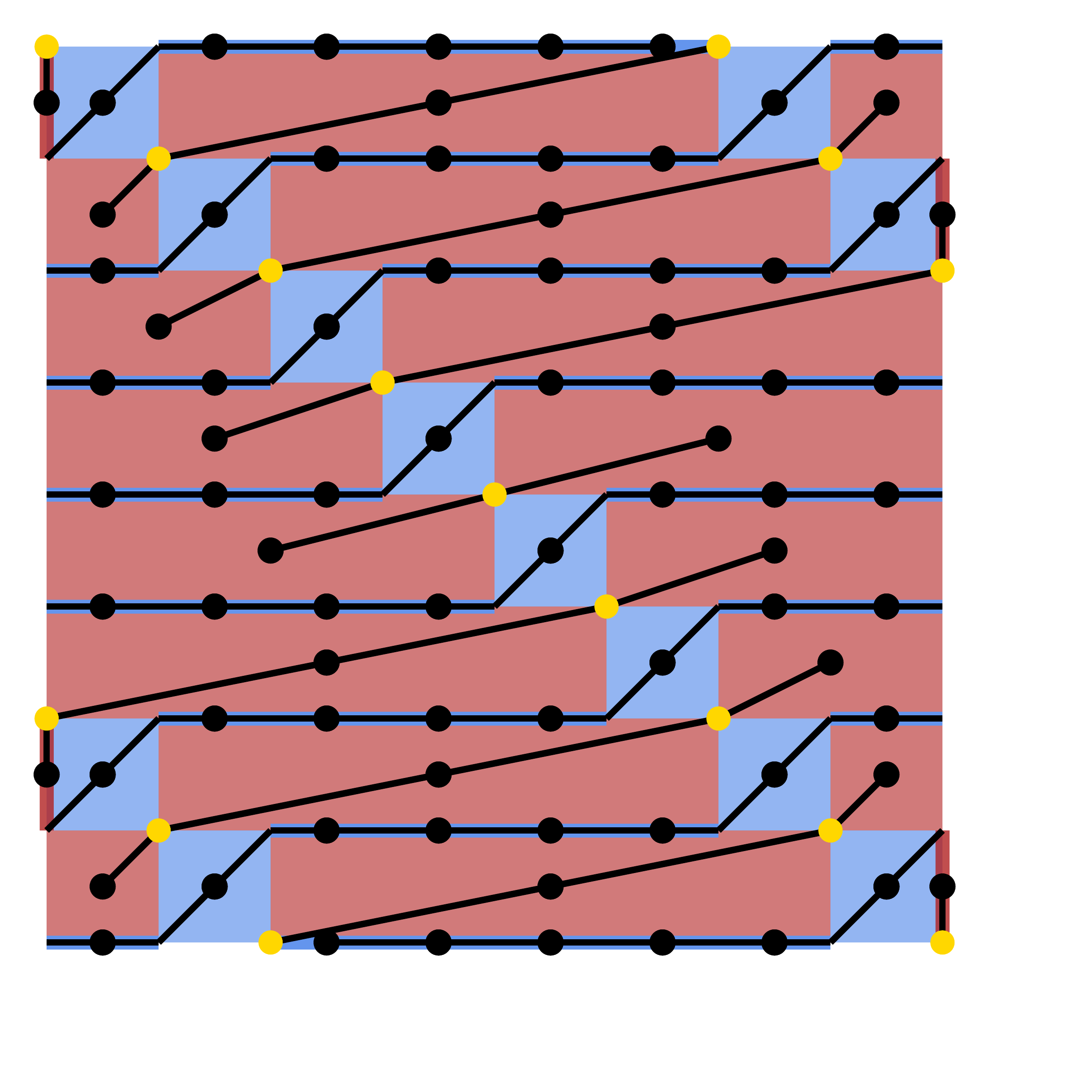}
        \label{fig:ZXXZonSq_l6_low}
    \end{subfigure}
    \hfill
    \begin{subfigure}{0.45\linewidth}
    \caption{}
        \includegraphics[width = \linewidth]{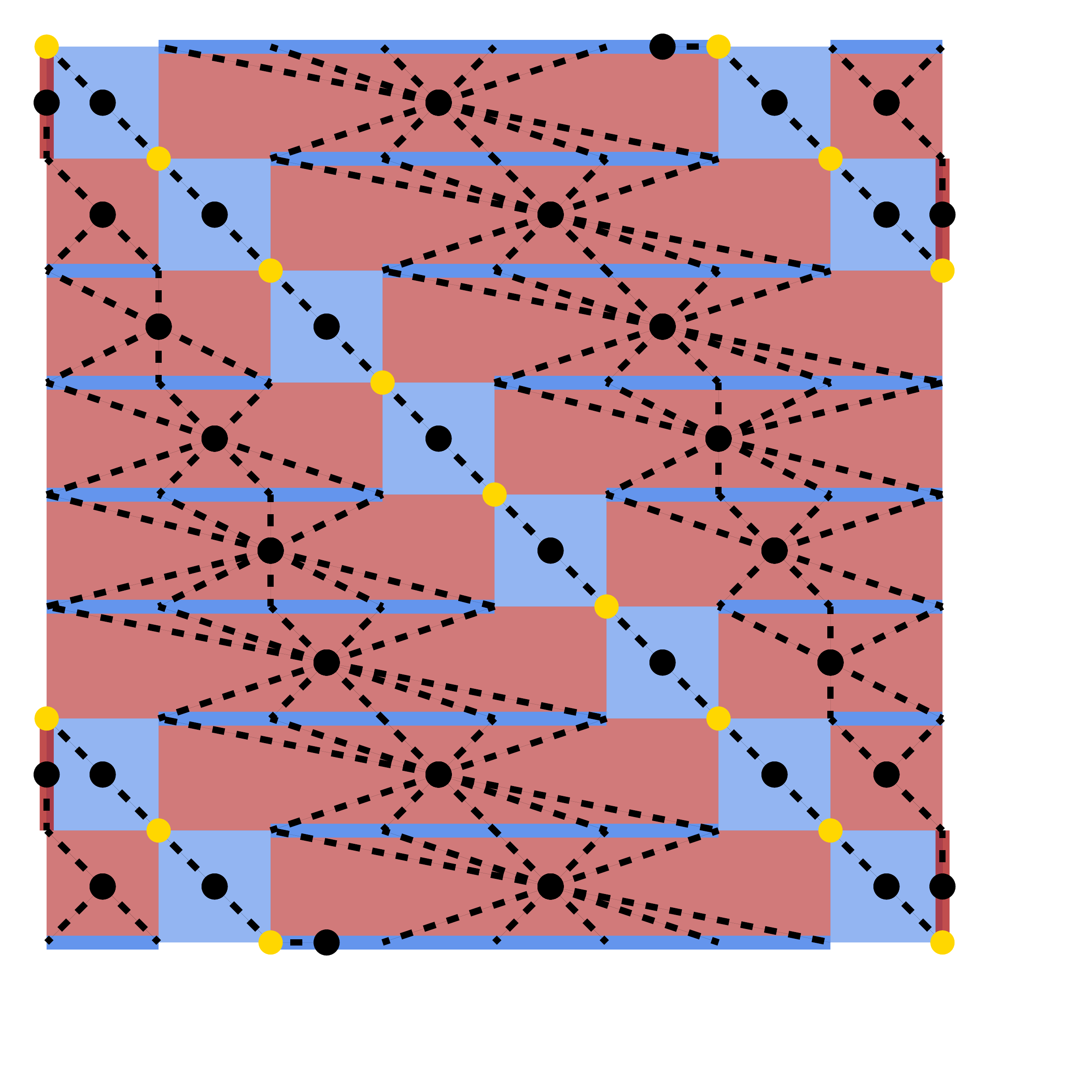}
        \label{fig:ZXXZonSq_l6_high}
    \end{subfigure}
    \caption{\textbf{XZZX$\square$ and ZXXZ$\square$ deformations} Graphs of low/high-weight edges of $\ell = 6$ elongated compass code with XZZX$\square$ deformation (\textbf{a}/\textbf{b}) and ZXXZ$\square$ deformation (\textbf{c}/\textbf{d}). Black dots represent stabilizers and yellow dots indicate the qubits that undergo a Hadamard transformation according to the deformation. \textbf{a}) The low-weight edges divide the lattice into disjoint regions. This restricts the spread of syndromes due to high-rate errors. \textbf{b}) The high-weight edges create a highly connected graph, allowing defects to spread across the lattice. \textbf{c}) In the case of the ZXXZ$\square$ deformation, the graph of low-weight edges is composed of disjoint strings that are easy to decode. \textbf{d}) The high-weight graph is divided into disjoint regions which restrict the spread of defects. Note that these graphs are easier to decode compared to the low-weight and high-weight graphs of the XZZX$\square$-deformed compass codes.}
    \label{fig:Deformations_l6}
\end{figure}

Clifford deformations do not change the location or support qubits of the stabilizers. As a result, a syndrome found on both the deformed and undeformed codes is caused by errors that are equivalent up to the Clifford transformations. We can understand this as follows. Suppose that $\ket{\Bar{\psi}}$ is the logical state of an elongated compass code and $\ket{\Bar{\psi'}}$ the logical state of a deformed version of the code. Then, $\ket{\Bar{\psi'}} = U\ket{\Bar{\psi}}$ where $U = \Pi_{i\in\mathcal{C}} H_i$ and $\mathcal{C}$ is the set of qubits that undergo a Hadamard transformation according to the Clifford deformation. Equivalently, $X (Z)$ errors on qubits that undergo a Hadamard transformation would satisfy $X \ket{\psi'} = H(Z\ket{\psi}) (Z \ket{\psi'} = H(X\ket{\psi}) )$ with probability $p_x (p_z)$. This highlights the fact that $X (Z)$ errors occurring on the deformed qubits with probability $p_x (p_z)$ translate to $Z (X)$ errors on the undeformed code with probability $p_x (p_z)$. Thus, decoding syndromes on the deformed code would be equivalent to decoding syndromes on the undeformed code under the following noise model:

\begin{align}\label{eq:newnoisemodel}
    p_{x,q} = \begin{dcases}
        p_x & U_q = I\\
        p_z & U_q = H 
    \end{dcases} 
    \qquad
    p_{z,q} = \begin{dcases}
        p_z & U_q = I\\
        p_x & U_q = H 
    \end{dcases} 
\end{align}
 
We decode the syndromes on the deformed code using the CSS decoder graphs of the undeformed code with weights modified according to the inhomogeneous noise model (see Eq. \ref{eq:newnoisemodel}). After decoding, we apply the Clifford transformations to the recovery operators to obtain the appropriate correction. 

We can see the effect of bias on our decoder by looking at the weights of the edges in our decoder graphs. The input to our decoder will be the $X$ and $Z$ decoder graphs of the CSS code. When $\eta = 0.5$, all probabilities of error are the same so the edges have the same weight. However, as the bias increases, the edges of the $X$($Z$) decoder graphs will have edges with weights determined by $p_z$($p_x$) corresponding to qubits that do not undergo a Clifford deformation and $p_x$($p_z$) for qubits that do. Using this, we can classify the edges on the $X$ and $Z$ decoder graphs as either having high weight (low probability of error) or low weight (high probability of error). 
 
In Figure \ref{fig:XZZX_l2}, we demonstrate how we classify the edges of the XZZX surface code. We start with the decoder graphs for the $X$ and $Z$ stabilizers of the surface code in Figures \ref{fig:l2_XZZX_Zstab} and \ref{fig:l2_XZZX_Xstab}. We distinguish between low and high weight edges by making them solid or dashed respectively. The low-weight edges from the $X$ and $Z$ matching graphs are combined in Figure \ref{fig:combinedlow_l2} to create a graph with only low-weight edges, and high-weight edges are combined in Figure \ref{fig:combinedhigh_l2}. We can use the same procedure to create high-weight and low-weight graphs for the deformed elongated compass codes. We show the resulting graphs in Figures \ref{fig:Deformations_l6}, \ref{fig:graphs_l3} and \ref{fig:graphs_l5}. 

We can see the structure that the Clifford deformations add to the codes in the low-weight and high-weight graphs. In the case of the XZZX surface code, the low-weight edges form parallel lines. The high-weight edges also form parallel lines, but these are in a direction perpendicular to the low-weight edges (Figure \ref{fig:XZZX_l2}). These figures illustrate why the XZZX surface code can be decoded as a set of disjoint repetition codes at infinite bias. 

The low-weight graphs of the XZZX$\square$-deformed compass codes are divided into sections by edges forming diagonal lines similar to those in the graphs of the XZZX surface code (Figure \ref{fig:XZZXonSq_l6_low}). A consequence of this is that the spread of syndromes due to high-rate errors is restricted to a particular section. These sections are not all one-dimensional as in the case of the XZZX surface code, but they will help the decoder correct the high-rate $Z$ errors in comparison to the CSS compass codes. The XZZX$\square$ deformation preserves many weight-2 $X$ stabilizers which gather more information on the high-rate $Z$ errors. We can see that this appears in Figure \ref{fig:XZZXonSq_l6_low} as repetition codes enclosed by diamonds. Thus, the vertices of the low-weight graph have degree of at most 4. The trade-off here is that the degree of the vertices in the high-weight graphs can be large (Figure \ref{fig:XZZXonSq_l6_high}). In general, the high-weight decoder graphs are non-local so the defects due to high-weight errors can spread across the entire lattice. As a result, the decoder will have a harder time decoding the low-rate errors ($X$ errors). 

The low-weight and high-weight decoder graphs corresponding to the ZXXZ$\square$ deformed compass codes are shown in Figures \ref{fig:ZXXZonSq_l6_low}-\ref{fig:ZXXZonSq_l6_high}. We observe that the low-weight graph is composed of disjoint strings which is desirable for the decoder. Additionally, it is useful to note the connectivity of the high-weight graphs is similar to that of the low-weight graphs of the XZZX$\square$ deformations. That is, the high-weight graphs are partitioned. This makes the ZXXZ$\square$-deformed compass codes more competitive at modest biases. 

\begin{figure*}[htbp!]
\includegraphics[width = 0.99\linewidth]{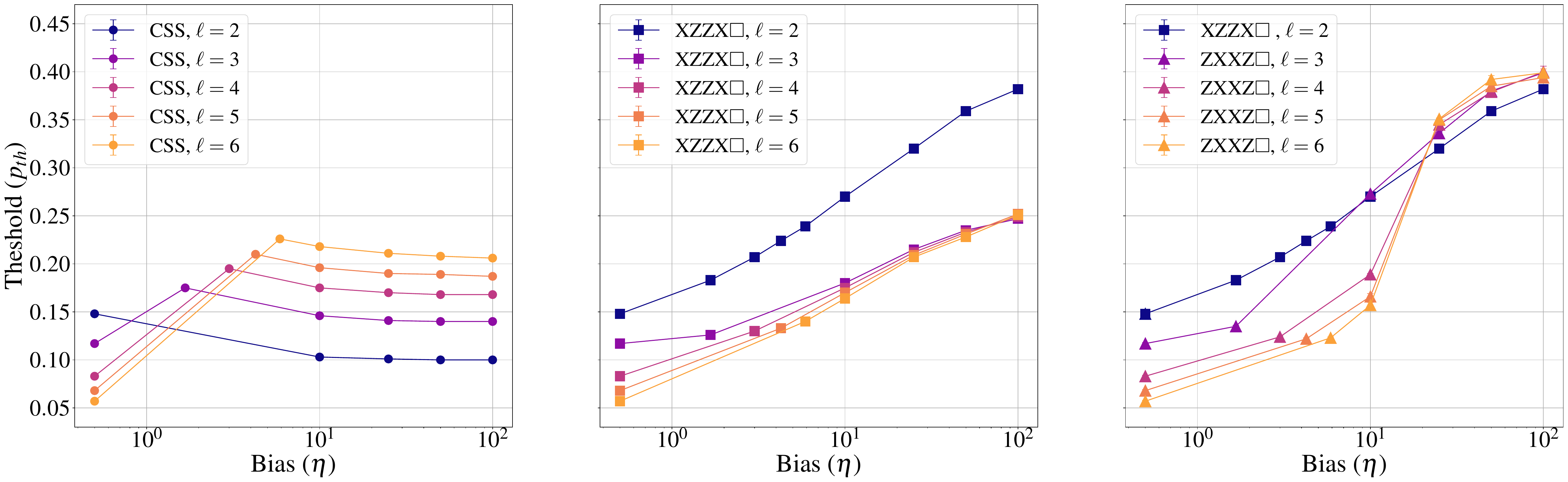}
    \caption{Thresholds for compass codes without deformation (CSS), XZZX$\square$-deformed and ZXXZ$\square$-deformed elongated compass codes going from left to right respectively. Thresholds are reported for codes with elongation parameters $\ell = 2,3,4,5,6$ under noise models with bias $\eta$. The values of bias are on the horizontal axis starting with $\eta = 0.5$, corresponding to no bias. Note that the CSS, XZZX$\square$-deformed and ZXXZ$\square$-deformed codes have the same thresholds at $\eta = 0.5$ since the codes are equivalent in our decoding scheme. We also note that the two deformations on the compass code with $\ell = 2$ correspond to the XZZX surface code and thus have the same thresholds. The CSS compass codes have a maximum threshold at $\eta_{\ell}^{opt}$ and flatten out as the bias approaches infinity. The thresholds of the XZZX$\square$-deformed compass codes increase with bias and the advantages of a higher $\ell$ reduce as the bias increases. The thresholds of the ZXXZ$\square$-deformed compass codes grow faster with bias compared to the XZZX$\square$-deformed codes. These thresholds exceeded the XZZX surface code thresholds for biases $10<\eta\leq 100$. Values for the thresholds shown here are recorded in Table \ref{tab:Totalthresh}.}
    \label{fig:TotalThresholds}
\end{figure*}

\section{\label{sec:methods}Methods}
We run Monte Carlo simulations of the CSS and Clifford deformed codes under code capacity and phenomenological noise models. In each shot, we create a noise vector, determine the corresponding syndrome, and decode the syndrome to get a correction. After decoding, we determine whether the residual error is trivial or if a logical error has occurred. Under phenomenological noise, the noise vectors and decoder graph are three-dimensional to include $L$ measurement rounds. We assume the last round of measurements is perfect.  

We evaluate the codes by calculating their total threshold at different bias values ($\eta$) for elongation parameters ($\ell$) from 2 to 6. The threshold values we report are estimated with finite-size scaling fits (see Figures \ref{fig:l_3_bias_10_XZZX_threshold_inset}-\ref{fig:l_3_bias_10_XZZX_FSS}). Namely, near the threshold $p_{th}$ we assume that the logical error rate is a quadratic function of $(p-p_{th})L^{1/\nu}$ where $p_{th}$ is the threshold, $L$ is the distance, and $\nu$ is a critical exponent \cite{wang2003confinement}. As expected, we observe stronger finite-size effects with increasing size of the unit cells of elongated compass codes. We observe numerically that the effective inhomogeneous error model due to Clifford deformations further increases the size scale needed to accurately determine the threshold.  As a result, the thresholds we present are accurate over the code distances presented, but some may not capture the thermodynamic limit. We discuss this in more detail in Appendix \ref{appendix:Thresholds}. 

We consider noise models with biases $\eta  \in \{0.5,\eta_{\ell}^{opt}, 10, 25, 50, 100\}$. Here, $\eta_{\ell}^{opt}$ are the optimal biases for the CSS elongated compass code with elongation parameter $\ell$ found in \cite{li20192d} and listed in Section \ref{sec:ElongatedCompassCodes}. We include $\eta_{\ell}^{opt}$ to compare the deformed compass codes to the optimal performance of the CSS compass codes. The higher biases are representative of the biases found in various quantum computing architectures \cite{nigg2014quantum,ballance2016high, taylor2005fault, aliferis2009fault, grimm2020stabilization,lescanne2020exponential, berdou2023one, bocquet2024quantum}. 

\section{\label{sec:results} Results and Discussion}
\begin{figure}[h!]
    \centering
    \begin{subfigure}{0.9\columnwidth}
    \caption{}
    \includegraphics[width=\linewidth]{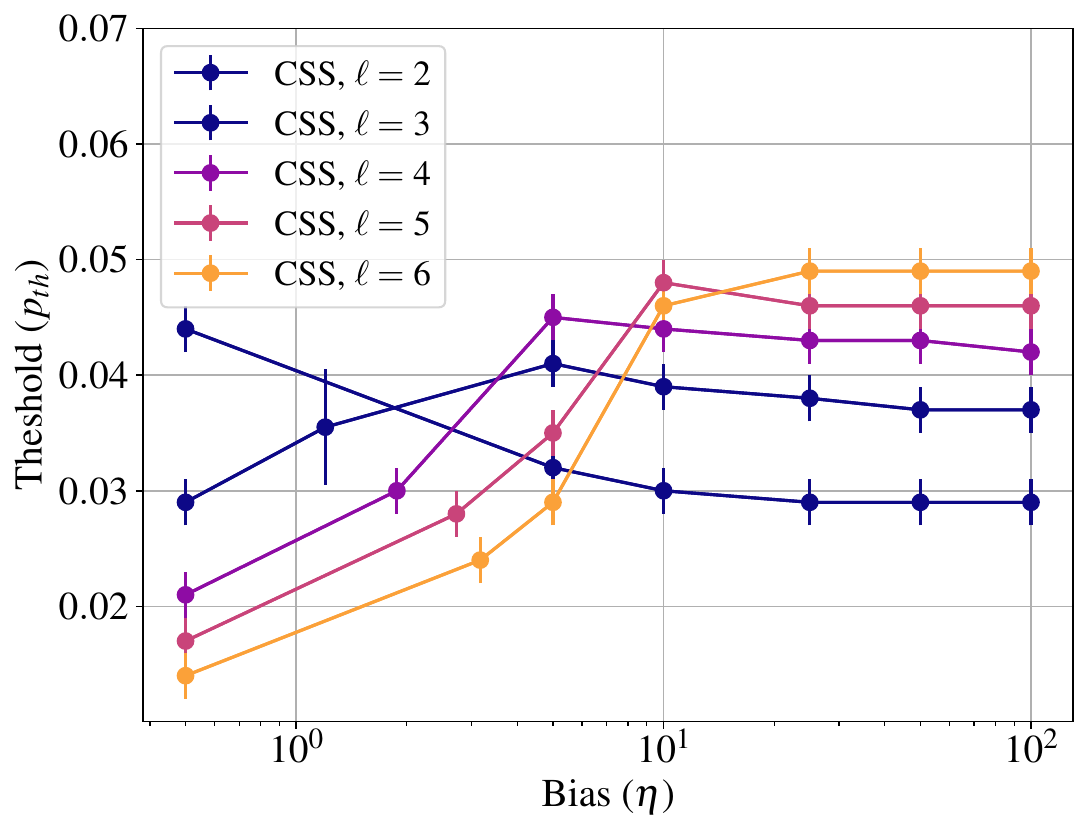}
    \label{fig:CSS_phenom}
    \end{subfigure}
    \\
    \begin{subfigure}{0.9\columnwidth}
    \caption{}
\includegraphics[width=\linewidth]{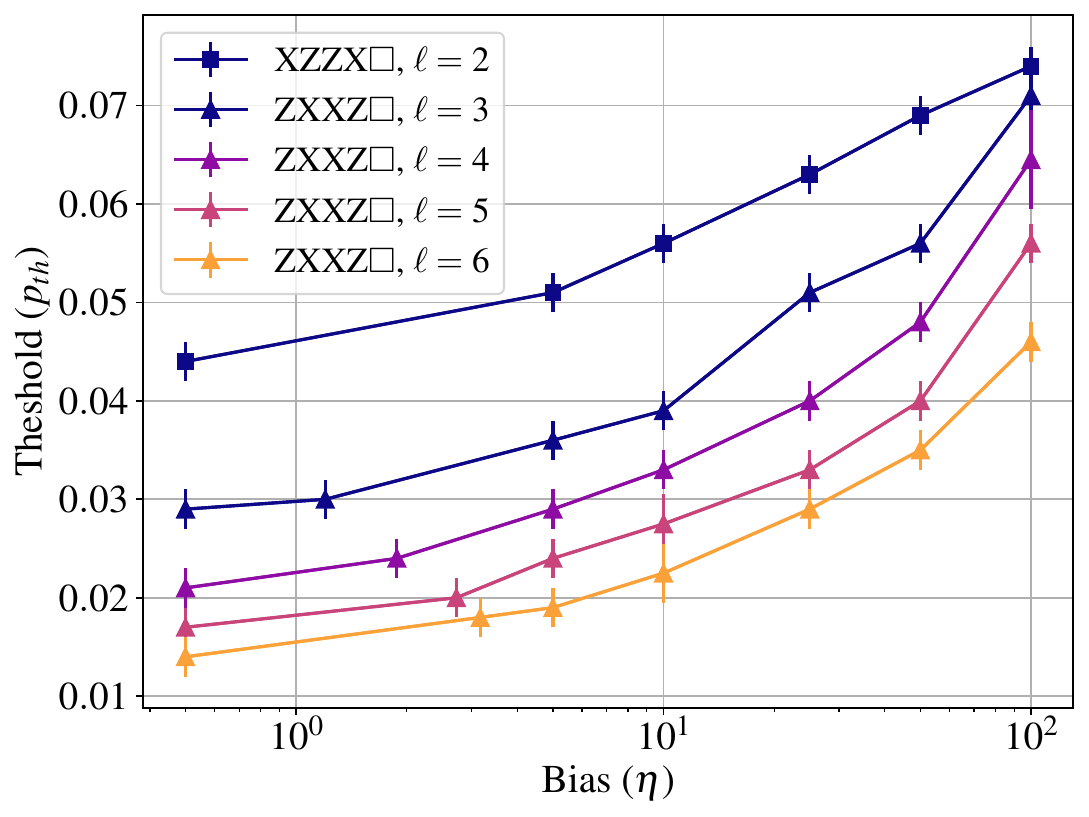}
    \label{fig:ZXXZ_Phenom}
    \end{subfigure}
    \caption{Phenomenological thresholds of a) CSS and b) $ZXXZ\square$-deformed elongated compass codes. Lines are drawn to guide the eye.}
    \label{fig:Phenom_thresholds}
\end{figure}

All threshold values from code capacity level simulations are listed in Table \ref{tab:Totalthresh} and shown in Figure \ref{fig:TotalThresholds}. Phenomenological thresholds of CSS and ZXXZ$\square$-deformed elongated compass codes are shown  in Figure \ref{fig:Phenom_thresholds}. When $\eta = 0.5$, the CSS, XZZX$\square$-deformed and ZXXZ$\square$-deformed compass codes are equivalent in our decoding scheme, so they have the same threshold. We note that finite-size effects are significant in the codes we consider, so not all reported thresholds should be interpreted as thresholds in the thermodynamic limit. For more details on this, see Appendix \ref{appendix:Thresholds}.

\begin{figure}[h!]
    \centering
    \begin{subfigure}{0.9\columnwidth}
    \caption{}
    \includegraphics[width=\linewidth]{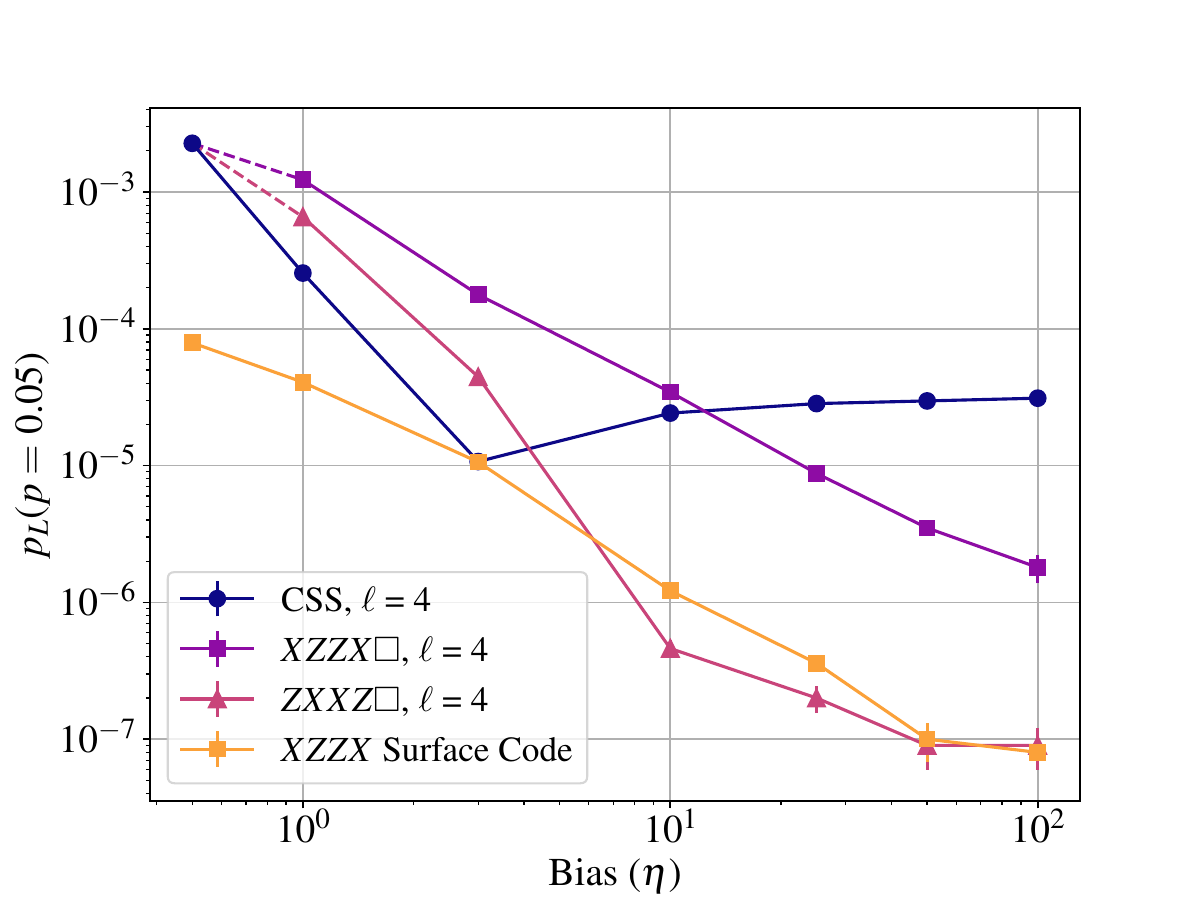}
    \label{fig:l_4_d19_0.05}
    \end{subfigure}
    \\
    \begin{subfigure}{0.9\columnwidth}
    \caption{}
\includegraphics[width=\linewidth]{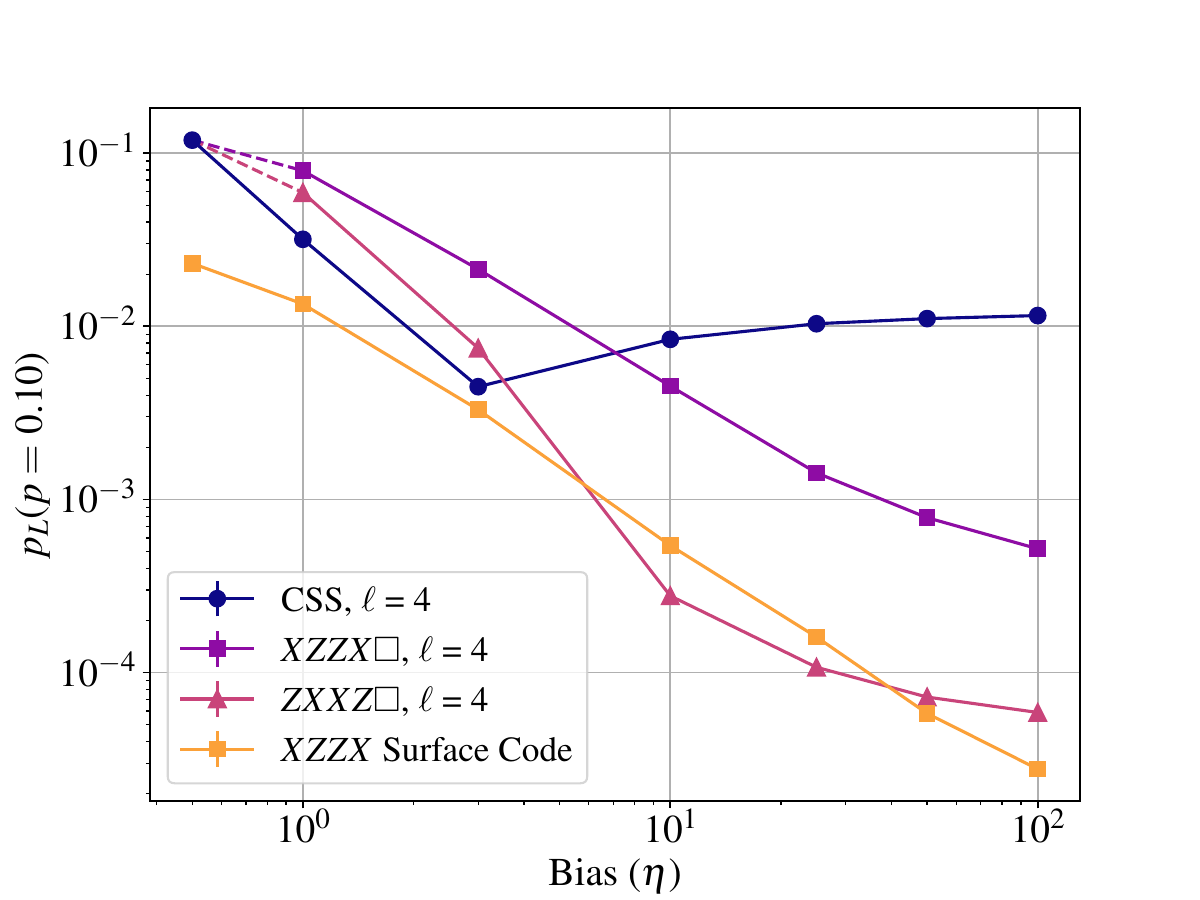}
    \label{fig:l_4_d19_0.10}
    \end{subfigure}
    \caption{Logical error rates of distance 19 CSS, XZZX$\square$-deformed and ZXXZ$\square$-deformed compass codes with $\ell = 4$ at \textbf{a}) $p=0.05$ and at \textbf{b}) $p=0.10$.  We also include logical error rates of XZZX surface code for comparison. We see that the ZXXZ$\square$ and XZZX$\square$-deformed codes suppress the logical error rate more than the CSS code for $\eta>10$. The ZXXZ$\square$-deformed code has the lowest logical error rates at moderate biases and achieves lower logical error rates than the XZZX surface code at some biases.}    \label{fig:l_4_d19_CDcomparisons}
\end{figure}

Under code capacity noise, the CSS compass codes reach a maximum threshold at the optimal biases $\eta_{\ell}^{opt}$ and these maximum thresholds increase with $\ell$ as expected. For each $\ell$, the thresholds of the CSS compass codes asymptote to the $Z$ threshold of the codes at depolarizing noise as the bias increases. Larger elongation parameters are preferable on CSS compass codes for noise models with biased $Z$ errors. We also observe that the thresholds of the XZZX surface code at the optimal biases are comparable to the thresholds of the CSS elongated compass codes at their respective optimal biases. 

The thresholds of the XZZX$\square$-deformed compass increase with bias for all elongation parameters considered. We can attribute this improvement to the fact that there are regions of the lattice to which the syndromes are confined. However, higher elongation parameters do not improve the thresholds further as the bias increases for codes with $\ell>2$. This is not surprising since the general structure of the low-weight decoder graphs for the XZZX$\square$-deformed compass codes with $\ell > 2$ looks similar to that shown in Figure \ref{fig:XZZXonSq_l6_low}. Namely, all graphs composed of lower-weight edges are partitioned by the diagonals formed by the XZZX stabilizers. Between these diagonals, there are chains of diamonds, each enclosing a string of length $\ell-2$. The graphs composed of higher-weight edges have a similar structure. However, as the elongation parameter increases, the vertex degree also increases, which may impede the growth of the thresholds with respect to bias.

The thresholds of the ZXXZ$\square$-deformed compass codes increase with bias and surpass the XZZX surface code thresholds for moderate biases (Figure \ref{fig:TotalThresholds}). This improvement begins between $\eta = 10$ and $\eta = 25$ and is still observed when $\eta = 100$. Increasing the elongation parameter on these codes does lead to further improvement, but it becomes less relevant as the bias gets higher. We can understand the rapid increase in the thresholds by noting that both the low-weight and high-weight graphs of the ZXXZ$\square$-deformed codes (Figure \ref{fig:ZXXZonSq_l6_low}-\ref{fig:ZXXZonSq_l6_high}) restrict the spread of defects, which is not the case for the high-weight graphs of the XZZX$\square$-deformed codes (Figure \ref{fig:XZZXonSq_l6_high}). The XZZX surface code wins at lower biases because the high-weight graphs of the ZXXZ$\square$-deformed codes have a higher degree than that of the XZZX surface code. As the bias increases, the high-weight graph becomes less relevant in the decoding process.

We also compare the logical error rates of the codes at physical error rates $p=0.05$ and $p=0.10$ to evaluate subthreshold behavior (see Figure \ref{fig:l_4_d19_CDcomparisons}). The ZXXZ$\square$-deformed compass codes have the lowest logical error rates of the codes we consider at biases $10 \leq \eta \leq 100$. We also compare the logical error rates to those of the XZZX surface code in Figure \ref{fig:l_4_d19_CDcomparisons} for codes with $\ell = 4$. We see similar behavior for higher elongation parameters. The CSS compass codes perform best at low biases, but their logical error rates begin to increase after a particular bias whereas those of the other codes continue decreasing as the bias increases. The logical error rates of the ZXXZ$\square$ deformed codes are comparable to those of the XZZX surface code for biases greater than 10. 

We expect that the better performance of the ZXXZ$\square$-deformed elongated compass codes relative to the XZZX surface code in the code-capacity error model does not translate into an improvement in the circuit error model. The large stabilizers will require more complicated syndrome extraction circuits. To avoid the complication of circuit timing and syndrome extraction choices, we use a phenomenological model with weighted measurements (see Sec. \ref{sec:noisemodel}) to capture the loss of relative performance. 

Thresholds of CSS and ZXXZ$\square$-deformed compass codes under phenomenological noise are shown in  Figure \ref{fig:Phenom_thresholds}. We find that the thresholds of the ZXXZ$\square$-deformed compass codes increase with bias as in the case of code capacity noise, but there is no advantage to using higher elongation parameters for any of the bias values we consider.
Consequently, the XZZX surface code achieves the highest thresholds under phenomenological noise. 

Data and source code related to this work can be accessed from \url{https://doi.org/10.7924/r4f47wc95} \cite{Campos_datarepo}.

\section{\label{sec:conclusions} Conclusion}
Clifford deformations \cite{bonilla2021xzzx, dua2022clifford,tiurev2023correcting} and compass codes \cite{li20192d,huang2020fault,pato2024logical} have both been studied in the context of biased noise models. Elongated compass codes are a particular class of compass codes that are created by fixing gauges according to a set of rules dictated by the elongation parameter $\ell$. Their performance improved in comparison to the surface code under noise models biased towards dephasing, but the asymmetry in the stabilizers is such that their performance is optimized at a particular bias. 
As a result, we considered the ZXXZ$\square$ and XZZX$\square$ deformations on elongated compass codes which preserve the weight-2 $X$ stabilizers while simplifying the structure of the decoding graphs. The resulting codes have thresholds that increase with bias. We also analyzed subthreshold behavior of the logical error rates and found that the deformed codes suppressed them more efficiently as bias increased in comparison to the CSS elongated compass codes.

We find that the thresholds of the ZXXZ$\square$-deformed compass codes surpass those of the XZZX surface code for experimentally relevant biases under code capacity noise. Furthermore, these codes exhibit lower logical error rates than the XZZX surface code (Figure \ref{fig:l_4_d19_CDcomparisons}). However, to make a fair comparison, we account for the increasing weight of the stabilizers of the Clifford deformed elongated compass codes in our phenomenological noise level simulations. Our results show that the ZXXZ$\square$-deformed compass codes do not achieve thresholds higher than those of the XZZX surface code. Nevertheless, we do accomplish our goal of modifying the elongated compass codes so that their thresholds increase with bias. 
 
A natural extension of this work is to study these codes under circuit-level noise. Such an investigation would require the design of efficient gate schedules and consideration of the degree to which individual gates preserve noise bias. Additionally, we expect improvements in performance by using decoders that incorporate noise correlations and exploit the structure of the code.

The success of Clifford deformed stabilizer codes under biased noise models has also been explored beyond the scope of circuit-based quantum computing. For example, bias-preserving XZZX cluster states exhibited high thresholds in comparison to the foliated surface code under biased noise models in measurement-based (MBQC) and fusion-based quantum computing \cite{claes2023tailored,sahay2023tailoring}. In a similar fashion, we can apply appropriate Clifford deformations to non-foliated cluster states \cite{nickerson2018measurement,newman2020generating} by looking at their effect on the decoder graphs of the cluster states. 

\section{Acknowledgements}
The authors thank S. Huang, B. Pato, Y. Lin, E. Takou and B. J. Brown for valuable discussions.  This work was supported by the NSF QLCI for Robust Quantum Simulation (OMA-2120757), and the ARO/LPS QCISS program (W911NF-21-1-0005), and the Office of the Director of National Intelligence (ODNI), Intelligence Advanced Research Projects Activity (IARPA), under the Entangled Logical Qubits program through Cooperative Agreement Number W911NF-23-2-0216.

\bibliography{apssamp_bib}

\providecommand{\noopsort}[1]{}\providecommand{\singleletter}[1]{#1}%
\begin{thebibliography}{60}%
\makeatletter
\providecommand \@ifxundefined [1]{%
 \@ifx{#1\undefined}
}%
\providecommand \@ifnum [1]{%
 \ifnum #1\expandafter \@firstoftwo
 \else \expandafter \@secondoftwo
 \fi
}%
\providecommand \@ifx [1]{%
 \ifx #1\expandafter \@firstoftwo
 \else \expandafter \@secondoftwo
 \fi
}%
\providecommand \natexlab [1]{#1}%
\providecommand \enquote  [1]{``#1''}%
\providecommand \bibnamefont  [1]{#1}%
\providecommand \bibfnamefont [1]{#1}%
\providecommand \citenamefont [1]{#1}%
\providecommand \href@noop [0]{\@secondoftwo}%
\providecommand \href [0]{\begingroup \@sanitize@url \@href}%
\providecommand \@href[1]{\@@startlink{#1}\@@href}%
\providecommand \@@href[1]{\endgroup#1\@@endlink}%
\providecommand \@sanitize@url [0]{\catcode `\\12\catcode `\$12\catcode `\&12\catcode `\#12\catcode `\^12\catcode `\_12\catcode `\%12\relax}%
\providecommand \@@startlink[1]{}%
\providecommand \@@endlink[0]{}%
\providecommand \url  [0]{\begingroup\@sanitize@url \@url }%
\providecommand \@url [1]{\endgroup\@href {#1}{\urlprefix }}%
\providecommand \urlprefix  [0]{URL }%
\providecommand \Eprint [0]{\href }%
\providecommand \doibase [0]{https://doi.org/}%
\providecommand \selectlanguage [0]{\@gobble}%
\providecommand \bibinfo  [0]{\@secondoftwo}%
\providecommand \bibfield  [0]{\@secondoftwo}%
\providecommand \translation [1]{[#1]}%
\providecommand \BibitemOpen [0]{}%
\providecommand \bibitemStop [0]{}%
\providecommand \bibitemNoStop [0]{.\EOS\space}%
\providecommand \EOS [0]{\spacefactor3000\relax}%
\providecommand \BibitemShut  [1]{\csname bibitem#1\endcsname}%
\let\auto@bib@innerbib\@empty
\bibitem [{\citenamefont {Calderbank}\ and\ \citenamefont {Shor}(1996)}]{calderbank1996good}%
  \BibitemOpen
  \bibfield  {author} {\bibinfo {author} {\bibfnamefont {A.~R.}\ \bibnamefont {Calderbank}}\ and\ \bibinfo {author} {\bibfnamefont {P.~W.}\ \bibnamefont {Shor}},\ }\bibfield  {title} {\bibinfo {title} {{\color{Black}Good quantum error-correcting codes exist}},\ }\href {https://doi.org/https://doi.org/10.1103/PhysRevA.54.1098} {\bibfield  {journal} {\bibinfo  {journal} {Phys. Rev. A}\ }\textbf {\bibinfo {volume} {54}},\ \bibinfo {pages} {1098} (\bibinfo {year} {1996})}\BibitemShut {NoStop}%
\bibitem [{\citenamefont {Knill}\ \emph {et~al.}(1998)\citenamefont {Knill}, \citenamefont {Laflamme},\ and\ \citenamefont {Zurek}}]{knill1998resilient}%
  \BibitemOpen
  \bibfield  {author} {\bibinfo {author} {\bibfnamefont {E.}~\bibnamefont {Knill}}, \bibinfo {author} {\bibfnamefont {R.}~\bibnamefont {Laflamme}},\ and\ \bibinfo {author} {\bibfnamefont {W.~H.}\ \bibnamefont {Zurek}},\ }\bibfield  {title} {\bibinfo {title} {{\color{Black}Resilient quantum computation}},\ }\href {https://doi.org/10.1126/science.279.5349.3} {\bibfield  {journal} {\bibinfo  {journal} {Science}\ }\textbf {\bibinfo {volume} {279}},\ \bibinfo {pages} {342} (\bibinfo {year} {1998})}\BibitemShut {NoStop}%
\bibitem [{\citenamefont {Aharonov}\ and\ \citenamefont {Ben-Or}(1997)}]{aharonov1997fault}%
  \BibitemOpen
  \bibfield  {author} {\bibinfo {author} {\bibfnamefont {D.}~\bibnamefont {Aharonov}}\ and\ \bibinfo {author} {\bibfnamefont {M.}~\bibnamefont {Ben-Or}},\ }\bibfield  {title} {\bibinfo {title} {{\color{Black}Fault-tolerant quantum computation with constant error}},\ }in\ \href {https://doi.org/https://doi.org/10.48550/arXiv.quant-ph/9906129} {\emph {\bibinfo {booktitle} {Proceedings of the twenty-ninth annual ACM symposium on Theory of computing}}}\ (\bibinfo {year} {1997})\ pp.\ \bibinfo {pages} {176--188}\BibitemShut {NoStop}%
\bibitem [{\citenamefont {Bombin}\ \emph {et~al.}(2012)\citenamefont {Bombin}, \citenamefont {Duclos-Cianci},\ and\ \citenamefont {Poulin}}]{bombin2012universal}%
  \BibitemOpen
  \bibfield  {author} {\bibinfo {author} {\bibfnamefont {H.}~\bibnamefont {Bombin}}, \bibinfo {author} {\bibfnamefont {G.}~\bibnamefont {Duclos-Cianci}},\ and\ \bibinfo {author} {\bibfnamefont {D.}~\bibnamefont {Poulin}},\ }\bibfield  {title} {\bibinfo {title} {{\color{Black}Universal topological phase of two-dimensional stabilizer codes}},\ }\href {https://doi.org/10.1088/1367-2630/14/7/073048} {\bibfield  {journal} {\bibinfo  {journal} {New J. of Physics.}\ }\textbf {\bibinfo {volume} {14}},\ \bibinfo {pages} {073048} (\bibinfo {year} {2012})}\BibitemShut {NoStop}%
\bibitem [{\citenamefont {Tomita}\ and\ \citenamefont {Svore}(2014)}]{tomita2014low}%
  \BibitemOpen
  \bibfield  {author} {\bibinfo {author} {\bibfnamefont {Y.}~\bibnamefont {Tomita}}\ and\ \bibinfo {author} {\bibfnamefont {K.~M.}\ \bibnamefont {Svore}},\ }\bibfield  {title} {\bibinfo {title} {{\color{Black}Low-distance surface codes under realistic quantum noise}},\ }\href {https://doi.org/https://doi.org/10.1103/PhysRevA.90.062320} {\bibfield  {journal} {\bibinfo  {journal} {Phys. Rev. A}\ }\textbf {\bibinfo {volume} {90}},\ \bibinfo {pages} {062320} (\bibinfo {year} {2014})}\BibitemShut {NoStop}%
\bibitem [{\citenamefont {Bravyi}\ \emph {et~al.}(2014)\citenamefont {Bravyi}, \citenamefont {Suchara},\ and\ \citenamefont {Vargo}}]{bravyi2014efficient}%
  \BibitemOpen
  \bibfield  {author} {\bibinfo {author} {\bibfnamefont {S.}~\bibnamefont {Bravyi}}, \bibinfo {author} {\bibfnamefont {M.}~\bibnamefont {Suchara}},\ and\ \bibinfo {author} {\bibfnamefont {A.}~\bibnamefont {Vargo}},\ }\bibfield  {title} {\bibinfo {title} {{\color{Black}Efficient algorithms for maximum likelihood decoding in the surface code}},\ }\href {https://doi.org/https://doi.org/10.1103/PhysRevA.90.032326} {\bibfield  {journal} {\bibinfo  {journal} {Phys. Rev. A}\ }\textbf {\bibinfo {volume} {90}},\ \bibinfo {pages} {032326} (\bibinfo {year} {2014})}\BibitemShut {NoStop}%
\bibitem [{\citenamefont {Sahay}\ and\ \citenamefont {Brown}(2022)}]{sahay2022decoder}%
  \BibitemOpen
  \bibfield  {author} {\bibinfo {author} {\bibfnamefont {K.}~\bibnamefont {Sahay}}\ and\ \bibinfo {author} {\bibfnamefont {B.~J.}\ \bibnamefont {Brown}},\ }\bibfield  {title} {\bibinfo {title} {{\color{Black}Decoder for the triangular color code by matching on a m{\"o}bius strip}},\ }\href {https://doi.org/https://doi.org/10.1103/PRXQuantum.3.010310} {\bibfield  {journal} {\bibinfo  {journal} {PRX Quantum}\ }\textbf {\bibinfo {volume} {3}},\ \bibinfo {pages} {010310} (\bibinfo {year} {2022})}\BibitemShut {NoStop}%
\bibitem [{\citenamefont {Grimm}\ \emph {et~al.}(2020)\citenamefont {Grimm}, \citenamefont {Frattini}, \citenamefont {Puri}, \citenamefont {Mundhada}, \citenamefont {Touzard}, \citenamefont {Mirrahimi}, \citenamefont {Girvin}, \citenamefont {Shankar},\ and\ \citenamefont {Devoret}}]{grimm2020stabilization}%
  \BibitemOpen
  \bibfield  {author} {\bibinfo {author} {\bibfnamefont {A.}~\bibnamefont {Grimm}}, \bibinfo {author} {\bibfnamefont {N.~E.}\ \bibnamefont {Frattini}}, \bibinfo {author} {\bibfnamefont {S.}~\bibnamefont {Puri}}, \bibinfo {author} {\bibfnamefont {S.~O.}\ \bibnamefont {Mundhada}}, \bibinfo {author} {\bibfnamefont {S.}~\bibnamefont {Touzard}}, \bibinfo {author} {\bibfnamefont {M.}~\bibnamefont {Mirrahimi}}, \bibinfo {author} {\bibfnamefont {S.~M.}\ \bibnamefont {Girvin}}, \bibinfo {author} {\bibfnamefont {S.}~\bibnamefont {Shankar}},\ and\ \bibinfo {author} {\bibfnamefont {M.~H.}\ \bibnamefont {Devoret}},\ }\bibfield  {title} {\bibinfo {title} {{\color{Black}Stabilization and operation of a Kerr-cat qubit}},\ }\href {https://doi.org/https://doi.org/10.1038/s41586-020-2587-z} {\bibfield  {journal} {\bibinfo  {journal} {Nature}\ }\textbf {\bibinfo {volume} {584}},\ \bibinfo {pages} {205} (\bibinfo {year} {2020})}\BibitemShut {NoStop}%
\bibitem [{\citenamefont {Lescanne}\ \emph {et~al.}(2020)\citenamefont {Lescanne}, \citenamefont {Villiers}, \citenamefont {Peronnin}, \citenamefont {Sarlette}, \citenamefont {Delbecq}, \citenamefont {Huard}, \citenamefont {Kontos}, \citenamefont {Mirrahimi},\ and\ \citenamefont {Leghtas}}]{lescanne2020exponential}%
  \BibitemOpen
  \bibfield  {author} {\bibinfo {author} {\bibfnamefont {R.}~\bibnamefont {Lescanne}}, \bibinfo {author} {\bibfnamefont {M.}~\bibnamefont {Villiers}}, \bibinfo {author} {\bibfnamefont {T.}~\bibnamefont {Peronnin}}, \bibinfo {author} {\bibfnamefont {A.}~\bibnamefont {Sarlette}}, \bibinfo {author} {\bibfnamefont {M.}~\bibnamefont {Delbecq}}, \bibinfo {author} {\bibfnamefont {B.}~\bibnamefont {Huard}}, \bibinfo {author} {\bibfnamefont {T.}~\bibnamefont {Kontos}}, \bibinfo {author} {\bibfnamefont {M.}~\bibnamefont {Mirrahimi}},\ and\ \bibinfo {author} {\bibfnamefont {Z.}~\bibnamefont {Leghtas}},\ }\bibfield  {title} {\bibinfo {title} {{\color{Black}Exponential suppression of bit-flips in a qubit encoded in an oscillator}},\ }\href {https://doi.org/https://doi.org/10.1038/s41567-020-0824-x} {\bibfield  {journal} {\bibinfo  {journal} {Nature}\ }\textbf {\bibinfo {volume} {16}},\ \bibinfo {pages} {509} (\bibinfo {year} {2020})}\BibitemShut {NoStop}%
\bibitem [{\citenamefont {Berdou}\ \emph {et~al.}(2023)\citenamefont {Berdou}, \citenamefont {Murani}, \citenamefont {Reglade}, \citenamefont {Smith}, \citenamefont {Villiers}, \citenamefont {Palomo}, \citenamefont {Rosticher}, \citenamefont {Denis}, \citenamefont {Morfin}, \citenamefont {Delbecq} \emph {et~al.}}]{berdou2023one}%
  \BibitemOpen
  \bibfield  {author} {\bibinfo {author} {\bibfnamefont {C.}~\bibnamefont {Berdou}}, \bibinfo {author} {\bibfnamefont {A.}~\bibnamefont {Murani}}, \bibinfo {author} {\bibfnamefont {U.}~\bibnamefont {Reglade}}, \bibinfo {author} {\bibfnamefont {W.~C.}\ \bibnamefont {Smith}}, \bibinfo {author} {\bibfnamefont {M.}~\bibnamefont {Villiers}}, \bibinfo {author} {\bibfnamefont {J.}~\bibnamefont {Palomo}}, \bibinfo {author} {\bibfnamefont {M.}~\bibnamefont {Rosticher}}, \bibinfo {author} {\bibfnamefont {A.}~\bibnamefont {Denis}}, \bibinfo {author} {\bibfnamefont {P.}~\bibnamefont {Morfin}}, \bibinfo {author} {\bibfnamefont {M.}~\bibnamefont {Delbecq}}, \emph {et~al.},\ }\bibfield  {title} {\bibinfo {title} {{\color{Black}One hundred second bit-flip time in a two-photon dissipative oscillator}},\ }\href {https://doi.org/https://doi.org/10.1103/PRXQuantum.4.020350} {\bibfield  {journal} {\bibinfo  {journal} {PRX Quantum}\ }\textbf {\bibinfo {volume} {4}},\ \bibinfo {pages} {020350} (\bibinfo {year} {2023})}\BibitemShut
  {NoStop}%
\bibitem [{\citenamefont {Bocquet}\ \emph {et~al.}(2024)\citenamefont {Bocquet}, \citenamefont {Leghtas}, \citenamefont {Reglade}, \citenamefont {Gautier}, \citenamefont {Cohen}, \citenamefont {Marquet}, \citenamefont {Albertinale}, \citenamefont {Pankratova}, \citenamefont {Hall{\'e}n}, \citenamefont {Rautschke} \emph {et~al.}}]{bocquet2024quantum}%
  \BibitemOpen
  \bibfield  {author} {\bibinfo {author} {\bibfnamefont {A.}~\bibnamefont {Bocquet}}, \bibinfo {author} {\bibfnamefont {Z.}~\bibnamefont {Leghtas}}, \bibinfo {author} {\bibfnamefont {U.}~\bibnamefont {Reglade}}, \bibinfo {author} {\bibfnamefont {R.}~\bibnamefont {Gautier}}, \bibinfo {author} {\bibfnamefont {J.}~\bibnamefont {Cohen}}, \bibinfo {author} {\bibfnamefont {A.}~\bibnamefont {Marquet}}, \bibinfo {author} {\bibfnamefont {E.}~\bibnamefont {Albertinale}}, \bibinfo {author} {\bibfnamefont {N.}~\bibnamefont {Pankratova}}, \bibinfo {author} {\bibfnamefont {M.}~\bibnamefont {Hall{\'e}n}}, \bibinfo {author} {\bibfnamefont {F.}~\bibnamefont {Rautschke}}, \emph {et~al.},\ }\bibfield  {title} {\bibinfo {title} {{\color{Black}Quantum control of a cat-qubit with bit-flip times exceeding ten seconds}},\ }\bibfield  {journal} {\bibinfo  {journal} {Bulletin of the American Physical Society}\ }\href {https://doi.org/https://doi.org/10.1038/s41586-024-07294-3} {https://doi.org/10.1038/s41586-024-07294-3} (\bibinfo
  {year} {2024})\BibitemShut {NoStop}%
\bibitem [{\citenamefont {Chou}\ \emph {et~al.}(2024)\citenamefont {Chou}, \citenamefont {Shemma}, \citenamefont {McCarrick}, \citenamefont {Chien}, \citenamefont {Teoh}, \citenamefont {Winkel}, \citenamefont {Anderson}, \citenamefont {Chen}, \citenamefont {Curtis}, \citenamefont {de~Graaf} \emph {et~al.}}]{chou2024superconducting}%
  \BibitemOpen
  \bibfield  {author} {\bibinfo {author} {\bibfnamefont {K.~S.}\ \bibnamefont {Chou}}, \bibinfo {author} {\bibfnamefont {T.}~\bibnamefont {Shemma}}, \bibinfo {author} {\bibfnamefont {H.}~\bibnamefont {McCarrick}}, \bibinfo {author} {\bibfnamefont {T.-C.}\ \bibnamefont {Chien}}, \bibinfo {author} {\bibfnamefont {J.~D.}\ \bibnamefont {Teoh}}, \bibinfo {author} {\bibfnamefont {P.}~\bibnamefont {Winkel}}, \bibinfo {author} {\bibfnamefont {A.}~\bibnamefont {Anderson}}, \bibinfo {author} {\bibfnamefont {J.}~\bibnamefont {Chen}}, \bibinfo {author} {\bibfnamefont {J.~C.}\ \bibnamefont {Curtis}}, \bibinfo {author} {\bibfnamefont {S.~J.}\ \bibnamefont {de~Graaf}}, \emph {et~al.},\ }\bibfield  {title} {\bibinfo {title} {{\color{Black}A superconducting dual-rail cavity qubit with erasure-detected logical measurements}},\ }\href {https://doi.org/https://doi.org/10.1038/s41567-024-02539-4} {\bibfield  {journal} {\bibinfo  {journal} {Nature Physics}\ ,\ \bibinfo {pages} {1}} (\bibinfo {year} {2024})}\BibitemShut {NoStop}%
\bibitem [{\citenamefont {Kubica}\ \emph {et~al.}(2023)\citenamefont {Kubica}, \citenamefont {Haim}, \citenamefont {Vaknin}, \citenamefont {Levine}, \citenamefont {Brand{\~a}o},\ and\ \citenamefont {Retzker}}]{kubica2023erasure}%
  \BibitemOpen
  \bibfield  {author} {\bibinfo {author} {\bibfnamefont {A.}~\bibnamefont {Kubica}}, \bibinfo {author} {\bibfnamefont {A.}~\bibnamefont {Haim}}, \bibinfo {author} {\bibfnamefont {Y.}~\bibnamefont {Vaknin}}, \bibinfo {author} {\bibfnamefont {H.}~\bibnamefont {Levine}}, \bibinfo {author} {\bibfnamefont {F.}~\bibnamefont {Brand{\~a}o}},\ and\ \bibinfo {author} {\bibfnamefont {A.}~\bibnamefont {Retzker}},\ }\bibfield  {title} {\bibinfo {title} {{\color{Black}Erasure qubits: Overcoming the T 1 limit in superconducting circuits}},\ }\href {https://doi.org/https://doi.org/10.1103/PhysRevX.13.041022} {\bibfield  {journal} {\bibinfo  {journal} {Physical Review X}\ }\textbf {\bibinfo {volume} {13}},\ \bibinfo {pages} {041022} (\bibinfo {year} {2023})}\BibitemShut {NoStop}%
\bibitem [{\citenamefont {Teoh}\ \emph {et~al.}(2023)\citenamefont {Teoh}, \citenamefont {Winkel}, \citenamefont {Babla}, \citenamefont {Chapman}, \citenamefont {Claes}, \citenamefont {de~Graaf}, \citenamefont {Garmon}, \citenamefont {Kalfus}, \citenamefont {Lu}, \citenamefont {Maiti} \emph {et~al.}}]{teoh2023dual}%
  \BibitemOpen
  \bibfield  {author} {\bibinfo {author} {\bibfnamefont {J.~D.}\ \bibnamefont {Teoh}}, \bibinfo {author} {\bibfnamefont {P.}~\bibnamefont {Winkel}}, \bibinfo {author} {\bibfnamefont {H.~K.}\ \bibnamefont {Babla}}, \bibinfo {author} {\bibfnamefont {B.~J.}\ \bibnamefont {Chapman}}, \bibinfo {author} {\bibfnamefont {J.}~\bibnamefont {Claes}}, \bibinfo {author} {\bibfnamefont {S.~J.}\ \bibnamefont {de~Graaf}}, \bibinfo {author} {\bibfnamefont {J.~W.}\ \bibnamefont {Garmon}}, \bibinfo {author} {\bibfnamefont {W.~D.}\ \bibnamefont {Kalfus}}, \bibinfo {author} {\bibfnamefont {Y.}~\bibnamefont {Lu}}, \bibinfo {author} {\bibfnamefont {A.}~\bibnamefont {Maiti}}, \emph {et~al.},\ }\bibfield  {title} {\bibinfo {title} {{\color{Black}Dual-rail encoding with superconducting cavities}},\ }\href {https://doi.org/https://doi.org/10.1073/pnas.222173612} {\bibfield  {journal} {\bibinfo  {journal} {Proceedings of the National Academy of Sciences}\ }\textbf {\bibinfo {volume} {120}},\ \bibinfo {pages} {e2221736120} (\bibinfo
  {year} {2023})}\BibitemShut {NoStop}%
\bibitem [{\citenamefont {Cong}\ \emph {et~al.}(2022)\citenamefont {Cong}, \citenamefont {Levine}, \citenamefont {Keesling}, \citenamefont {Bluvstein}, \citenamefont {Wang},\ and\ \citenamefont {Lukin}}]{cong2022hardware_neutralatoms}%
  \BibitemOpen
  \bibfield  {author} {\bibinfo {author} {\bibfnamefont {I.}~\bibnamefont {Cong}}, \bibinfo {author} {\bibfnamefont {H.}~\bibnamefont {Levine}}, \bibinfo {author} {\bibfnamefont {A.}~\bibnamefont {Keesling}}, \bibinfo {author} {\bibfnamefont {D.}~\bibnamefont {Bluvstein}}, \bibinfo {author} {\bibfnamefont {S.-T.}\ \bibnamefont {Wang}},\ and\ \bibinfo {author} {\bibfnamefont {M.~D.}\ \bibnamefont {Lukin}},\ }\bibfield  {title} {\bibinfo {title} {{\color{Black}Hardware-efficient, fault-tolerant quantum computation with rydberg atoms}},\ }\href {https://doi.org/https://doi.org/10.1103/PhysRevX.12.021049} {\bibfield  {journal} {\bibinfo  {journal} {Phys. Rev. X}\ }\textbf {\bibinfo {volume} {12}},\ \bibinfo {pages} {021049} (\bibinfo {year} {2022})}\BibitemShut {NoStop}%
\bibitem [{\citenamefont {Wu}\ \emph {et~al.}(2022)\citenamefont {Wu}, \citenamefont {Kolkowitz}, \citenamefont {Puri},\ and\ \citenamefont {Thompson}}]{wu2022erasure}%
  \BibitemOpen
  \bibfield  {author} {\bibinfo {author} {\bibfnamefont {Y.}~\bibnamefont {Wu}}, \bibinfo {author} {\bibfnamefont {S.}~\bibnamefont {Kolkowitz}}, \bibinfo {author} {\bibfnamefont {S.}~\bibnamefont {Puri}},\ and\ \bibinfo {author} {\bibfnamefont {J.~D.}\ \bibnamefont {Thompson}},\ }\bibfield  {title} {\bibinfo {title} {{\color{Black}Erasure conversion for fault-tolerant quantum computing in alkaline earth Rydberg atom arrays}},\ }\href {https://doi.org/https://doi.org/10.1038/s41467-022-32094-6} {\bibfield  {journal} {\bibinfo  {journal} {Nature}\ }\textbf {\bibinfo {volume} {13}},\ \bibinfo {pages} {4657} (\bibinfo {year} {2022})}\BibitemShut {NoStop}%
\bibitem [{\citenamefont {Kang}\ \emph {et~al.}(2023)\citenamefont {Kang}, \citenamefont {Campbell},\ and\ \citenamefont {Brown}}]{kang2023quantum}%
  \BibitemOpen
  \bibfield  {author} {\bibinfo {author} {\bibfnamefont {M.}~\bibnamefont {Kang}}, \bibinfo {author} {\bibfnamefont {W.~C.}\ \bibnamefont {Campbell}},\ and\ \bibinfo {author} {\bibfnamefont {K.~R.}\ \bibnamefont {Brown}},\ }\bibfield  {title} {\bibinfo {title} {{\color{Black}Quantum error correction with metastable states of trapped ions using erasure conversion}},\ }\href {https://doi.org/https://doi.org/10.1103/PRXQuantum.4.020358} {\bibfield  {journal} {\bibinfo  {journal} {PRX Quantum}\ }\textbf {\bibinfo {volume} {4}},\ \bibinfo {pages} {020358} (\bibinfo {year} {2023})}\BibitemShut {NoStop}%
\bibitem [{\citenamefont {Nigg}\ \emph {et~al.}(2014)\citenamefont {Nigg}, \citenamefont {Mueller}, \citenamefont {Martinez}, \citenamefont {Schindler}, \citenamefont {Hennrich}, \citenamefont {Monz}, \citenamefont {Martin-Delgado},\ and\ \citenamefont {Blatt}}]{nigg2014quantum}%
  \BibitemOpen
  \bibfield  {author} {\bibinfo {author} {\bibfnamefont {D.}~\bibnamefont {Nigg}}, \bibinfo {author} {\bibfnamefont {M.}~\bibnamefont {Mueller}}, \bibinfo {author} {\bibfnamefont {E.~A.}\ \bibnamefont {Martinez}}, \bibinfo {author} {\bibfnamefont {P.}~\bibnamefont {Schindler}}, \bibinfo {author} {\bibfnamefont {M.}~\bibnamefont {Hennrich}}, \bibinfo {author} {\bibfnamefont {T.}~\bibnamefont {Monz}}, \bibinfo {author} {\bibfnamefont {M.~A.}\ \bibnamefont {Martin-Delgado}},\ and\ \bibinfo {author} {\bibfnamefont {R.}~\bibnamefont {Blatt}},\ }\bibfield  {title} {\bibinfo {title} {{\color{Black}Quantum computations on a topologically encoded qubit}},\ }\href {https://doi.org/10.1126/science.12537} {\bibfield  {journal} {\bibinfo  {journal} {Science}\ }\textbf {\bibinfo {volume} {345}},\ \bibinfo {pages} {302} (\bibinfo {year} {2014})}\BibitemShut {NoStop}%
\bibitem [{\citenamefont {Ballance}\ \emph {et~al.}(2016)\citenamefont {Ballance}, \citenamefont {Harty}, \citenamefont {Linke}, \citenamefont {Sepiol},\ and\ \citenamefont {Lucas}}]{ballance2016high}%
  \BibitemOpen
  \bibfield  {author} {\bibinfo {author} {\bibfnamefont {C.}~\bibnamefont {Ballance}}, \bibinfo {author} {\bibfnamefont {T.}~\bibnamefont {Harty}}, \bibinfo {author} {\bibfnamefont {N.}~\bibnamefont {Linke}}, \bibinfo {author} {\bibfnamefont {M.}~\bibnamefont {Sepiol}},\ and\ \bibinfo {author} {\bibfnamefont {D.}~\bibnamefont {Lucas}},\ }\bibfield  {title} {\bibinfo {title} {{\color{Black}High-fidelity quantum logic gates using trapped-ion hyperfine qubits}},\ }\href {https://doi.org/https://doi.org/10.1103/PhysRevLett.117.060504} {\bibfield  {journal} {\bibinfo  {journal} {Phys. Rev. Lett.}\ }\textbf {\bibinfo {volume} {117}},\ \bibinfo {pages} {060504} (\bibinfo {year} {2016})}\BibitemShut {NoStop}%
\bibitem [{\citenamefont {Taylor}\ \emph {et~al.}(2005)\citenamefont {Taylor}, \citenamefont {Engel}, \citenamefont {D{\"u}r}, \citenamefont {Yacoby}, \citenamefont {Marcus}, \citenamefont {Zoller},\ and\ \citenamefont {Lukin}}]{taylor2005fault}%
  \BibitemOpen
  \bibfield  {author} {\bibinfo {author} {\bibfnamefont {J.}~\bibnamefont {Taylor}}, \bibinfo {author} {\bibfnamefont {H.-A.}\ \bibnamefont {Engel}}, \bibinfo {author} {\bibfnamefont {W.}~\bibnamefont {D{\"u}r}}, \bibinfo {author} {\bibfnamefont {A.}~\bibnamefont {Yacoby}}, \bibinfo {author} {\bibfnamefont {C.}~\bibnamefont {Marcus}}, \bibinfo {author} {\bibfnamefont {P.}~\bibnamefont {Zoller}},\ and\ \bibinfo {author} {\bibfnamefont {M.}~\bibnamefont {Lukin}},\ }\bibfield  {title} {\bibinfo {title} {{\color{Black}Fault-tolerant architecture for quantum computation using electrically controlled semiconductor spins}},\ }\href {https://doi.org/https://doi.org/10.1038/nphys174} {\bibfield  {journal} {\bibinfo  {journal} {Nature}\ }\textbf {\bibinfo {volume} {1}},\ \bibinfo {pages} {177} (\bibinfo {year} {2005})}\BibitemShut {NoStop}%
\bibitem [{\citenamefont {Aliferis}\ \emph {et~al.}(2009)\citenamefont {Aliferis}, \citenamefont {Brito}, \citenamefont {DiVincenzo}, \citenamefont {Preskill}, \citenamefont {Steffen},\ and\ \citenamefont {Terhal}}]{aliferis2009fault}%
  \BibitemOpen
  \bibfield  {author} {\bibinfo {author} {\bibfnamefont {P.}~\bibnamefont {Aliferis}}, \bibinfo {author} {\bibfnamefont {F.}~\bibnamefont {Brito}}, \bibinfo {author} {\bibfnamefont {D.~P.}\ \bibnamefont {DiVincenzo}}, \bibinfo {author} {\bibfnamefont {J.}~\bibnamefont {Preskill}}, \bibinfo {author} {\bibfnamefont {M.}~\bibnamefont {Steffen}},\ and\ \bibinfo {author} {\bibfnamefont {B.~M.}\ \bibnamefont {Terhal}},\ }\bibfield  {title} {\bibinfo {title} {{\color{Black}Fault-tolerant computing with biased-noise superconducting qubits: a case study}},\ }\href {https://doi.org/10.1088/1367-2630/11/1/013061} {\bibfield  {journal} {\bibinfo  {journal} {New J. of Physics.}\ }\textbf {\bibinfo {volume} {11}},\ \bibinfo {pages} {013061} (\bibinfo {year} {2009})}\BibitemShut {NoStop}%
\bibitem [{\citenamefont {Rosenblum}\ \emph {et~al.}(2018)\citenamefont {Rosenblum}, \citenamefont {Reinhold}, \citenamefont {Mirrahimi}, \citenamefont {Jiang}, \citenamefont {Frunzio},\ and\ \citenamefont {Schoelkopf}}]{rosenblum2018fault}%
  \BibitemOpen
  \bibfield  {author} {\bibinfo {author} {\bibfnamefont {S.}~\bibnamefont {Rosenblum}}, \bibinfo {author} {\bibfnamefont {P.}~\bibnamefont {Reinhold}}, \bibinfo {author} {\bibfnamefont {M.}~\bibnamefont {Mirrahimi}}, \bibinfo {author} {\bibfnamefont {L.}~\bibnamefont {Jiang}}, \bibinfo {author} {\bibfnamefont {L.}~\bibnamefont {Frunzio}},\ and\ \bibinfo {author} {\bibfnamefont {R.~J.}\ \bibnamefont {Schoelkopf}},\ }\bibfield  {title} {\bibinfo {title} {{\color{Black}Fault-tolerant detection of a quantum error}},\ }\href {https://doi.org/10.1126/science.aat39} {\bibfield  {journal} {\bibinfo  {journal} {Science}\ }\textbf {\bibinfo {volume} {361}},\ \bibinfo {pages} {266} (\bibinfo {year} {2018})}\BibitemShut {NoStop}%
\bibitem [{\citenamefont {Tuckett}\ \emph {et~al.}(2018)\citenamefont {Tuckett}, \citenamefont {Bartlett},\ and\ \citenamefont {Flammia}}]{tuckett2018ultrahigh}%
  \BibitemOpen
  \bibfield  {author} {\bibinfo {author} {\bibfnamefont {D.~K.}\ \bibnamefont {Tuckett}}, \bibinfo {author} {\bibfnamefont {S.~D.}\ \bibnamefont {Bartlett}},\ and\ \bibinfo {author} {\bibfnamefont {S.~T.}\ \bibnamefont {Flammia}},\ }\bibfield  {title} {\bibinfo {title} {{\color{Black}Ultrahigh error threshold for surface codes with biased noise}},\ }\href {https://doi.org/10.1103/PhysRevLett.120.050505} {\bibfield  {journal} {\bibinfo  {journal} {Phys. Rev. Lett.}\ }\textbf {\bibinfo {volume} {120}},\ \bibinfo {pages} {050505} (\bibinfo {year} {2018})}\BibitemShut {NoStop}%
\bibitem [{\citenamefont {Tuckett}\ \emph {et~al.}(2019)\citenamefont {Tuckett}, \citenamefont {Darmawan}, \citenamefont {Chubb}, \citenamefont {Bravyi}, \citenamefont {Bartlett},\ and\ \citenamefont {Flammia}}]{tuckett2019tailoring}%
  \BibitemOpen
  \bibfield  {author} {\bibinfo {author} {\bibfnamefont {D.~K.}\ \bibnamefont {Tuckett}}, \bibinfo {author} {\bibfnamefont {A.~S.}\ \bibnamefont {Darmawan}}, \bibinfo {author} {\bibfnamefont {C.~T.}\ \bibnamefont {Chubb}}, \bibinfo {author} {\bibfnamefont {S.}~\bibnamefont {Bravyi}}, \bibinfo {author} {\bibfnamefont {S.~D.}\ \bibnamefont {Bartlett}},\ and\ \bibinfo {author} {\bibfnamefont {S.~T.}\ \bibnamefont {Flammia}},\ }\bibfield  {title} {\bibinfo {title} {{\color{Black}Tailoring surface codes for highly biased noise}},\ }\href {https://doi.org/https://doi.org/10.1103/PhysRevX.9.041031} {\bibfield  {journal} {\bibinfo  {journal} {Phys. Rev. X}\ }\textbf {\bibinfo {volume} {9}},\ \bibinfo {pages} {041031} (\bibinfo {year} {2019})}\BibitemShut {NoStop}%
\bibitem [{\citenamefont {Stephens}\ \emph {et~al.}(2013)\citenamefont {Stephens}, \citenamefont {Munro},\ and\ \citenamefont {Nemoto}}]{stephens2013high}%
  \BibitemOpen
  \bibfield  {author} {\bibinfo {author} {\bibfnamefont {A.~M.}\ \bibnamefont {Stephens}}, \bibinfo {author} {\bibfnamefont {W.~J.}\ \bibnamefont {Munro}},\ and\ \bibinfo {author} {\bibfnamefont {K.}~\bibnamefont {Nemoto}},\ }\bibfield  {title} {\bibinfo {title} {{\color{Black}High-threshold topological quantum error correction against biased noise}},\ }\href {https://doi.org/https://doi.org/10.1103/PhysRevA.88.060301} {\bibfield  {journal} {\bibinfo  {journal} {Phys. Rev. A}\ }\textbf {\bibinfo {volume} {88}},\ \bibinfo {pages} {060301} (\bibinfo {year} {2013})}\BibitemShut {NoStop}%
\bibitem [{\citenamefont {Li}\ \emph {et~al.}(2019)\citenamefont {Li}, \citenamefont {Miller}, \citenamefont {Newman}, \citenamefont {Wu},\ and\ \citenamefont {Brown}}]{li20192d}%
  \BibitemOpen
  \bibfield  {author} {\bibinfo {author} {\bibfnamefont {M.}~\bibnamefont {Li}}, \bibinfo {author} {\bibfnamefont {D.}~\bibnamefont {Miller}}, \bibinfo {author} {\bibfnamefont {M.}~\bibnamefont {Newman}}, \bibinfo {author} {\bibfnamefont {Y.}~\bibnamefont {Wu}},\ and\ \bibinfo {author} {\bibfnamefont {K.~R.}\ \bibnamefont {Brown}},\ }\bibfield  {title} {\bibinfo {title} {{\color{Black}2d compass codes}},\ }\href {https://doi.org/https://doi.org/10.1103/PhysRevX.9.021041} {\bibfield  {journal} {\bibinfo  {journal} {Phys. Rev. X}\ }\textbf {\bibinfo {volume} {9}},\ \bibinfo {pages} {021041} (\bibinfo {year} {2019})}\BibitemShut {NoStop}%
\bibitem [{\citenamefont {Bonilla~Ataides}\ \emph {et~al.}(2021)\citenamefont {Bonilla~Ataides}, \citenamefont {Tuckett}, \citenamefont {Bartlett}, \citenamefont {Flammia},\ and\ \citenamefont {Brown}}]{bonilla2021xzzx}%
  \BibitemOpen
  \bibfield  {author} {\bibinfo {author} {\bibfnamefont {J.~P.}\ \bibnamefont {Bonilla~Ataides}}, \bibinfo {author} {\bibfnamefont {D.~K.}\ \bibnamefont {Tuckett}}, \bibinfo {author} {\bibfnamefont {S.~D.}\ \bibnamefont {Bartlett}}, \bibinfo {author} {\bibfnamefont {S.~T.}\ \bibnamefont {Flammia}},\ and\ \bibinfo {author} {\bibfnamefont {B.~J.}\ \bibnamefont {Brown}},\ }\bibfield  {title} {\bibinfo {title} {{\color{Black}The XZZX surface code}},\ }\href {https://doi.org/https://doi.org/10.1038/s41467-021-22274-1} {\bibfield  {journal} {\bibinfo  {journal} {Nature Commun.}\ }\textbf {\bibinfo {volume} {12}},\ \bibinfo {pages} {2172} (\bibinfo {year} {2021})}\BibitemShut {NoStop}%
\bibitem [{\citenamefont {Sahay}\ \emph {et~al.}(2023{\natexlab{a}})\citenamefont {Sahay}, \citenamefont {Jin}, \citenamefont {Claes}, \citenamefont {Thompson},\ and\ \citenamefont {Puri}}]{Sahay_Neutralatomswithbiasederasure}%
  \BibitemOpen
  \bibfield  {author} {\bibinfo {author} {\bibfnamefont {K.}~\bibnamefont {Sahay}}, \bibinfo {author} {\bibfnamefont {J.}~\bibnamefont {Jin}}, \bibinfo {author} {\bibfnamefont {J.}~\bibnamefont {Claes}}, \bibinfo {author} {\bibfnamefont {J.~D.}\ \bibnamefont {Thompson}},\ and\ \bibinfo {author} {\bibfnamefont {S.}~\bibnamefont {Puri}},\ }\bibfield  {title} {\bibinfo {title} {{\color{Black}High-Threshold Codes for Neutral-Atom Qubits with Biased Erasure Errors}},\ }\href {https://doi.org/10.1103/PhysRevX.13.041013} {\bibfield  {journal} {\bibinfo  {journal} {Phys. Rev. X}\ }\textbf {\bibinfo {volume} {13}},\ \bibinfo {pages} {041013} (\bibinfo {year} {2023}{\natexlab{a}})}\BibitemShut {NoStop}%
\bibitem [{\citenamefont {Dua}\ \emph {et~al.}(2024)\citenamefont {Dua}, \citenamefont {Kubica}, \citenamefont {Jiang}, \citenamefont {Flammia},\ and\ \citenamefont {Gullans}}]{dua2022clifford}%
  \BibitemOpen
  \bibfield  {author} {\bibinfo {author} {\bibfnamefont {A.}~\bibnamefont {Dua}}, \bibinfo {author} {\bibfnamefont {A.}~\bibnamefont {Kubica}}, \bibinfo {author} {\bibfnamefont {L.}~\bibnamefont {Jiang}}, \bibinfo {author} {\bibfnamefont {S.~T.}\ \bibnamefont {Flammia}},\ and\ \bibinfo {author} {\bibfnamefont {M.~J.}\ \bibnamefont {Gullans}},\ }\bibfield  {title} {\bibinfo {title} {{\color{Black}Clifford-Deformed Surface Codes}},\ }\href {https://doi.org/10.1103/PRXQuantum.5.010347} {\bibfield  {journal} {\bibinfo  {journal} {PRX Quantum}\ }\textbf {\bibinfo {volume} {5}},\ \bibinfo {pages} {010347} (\bibinfo {year} {2024})}\BibitemShut {NoStop}%
\bibitem [{\citenamefont {Tiurev}\ \emph {et~al.}(2023)\citenamefont {Tiurev}, \citenamefont {Derks}, \citenamefont {Roffe}, \citenamefont {Eisert},\ and\ \citenamefont {Reiner}}]{tiurev2023correcting}%
  \BibitemOpen
  \bibfield  {author} {\bibinfo {author} {\bibfnamefont {K.}~\bibnamefont {Tiurev}}, \bibinfo {author} {\bibfnamefont {P.-J.~H.}\ \bibnamefont {Derks}}, \bibinfo {author} {\bibfnamefont {J.}~\bibnamefont {Roffe}}, \bibinfo {author} {\bibfnamefont {J.}~\bibnamefont {Eisert}},\ and\ \bibinfo {author} {\bibfnamefont {J.-M.}\ \bibnamefont {Reiner}},\ }\bibfield  {title} {\bibinfo {title} {{\color{Black}Correcting non-independent and non-identically distributed errors with surface codes}},\ }\href {https://doi.org/https://doi.org/10.22331/q-2023-09-26-1123} {\bibfield  {journal} {\bibinfo  {journal} {Quantum}\ }\textbf {\bibinfo {volume} {7}},\ \bibinfo {pages} {1123} (\bibinfo {year} {2023})}\BibitemShut {NoStop}%
\bibitem [{\citenamefont {Huang}\ and\ \citenamefont {Brown}(2020)}]{huang2020fault}%
  \BibitemOpen
  \bibfield  {author} {\bibinfo {author} {\bibfnamefont {S.}~\bibnamefont {Huang}}\ and\ \bibinfo {author} {\bibfnamefont {K.~R.}\ \bibnamefont {Brown}},\ }\bibfield  {title} {\bibinfo {title} {{\color{Black}Fault-tolerant compass codes}},\ }\href {https://doi.org/https://doi.org/10.1103/PhysRevA.101.042312} {\bibfield  {journal} {\bibinfo  {journal} {Phys. Rev. A}\ }\textbf {\bibinfo {volume} {101}},\ \bibinfo {pages} {042312} (\bibinfo {year} {2020})}\BibitemShut {NoStop}%
\bibitem [{\citenamefont {Paetznick}\ and\ \citenamefont {Reichardt}(2013)}]{paetznick2013universal}%
  \BibitemOpen
  \bibfield  {author} {\bibinfo {author} {\bibfnamefont {A.}~\bibnamefont {Paetznick}}\ and\ \bibinfo {author} {\bibfnamefont {B.~W.}\ \bibnamefont {Reichardt}},\ }\bibfield  {title} {\bibinfo {title} {{\color{Black}Universal fault-tolerant quantum computation with only transversal gates and error correction}},\ }\href {https://doi.org/https://doi.org/10.1103/PhysRevLett.111.090505} {\bibfield  {journal} {\bibinfo  {journal} {Phys. Rev. Lett.}\ }\textbf {\bibinfo {volume} {111}},\ \bibinfo {pages} {090505} (\bibinfo {year} {2013})}\BibitemShut {NoStop}%
\bibitem [{\citenamefont {Bomb{\'\i}n}(2015)}]{bombin2015gauge}%
  \BibitemOpen
  \bibfield  {author} {\bibinfo {author} {\bibfnamefont {H.}~\bibnamefont {Bomb{\'\i}n}},\ }\bibfield  {title} {\bibinfo {title} {{\color{Black}Gauge color codes: optimal transversal gates and gauge fixing in topological stabilizer codes}},\ }\href {https://doi.org/10.1088/1367-2630/17/8/083002} {\bibfield  {journal} {\bibinfo  {journal} {New J. of Physics.}\ }\textbf {\bibinfo {volume} {17}},\ \bibinfo {pages} {083002} (\bibinfo {year} {2015})}\BibitemShut {NoStop}%
\bibitem [{\citenamefont {Pato}\ \emph {et~al.}(2025)\citenamefont {Pato}, \citenamefont {Staples},\ and\ \citenamefont {Brown}}]{pato2024logical}%
  \BibitemOpen
  \bibfield  {author} {\bibinfo {author} {\bibfnamefont {B.}~\bibnamefont {Pato}}, \bibinfo {author} {\bibfnamefont {J.~W.}\ \bibnamefont {Staples}},\ and\ \bibinfo {author} {\bibfnamefont {K.~R.}\ \bibnamefont {Brown}},\ }\bibfield  {title} {\bibinfo {title} {{\color{Black}Logical coherence in two-dimensional compass codes}},\ }\href {https://doi.org/https://doi.org/10.1103/PhysRevA.111.032424} {\bibfield  {journal} {\bibinfo  {journal} {Physical Review A}\ }\textbf {\bibinfo {volume} {111}},\ \bibinfo {pages} {032424} (\bibinfo {year} {2025})}\BibitemShut {NoStop}%
\bibitem [{\citenamefont {Kitaev}(2003)}]{kitaev2003fault}%
  \BibitemOpen
  \bibfield  {author} {\bibinfo {author} {\bibfnamefont {A.~Y.}\ \bibnamefont {Kitaev}},\ }\bibfield  {title} {\bibinfo {title} {{\color{Black}Fault-tolerant quantum computation by anyons}},\ }\href {https://doi.org/https://doi.org/10.1016/S0003-4916(02)00018-} {\bibfield  {journal} {\bibinfo  {journal} {Annals of physics}\ }\textbf {\bibinfo {volume} {303}},\ \bibinfo {pages} {2} (\bibinfo {year} {2003})}\BibitemShut {NoStop}%
\bibitem [{\citenamefont {Dennis}\ \emph {et~al.}(2002)\citenamefont {Dennis}, \citenamefont {Kitaev}, \citenamefont {Landahl},\ and\ \citenamefont {Preskill}}]{dennis2002topological}%
  \BibitemOpen
  \bibfield  {author} {\bibinfo {author} {\bibfnamefont {E.}~\bibnamefont {Dennis}}, \bibinfo {author} {\bibfnamefont {A.}~\bibnamefont {Kitaev}}, \bibinfo {author} {\bibfnamefont {A.}~\bibnamefont {Landahl}},\ and\ \bibinfo {author} {\bibfnamefont {J.}~\bibnamefont {Preskill}},\ }\bibfield  {title} {\bibinfo {title} {{\color{Black}Topological quantum memory}},\ }\href {https://doi.org/https://doi.org/10.1063/1.1499754} {\bibfield  {journal} {\bibinfo  {journal} {Journal of Mathematical Physics}\ }\textbf {\bibinfo {volume} {43}},\ \bibinfo {pages} {4452} (\bibinfo {year} {2002})}\BibitemShut {NoStop}%
\bibitem [{\citenamefont {Fowler}\ \emph {et~al.}(2012)\citenamefont {Fowler}, \citenamefont {Whiteside},\ and\ \citenamefont {Hollenberg}}]{fowler2012towards}%
  \BibitemOpen
  \bibfield  {author} {\bibinfo {author} {\bibfnamefont {A.~G.}\ \bibnamefont {Fowler}}, \bibinfo {author} {\bibfnamefont {A.~C.}\ \bibnamefont {Whiteside}},\ and\ \bibinfo {author} {\bibfnamefont {L.~C.}\ \bibnamefont {Hollenberg}},\ }\bibfield  {title} {\bibinfo {title} {{\color{Black}Towards practical classical processing for the surface code}},\ }\href {https://doi.org/https://doi.org/10.1103/PhysRevLett.108.180501} {\bibfield  {journal} {\bibinfo  {journal} {Phys. Rev. Lett.}\ }\textbf {\bibinfo {volume} {108}},\ \bibinfo {pages} {180501} (\bibinfo {year} {2012})}\BibitemShut {NoStop}%
\bibitem [{\citenamefont {San~Miguel}\ \emph {et~al.}(2023)\citenamefont {San~Miguel}, \citenamefont {Williamson},\ and\ \citenamefont {Brown}}]{san2023cellular}%
  \BibitemOpen
  \bibfield  {author} {\bibinfo {author} {\bibfnamefont {J.~F.}\ \bibnamefont {San~Miguel}}, \bibinfo {author} {\bibfnamefont {D.~J.}\ \bibnamefont {Williamson}},\ and\ \bibinfo {author} {\bibfnamefont {B.~J.}\ \bibnamefont {Brown}},\ }\bibfield  {title} {\bibinfo {title} {{\color{Black}A cellular automaton decoder for a noise-bias tailored color code}},\ }\href {https://doi.org/https://doi.org/10.22331/q-2023-03-09-940} {\bibfield  {journal} {\bibinfo  {journal} {Quantum}\ }\textbf {\bibinfo {volume} {7}},\ \bibinfo {pages} {940} (\bibinfo {year} {2023})}\BibitemShut {NoStop}%
\bibitem [{\citenamefont {Tiurev}\ \emph {et~al.}(2024)\citenamefont {Tiurev}, \citenamefont {Pesah}, \citenamefont {Derks}, \citenamefont {Roffe}, \citenamefont {Eisert}, \citenamefont {Kesselring},\ and\ \citenamefont {Reiner}}]{tiurev2024domain}%
  \BibitemOpen
  \bibfield  {author} {\bibinfo {author} {\bibfnamefont {K.}~\bibnamefont {Tiurev}}, \bibinfo {author} {\bibfnamefont {A.}~\bibnamefont {Pesah}}, \bibinfo {author} {\bibfnamefont {P.-J.~H.}\ \bibnamefont {Derks}}, \bibinfo {author} {\bibfnamefont {J.}~\bibnamefont {Roffe}}, \bibinfo {author} {\bibfnamefont {J.}~\bibnamefont {Eisert}}, \bibinfo {author} {\bibfnamefont {M.~S.}\ \bibnamefont {Kesselring}},\ and\ \bibinfo {author} {\bibfnamefont {J.-M.}\ \bibnamefont {Reiner}},\ }\bibfield  {title} {\bibinfo {title} {{\color{Black}Domain wall color code}},\ }\href {https://doi.org/https://doi.org/10.1103/PhysRevLett.133.110601} {\bibfield  {journal} {\bibinfo  {journal} {Physical Review Letters}\ }\textbf {\bibinfo {volume} {133}},\ \bibinfo {pages} {110601} (\bibinfo {year} {2024})}\BibitemShut {NoStop}%
\bibitem [{\citenamefont {Setiawan}\ and\ \citenamefont {McLauchlan}(2025)}]{setiawan2025tailoring}%
  \BibitemOpen
  \bibfield  {author} {\bibinfo {author} {\bibfnamefont {F.}~\bibnamefont {Setiawan}}\ and\ \bibinfo {author} {\bibfnamefont {C.}~\bibnamefont {McLauchlan}},\ }\bibfield  {title} {\bibinfo {title} {{\color{Black}Tailoring dynamical codes for biased noise: the X3Z3 Floquet code}},\ }\href {https://doi.org/https://doi.org/10.1038/s41534-025-01074-1} {\bibfield  {journal} {\bibinfo  {journal} {npj Quantum Information}\ }\textbf {\bibinfo {volume} {11}},\ \bibinfo {pages} {149} (\bibinfo {year} {2025})}\BibitemShut {NoStop}%
\bibitem [{\citenamefont {Debroy}\ \emph {et~al.}(2021)\citenamefont {Debroy}, \citenamefont {Egan}, \citenamefont {Noel}, \citenamefont {Risinger}, \citenamefont {Zhu}, \citenamefont {Biswas}, \citenamefont {Cetina}, \citenamefont {Monroe},\ and\ \citenamefont {Brown}}]{debroy2021optimizing}%
  \BibitemOpen
  \bibfield  {author} {\bibinfo {author} {\bibfnamefont {D.~M.}\ \bibnamefont {Debroy}}, \bibinfo {author} {\bibfnamefont {L.}~\bibnamefont {Egan}}, \bibinfo {author} {\bibfnamefont {C.}~\bibnamefont {Noel}}, \bibinfo {author} {\bibfnamefont {A.}~\bibnamefont {Risinger}}, \bibinfo {author} {\bibfnamefont {D.}~\bibnamefont {Zhu}}, \bibinfo {author} {\bibfnamefont {D.}~\bibnamefont {Biswas}}, \bibinfo {author} {\bibfnamefont {M.}~\bibnamefont {Cetina}}, \bibinfo {author} {\bibfnamefont {C.}~\bibnamefont {Monroe}},\ and\ \bibinfo {author} {\bibfnamefont {K.~R.}\ \bibnamefont {Brown}},\ }\bibfield  {title} {\bibinfo {title} {{\color{Black}Optimizing stabilizer parities for improved logical qubit memories}},\ }\href {https://doi.org/https://doi.org/10.1103/PhysRevLett.127.240501} {\bibfield  {journal} {\bibinfo  {journal} {Physical Review Letters}\ }\textbf {\bibinfo {volume} {127}},\ \bibinfo {pages} {240501} (\bibinfo {year} {2021})}\BibitemShut {NoStop}%
\bibitem [{goo(2023)}]{google2023suppressing}%
  \BibitemOpen
  \bibfield  {title} {\bibinfo {title} {{\color{Black}Suppressing quantum errors by scaling a surface code logical qubit}},\ }\href {https://doi.org/https://doi.org/10.1038/s41586-022-05434-1} {\bibfield  {journal} {\bibinfo  {journal} {Nature}\ }\textbf {\bibinfo {volume} {614}},\ \bibinfo {pages} {676} (\bibinfo {year} {2023})}\BibitemShut {NoStop}%
\bibitem [{\citenamefont {Kribs}\ \emph {et~al.}(2006)\citenamefont {Kribs}, \citenamefont {Laflamme}, \citenamefont {Poulin},\ and\ \citenamefont {Lesosky}}]{kribs2005operator}%
  \BibitemOpen
  \bibfield  {author} {\bibinfo {author} {\bibfnamefont {D.~W.}\ \bibnamefont {Kribs}}, \bibinfo {author} {\bibfnamefont {R.}~\bibnamefont {Laflamme}}, \bibinfo {author} {\bibfnamefont {D.}~\bibnamefont {Poulin}},\ and\ \bibinfo {author} {\bibfnamefont {M.}~\bibnamefont {Lesosky}},\ }\bibfield  {title} {\bibinfo {title} {{\color{Black}Operator quantum error correction}},\ }\href {https://doi.org/https://doi.org/10.48550/arXiv.quant-ph/0504189} {\bibfield  {journal} {\bibinfo  {journal} {Quant. Inf. Comput.}\ }\textbf {\bibinfo {volume} {6}},\ \bibinfo {pages} {382} (\bibinfo {year} {2006})}\BibitemShut {NoStop}%
\bibitem [{\citenamefont {Poulin}(2005)}]{poulin2005stabilizer}%
  \BibitemOpen
  \bibfield  {author} {\bibinfo {author} {\bibfnamefont {D.}~\bibnamefont {Poulin}},\ }\bibfield  {title} {\bibinfo {title} {{\color{Black}Stabilizer formalism for operator quantum error correction}},\ }\href {https://doi.org/https://doi.org/10.1103/PhysRevLett.95.230504} {\bibfield  {journal} {\bibinfo  {journal} {Phys. Rev. Lett.}\ }\textbf {\bibinfo {volume} {95}},\ \bibinfo {pages} {230504} (\bibinfo {year} {2005})}\BibitemShut {NoStop}%
\bibitem [{\citenamefont {Gottesman}(1997)}]{gottesman1997stabilizer}%
  \BibitemOpen
  \bibfield  {author} {\bibinfo {author} {\bibfnamefont {D.}~\bibnamefont {Gottesman}},\ }\emph {\bibinfo {title} {Stabilizer codes and quantum error correction}},\ \href {https://doi.org/https://doi.org/10.48550/arXiv.quant-ph/9705052} {\bibinfo {type} {Ph{D} thesis}} (\bibinfo {year} {1997}),\ \Eprint {https://arxiv.org/abs/arxiv:9705052 [quant-ph]} {arxiv:9705052 [quant-ph]} \BibitemShut {NoStop}%
\bibitem [{\citenamefont {Kugel}\ and\ \citenamefont {Khomskii}(1973)}]{kugel1973crystal}%
  \BibitemOpen
  \bibfield  {author} {\bibinfo {author} {\bibfnamefont {K.}~\bibnamefont {Kugel}}\ and\ \bibinfo {author} {\bibfnamefont {D.}~\bibnamefont {Khomskii}},\ }\bibfield  {title} {\bibinfo {title} {{\color{Black}Crystal-structure and magnetic properties of substances with orbital degeneracy}},\ }\href@noop {} {\bibfield  {journal} {\bibinfo  {journal} {Zh. Eksp. Teor. Fiz}\ }\textbf {\bibinfo {volume} {64}},\ \bibinfo {pages} {1429} (\bibinfo {year} {1973})}\BibitemShut {NoStop}%
\bibitem [{\citenamefont {Dorier}\ \emph {et~al.}(2005)\citenamefont {Dorier}, \citenamefont {Becca},\ and\ \citenamefont {Mila}}]{dorier2005quantum}%
  \BibitemOpen
  \bibfield  {author} {\bibinfo {author} {\bibfnamefont {J.}~\bibnamefont {Dorier}}, \bibinfo {author} {\bibfnamefont {F.}~\bibnamefont {Becca}},\ and\ \bibinfo {author} {\bibfnamefont {F.}~\bibnamefont {Mila}},\ }\bibfield  {title} {\bibinfo {title} {{\color{Black}Quantum compass model on the square lattice}},\ }\href {https://doi.org/https://doi.org/10.1103/PhysRevB.72.024448} {\bibfield  {journal} {\bibinfo  {journal} {Phys. Rev. B}\ }\textbf {\bibinfo {volume} {72}},\ \bibinfo {pages} {024448} (\bibinfo {year} {2005})}\BibitemShut {NoStop}%
\bibitem [{\citenamefont {Dou{\c{c}}ot}\ \emph {et~al.}(2005)\citenamefont {Dou{\c{c}}ot}, \citenamefont {Feigel’Man}, \citenamefont {Ioffe},\ and\ \citenamefont {Ioselevich}}]{douccot2005protected}%
  \BibitemOpen
  \bibfield  {author} {\bibinfo {author} {\bibfnamefont {B.}~\bibnamefont {Dou{\c{c}}ot}}, \bibinfo {author} {\bibfnamefont {M.}~\bibnamefont {Feigel’Man}}, \bibinfo {author} {\bibfnamefont {L.}~\bibnamefont {Ioffe}},\ and\ \bibinfo {author} {\bibfnamefont {A.}~\bibnamefont {Ioselevich}},\ }\bibfield  {title} {\bibinfo {title} {{\color{Black}Protected qubits and Chern-Simons theories in Josephson junction arrays}},\ }\href {https://doi.org/https://doi.org/10.1103/PhysRevB.71.024505} {\bibfield  {journal} {\bibinfo  {journal} {Phys. Rev. B}\ }\textbf {\bibinfo {volume} {71}},\ \bibinfo {pages} {024505} (\bibinfo {year} {2005})}\BibitemShut {NoStop}%
\bibitem [{\citenamefont {Shor}(1995)}]{shor1995scheme}%
  \BibitemOpen
  \bibfield  {author} {\bibinfo {author} {\bibfnamefont {P.~W.}\ \bibnamefont {Shor}},\ }\bibfield  {title} {\bibinfo {title} {{\color{Black}Scheme for reducing decoherence in quantum computer memory}},\ }\href {https://doi.org/https://doi.org/10.1103/PhysRevA.52.R2493} {\bibfield  {journal} {\bibinfo  {journal} {Phys. Rev. A}\ }\textbf {\bibinfo {volume} {52}},\ \bibinfo {pages} {R2493} (\bibinfo {year} {1995})}\BibitemShut {NoStop}%
\bibitem [{\citenamefont {Bacon}(2006)}]{bacon2006operator}%
  \BibitemOpen
  \bibfield  {author} {\bibinfo {author} {\bibfnamefont {D.}~\bibnamefont {Bacon}},\ }\bibfield  {title} {\bibinfo {title} {{\color{Black}Operator quantum error-correcting subsystems for self-correcting quantum memories}},\ }\href {https://doi.org/https://doi.org/10.1103/PhysRevA.73.012340} {\bibfield  {journal} {\bibinfo  {journal} {Phys. Rev. A}\ }\textbf {\bibinfo {volume} {73}},\ \bibinfo {pages} {012340} (\bibinfo {year} {2006})}\BibitemShut {NoStop}%
\bibitem [{\citenamefont {Steane}(1996)}]{steane1996multiple}%
  \BibitemOpen
  \bibfield  {author} {\bibinfo {author} {\bibfnamefont {A.}~\bibnamefont {Steane}},\ }\bibfield  {title} {\bibinfo {title} {{\color{Black}Multiple-particle interference and quantum error correction}},\ }\href {https://doi.org/https://doi.org/10.1098/rspa.1996.0136} {\bibfield  {journal} {\bibinfo  {journal} {Proceedings of the Royal Society of London. Series A: Mathematical, Physical and Engineering Sciences}\ }\textbf {\bibinfo {volume} {452}},\ \bibinfo {pages} {2551} (\bibinfo {year} {1996})}\BibitemShut {NoStop}%
\bibitem [{\citenamefont {Edmonds}(1965)}]{edmonds1965paths}%
  \BibitemOpen
  \bibfield  {author} {\bibinfo {author} {\bibfnamefont {J.}~\bibnamefont {Edmonds}},\ }\bibfield  {title} {\bibinfo {title} {{\color{Black}Paths, trees, and flowers}},\ }\href {https://doi.org/https://doi.org/10.4153/CJM-1965-045-4} {\bibfield  {journal} {\bibinfo  {journal} {Canadian Journal of mathematics}\ }\textbf {\bibinfo {volume} {17}},\ \bibinfo {pages} {449} (\bibinfo {year} {1965})}\BibitemShut {NoStop}%
\bibitem [{\citenamefont {Higgott}(2022)}]{higgott2022pymatching}%
  \BibitemOpen
  \bibfield  {author} {\bibinfo {author} {\bibfnamefont {O.}~\bibnamefont {Higgott}},\ }\bibfield  {title} {\bibinfo {title} {{\color{Black}PyMatching: A Python package for decoding quantum codes with minimum-weight perfect matching}},\ }\href {https://doi.org/10.1145/3505637} {\bibfield  {journal} {\bibinfo  {journal} {ACM Transactions on Quantum Computing}\ }\textbf {\bibinfo {volume} {3}},\ \bibinfo {pages} {1} (\bibinfo {year} {2022})}\BibitemShut {NoStop}%
\bibitem [{\citenamefont {Wang}\ \emph {et~al.}(2003)\citenamefont {Wang}, \citenamefont {Harrington},\ and\ \citenamefont {Preskill}}]{wang2003confinement}%
  \BibitemOpen
  \bibfield  {author} {\bibinfo {author} {\bibfnamefont {C.}~\bibnamefont {Wang}}, \bibinfo {author} {\bibfnamefont {J.}~\bibnamefont {Harrington}},\ and\ \bibinfo {author} {\bibfnamefont {J.}~\bibnamefont {Preskill}},\ }\bibfield  {title} {\bibinfo {title} {{\color{Black}Confinement-Higgs transition in a disordered gauge theory and the accuracy threshold for quantum memory}},\ }\href {https://doi.org/https://doi.org/10.1016/S0003-4916(02)00019-2} {\bibfield  {journal} {\bibinfo  {journal} {Annals of Physics}\ }\textbf {\bibinfo {volume} {303}},\ \bibinfo {pages} {31} (\bibinfo {year} {2003})}\BibitemShut {NoStop}%
\bibitem [{\citenamefont {Campos}\ and\ \citenamefont {Brown}(2024)}]{Campos_datarepo}%
  \BibitemOpen
  \bibfield  {author} {\bibinfo {author} {\bibfnamefont {J.~A.}\ \bibnamefont {Campos}}\ and\ \bibinfo {author} {\bibfnamefont {K.~R.}\ \bibnamefont {Brown}},\ }\href@noop {} {\bibinfo {title} {{\color{Black}Data and code from: Clifford-deformed compass codes}}} (\bibinfo {year} {2024}),\ \bibinfo {note} {\href{https://doi.org/10.7924/r4f47wc95}{https://doi.org/10.7924/r4f47wc95}}\BibitemShut {NoStop}%
\bibitem [{\citenamefont {Claes}\ \emph {et~al.}(2023)\citenamefont {Claes}, \citenamefont {Bourassa},\ and\ \citenamefont {Puri}}]{claes2023tailored}%
  \BibitemOpen
  \bibfield  {author} {\bibinfo {author} {\bibfnamefont {J.}~\bibnamefont {Claes}}, \bibinfo {author} {\bibfnamefont {J.~E.}\ \bibnamefont {Bourassa}},\ and\ \bibinfo {author} {\bibfnamefont {S.}~\bibnamefont {Puri}},\ }\bibfield  {title} {\bibinfo {title} {{\color{Black}Tailored cluster states with high threshold under biased noise}},\ }\href {https://doi.org/https://doi.org/10.1038/s41534-023-00677-w} {\bibfield  {journal} {\bibinfo  {journal} {npj Quantum Information}\ }\textbf {\bibinfo {volume} {9}},\ \bibinfo {pages} {9} (\bibinfo {year} {2023})}\BibitemShut {NoStop}%
\bibitem [{\citenamefont {Sahay}\ \emph {et~al.}(2023{\natexlab{b}})\citenamefont {Sahay}, \citenamefont {Claes},\ and\ \citenamefont {Puri}}]{sahay2023tailoring}%
  \BibitemOpen
  \bibfield  {author} {\bibinfo {author} {\bibfnamefont {K.}~\bibnamefont {Sahay}}, \bibinfo {author} {\bibfnamefont {J.}~\bibnamefont {Claes}},\ and\ \bibinfo {author} {\bibfnamefont {S.}~\bibnamefont {Puri}},\ }\bibfield  {title} {\bibinfo {title} {{\color{Black}Tailoring fusion-based error correction for high thresholds to biased fusion failures}},\ }\href {https://doi.org/https://doi.org/10.1103/PhysRevLett.131.120604} {\bibfield  {journal} {\bibinfo  {journal} {Physical Review Letters}\ }\textbf {\bibinfo {volume} {131}},\ \bibinfo {pages} {120604} (\bibinfo {year} {2023}{\natexlab{b}})}\BibitemShut {NoStop}%
\bibitem [{\citenamefont {Nickerson}\ and\ \citenamefont {Bomb{\'\i}n}(2018)}]{nickerson2018measurement}%
  \BibitemOpen
  \bibfield  {author} {\bibinfo {author} {\bibfnamefont {N.}~\bibnamefont {Nickerson}}\ and\ \bibinfo {author} {\bibfnamefont {H.}~\bibnamefont {Bomb{\'\i}n}},\ }\bibfield  {title} {\bibinfo {title} {{\color{Black}Measurement based fault tolerance beyond foliation}},\ }\Eprint {https://arxiv.org/abs/arXiv:1810.09621} {arXiv:1810.09621}  (\bibinfo {year} {2018})\BibitemShut {NoStop}%
\bibitem [{\citenamefont {Newman}\ \emph {et~al.}(2020)\citenamefont {Newman}, \citenamefont {de~Castro},\ and\ \citenamefont {Brown}}]{newman2020generating}%
  \BibitemOpen
  \bibfield  {author} {\bibinfo {author} {\bibfnamefont {M.}~\bibnamefont {Newman}}, \bibinfo {author} {\bibfnamefont {L.~A.}\ \bibnamefont {de~Castro}},\ and\ \bibinfo {author} {\bibfnamefont {K.~R.}\ \bibnamefont {Brown}},\ }\bibfield  {title} {\bibinfo {title} {{\color{Black}Generating fault-tolerant cluster states from crystal structures}},\ }\href {https://doi.org/https://doi.org/10.22331/q-2020-07-13-295} {\bibfield  {journal} {\bibinfo  {journal} {Quantum}\ }\textbf {\bibinfo {volume} {4}},\ \bibinfo {pages} {295} (\bibinfo {year} {2020})}\BibitemShut {NoStop}%
\bibitem [{\citenamefont {Xiao}\ \emph {et~al.}(2024)\citenamefont {Xiao}, \citenamefont {Srivastava},\ and\ \citenamefont {Granath}}]{xiao2024exact}%
  \BibitemOpen
  \bibfield  {author} {\bibinfo {author} {\bibfnamefont {Y.}~\bibnamefont {Xiao}}, \bibinfo {author} {\bibfnamefont {B.}~\bibnamefont {Srivastava}},\ and\ \bibinfo {author} {\bibfnamefont {M.}~\bibnamefont {Granath}},\ }\bibfield  {title} {\bibinfo {title} {{\color{Black}Exact results on finite size corrections for surface codes tailored to biased noise}},\ }\href {https://doi.org/https://doi.org/10.22331/q-2024-09-11-1468} {\bibfield  {journal} {\bibinfo  {journal} {Quantum}\ }\textbf {\bibinfo {volume} {8}},\ \bibinfo {pages} {1468} (\bibinfo {year} {2024})}\BibitemShut {NoStop}%
\end{thebibliography}%
\newpage
\onecolumngrid
\appendix
\section{Table of Thresholds}
\begin{table*}[h!]
    \centering
\begin{tabular}{|c?c|c|c|c|c|}
\hline
  \multicolumn{6}{|c|}{Thresholds (CSS $|$ XZZX$\square$ $|$ ZXXZ$\square$)}
  \\
  \hline
  \diagbox{$\eta$}{$\ell$} & 2 & 3 & 4 & 5 & 6 \\
\Xhline{3\arrayrulewidth}
0.5 & 14.8 & 11.7 & 8.3 & 6.8 & 5.7 
\\
 \hline
$\eta_{\ell}^{opt}$ & - & 17.5 $|$ 12.6 $|$ 13.5 & 19.5 $|$ 13.0 $|$ 12.4 & 21.0 $|$ 13.3 $|$ 12.2 & 22.6 $|$ 14.0 $|$ 12.3
\\
\hline
10 & 10.3 $|$ 27.0 $|$ - & 14.6 $|$ 18.0 $|$ 27.3& 17.5 $|$ 17.5 $|$ 18.9 & 19.6 $|$ 17.0 $|$ 16.6 & 21.8 $|$ 16.4 $|$ 15.7
\\
\hline
25 &  10.1 $|$ 32.0 $|$ - & 14.1 $|$ 21.5 $|$ 33.6 &  17.0 $|$ 21.2 $|$ 34.5 &  19.0 $|$ 20.9 $|$ 35.0 & 21.1 $|$ 20.7 $|$ 35.1
\\
\hline
50 & 10.0 $|$ 35.9 $|$ - & 14.0 $|$ 23.5 $|$ 38.0 & 16.8 $|$ 23.3 $|$ 37.9 & 18.9 $|$ 23.1 $|$ 38.5 & 20.8 $|$ 22.8 $|$ 39.2
\\
\hline
100 & 10.0 $|$ 38.2 $|$ - & 14.0 $|$ 24.7 $|$ 39.9 & 16.8 $|$ 24.9 $|$ 40.0 & 18.7 $|$ 25.2 $|$ 39.4 & 20.6 $|$ 25.1 $|$ 39.9
\\
\hline
\end{tabular}
    \caption{Thresholds for CSS, XZZX$\square$-deformed and ZXXZ$\square$-deformed compass codes at all elongation parameters and biases under code capacity noise. When $\eta = 0.5$, the thresholds of codes are the same since $p_x = p_z$ and thus the deformations do not change the weights. Also, when $\ell = 2$, the XZZX$\square$ and ZXXZ$\square$ deformations are equivalent. We record the remaining thresholds in the following way: CSS $|$ XZZX$\square$ $|$ ZXXZ$\square$. Numerical uncertainty of these threshold values do not exceed 0.8\%. Thresholds are estimated using finite-size scaling fits near threshold. See Appendix \ref{appendix:Thresholds} for more information.}
    \label{tab:Totalthresh}
\end{table*} 

\section{Additional Decoder Graphs}\label{appendix:graphs}
We include examples of graphs containing the low and high weight edges of the decoder graphs to motivate our choices for the locations of the Hadamard transformations. An demonstration of how we obtain low and high-weight graphs is shown in Figure \ref{fig:XZZX_l2}. 

\begin{figure}[tp!]
    \centering
    \begin{subfigure}{0.32\textwidth}
    \caption{}
        \includegraphics[width = 0.9\linewidth]{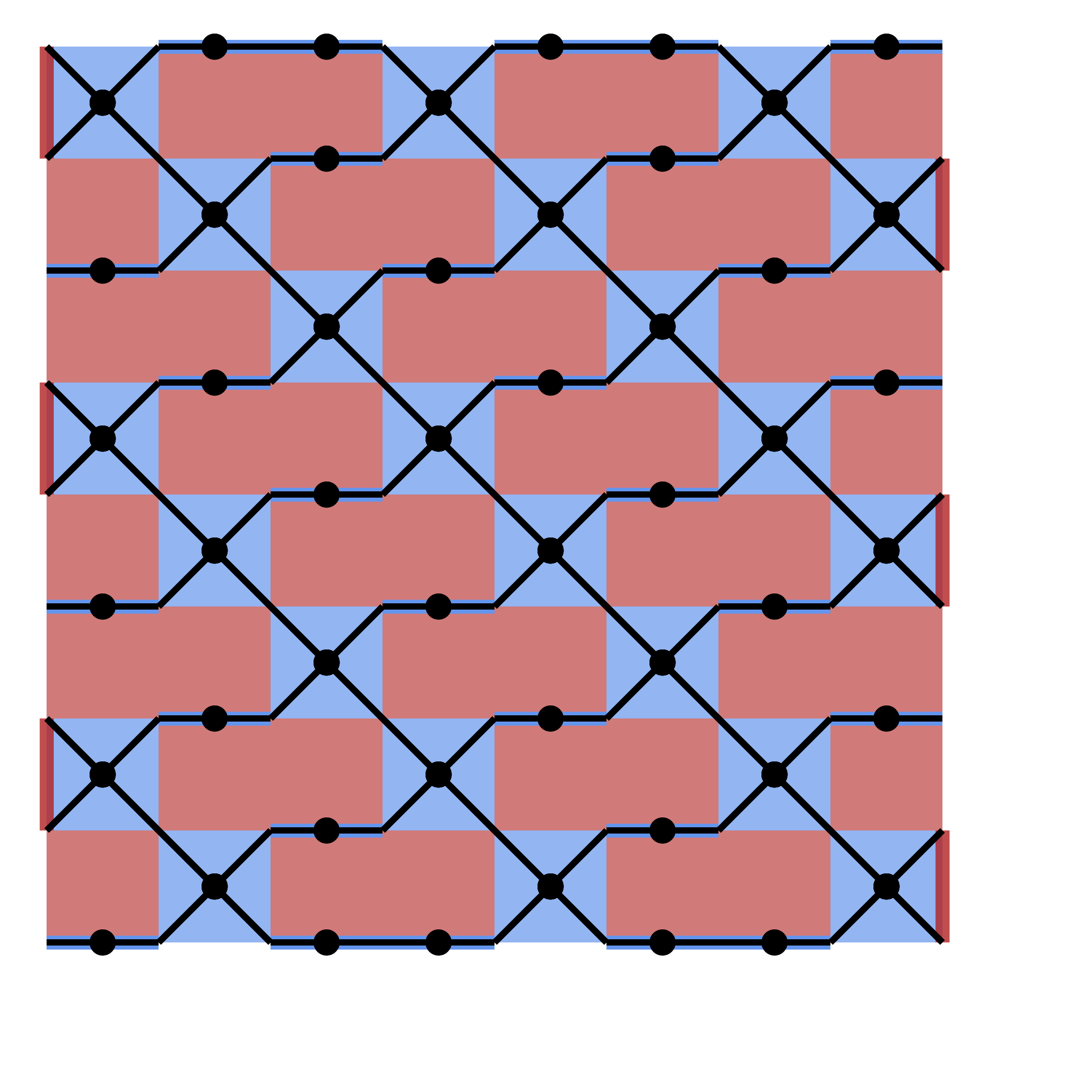}
        \label{fig:l3_lowweight_CSS}
    \end{subfigure}
    \hfill
    \begin{subfigure}{0.32\textwidth}
    \caption{}
        \includegraphics[width = 0.9\linewidth]{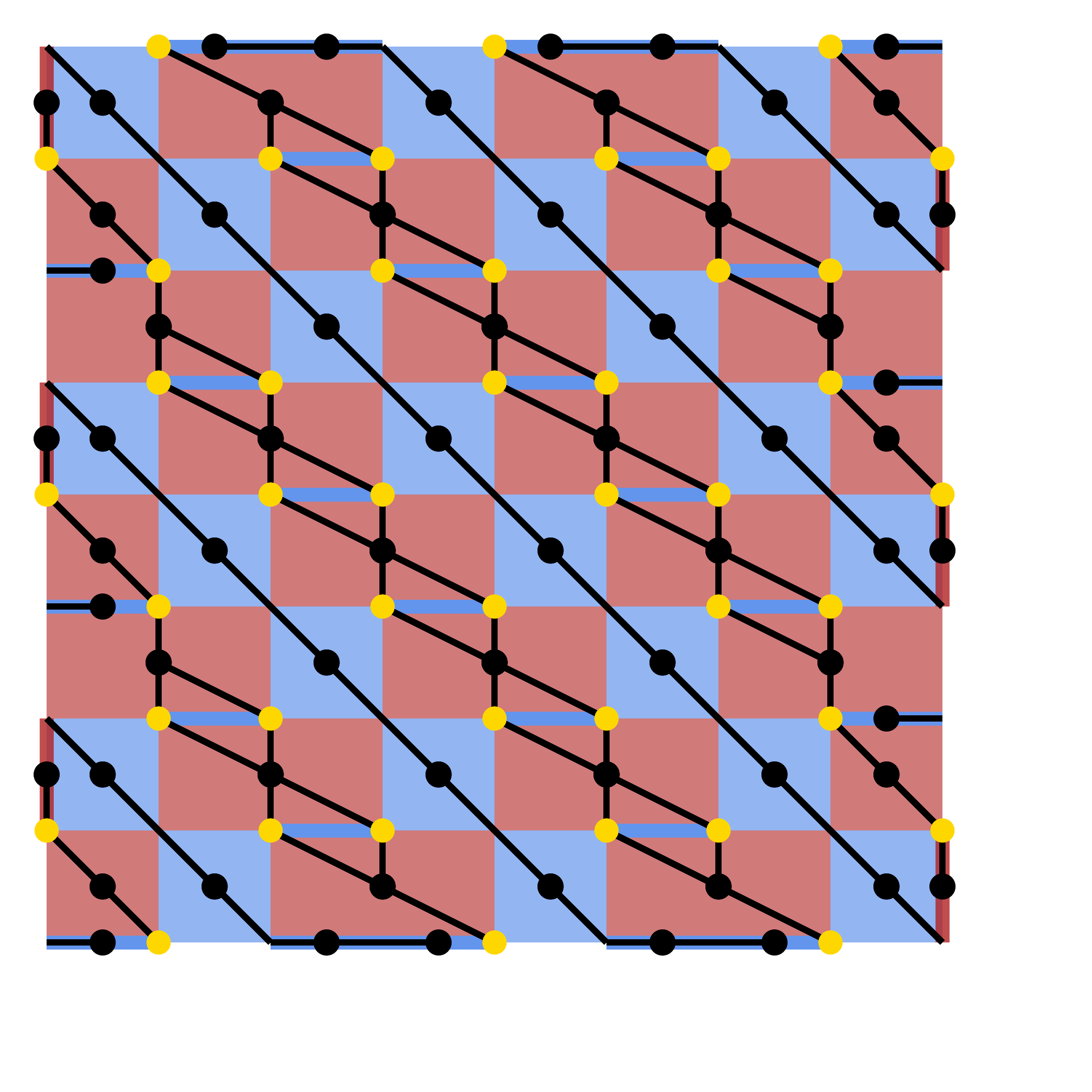}
        \label{fig:l3_lowweight_XZZX}
    \end{subfigure}
    \hfill
    \begin{subfigure}{0.32\textwidth}
    \caption{}
        \includegraphics[width = 0.9\linewidth]{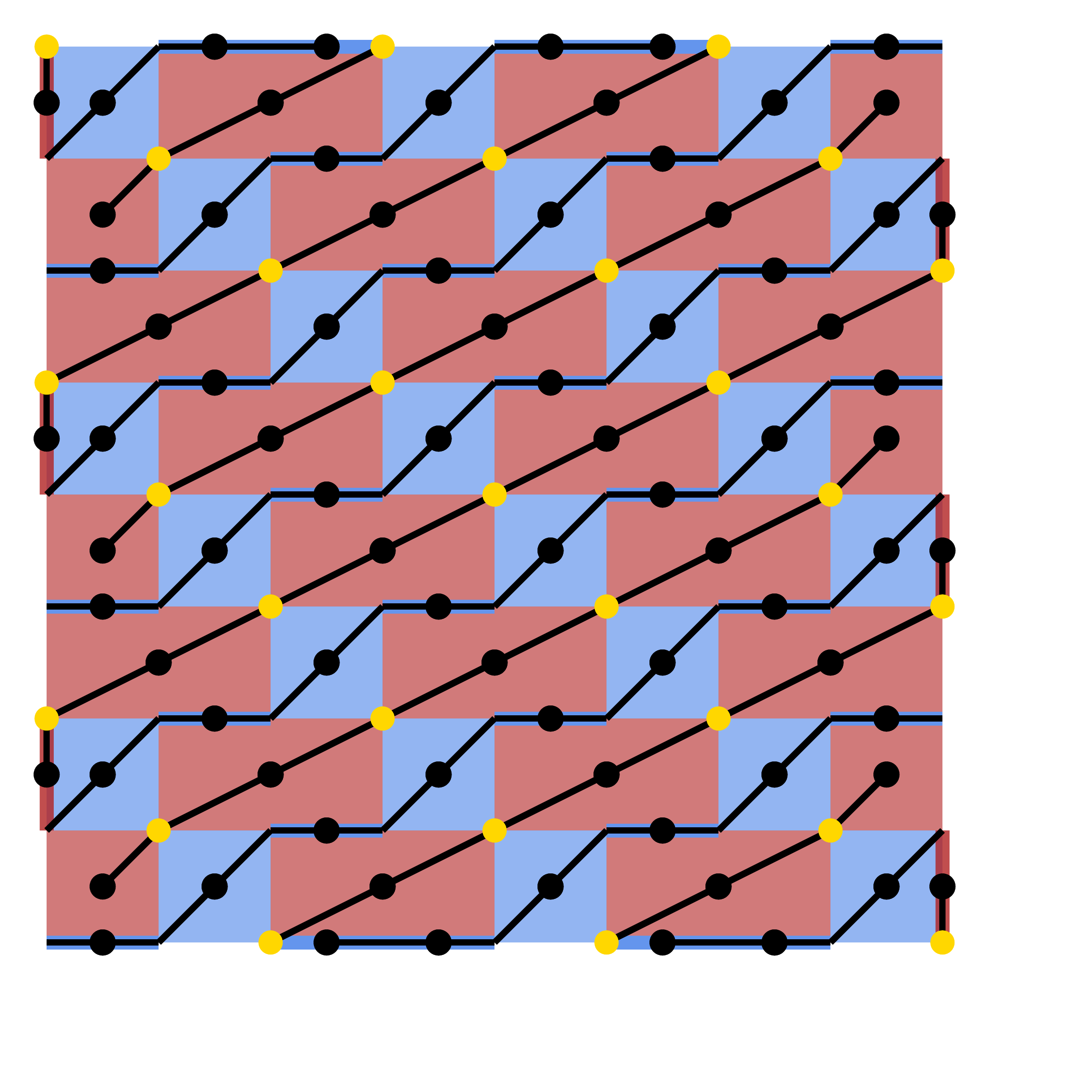}
        \label{fig:l3_lowweight_ZXXZ}
    \end{subfigure}
    \\
    \begin{subfigure}{0.32\textwidth}
    \caption{}
        \includegraphics[width = 0.9\linewidth]{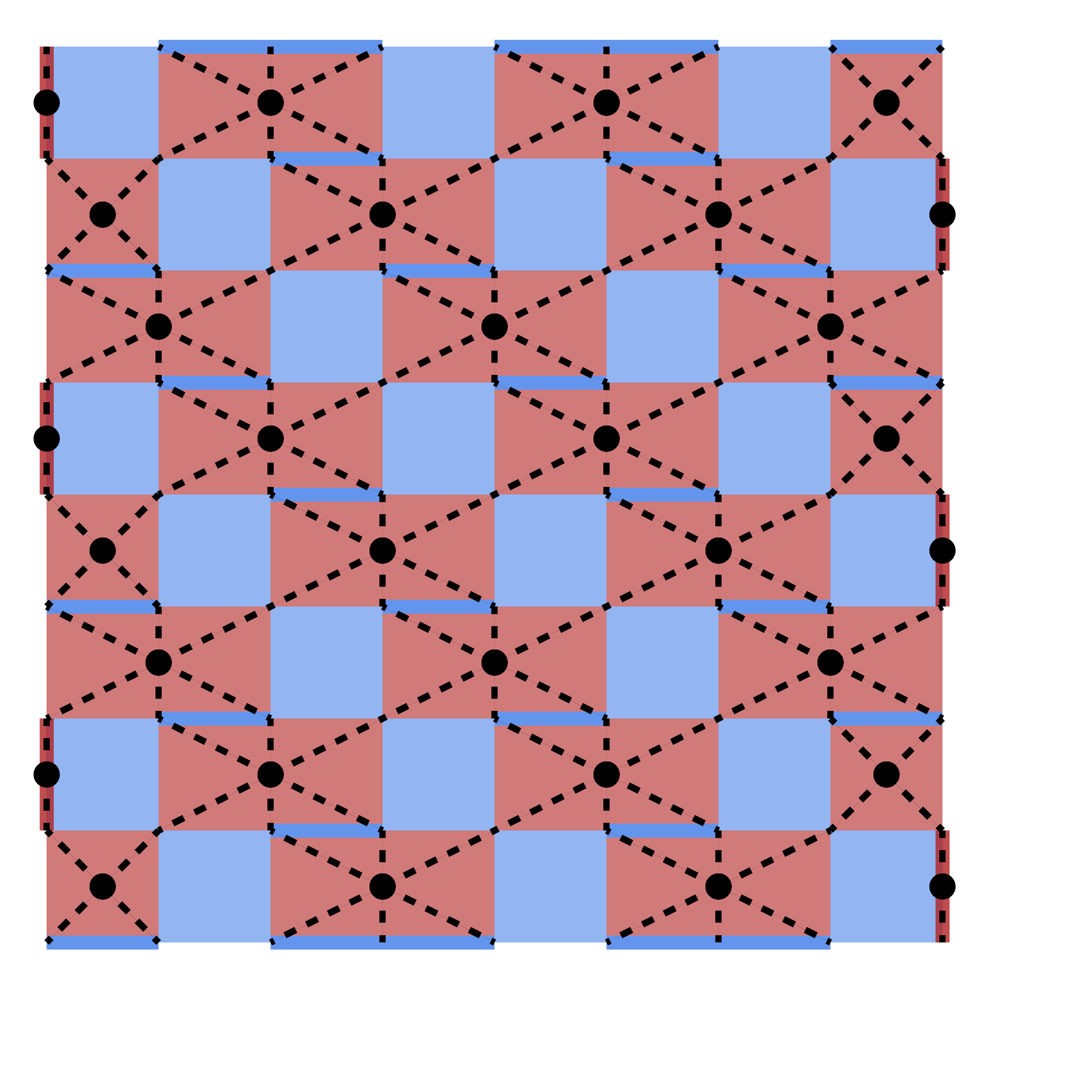}
        \label{fig:l3_highweight_CSS}
    \end{subfigure}
\hfill  
\begin{subfigure}{0.32\textwidth}
    \caption{}
        \includegraphics[width = 0.9\linewidth]{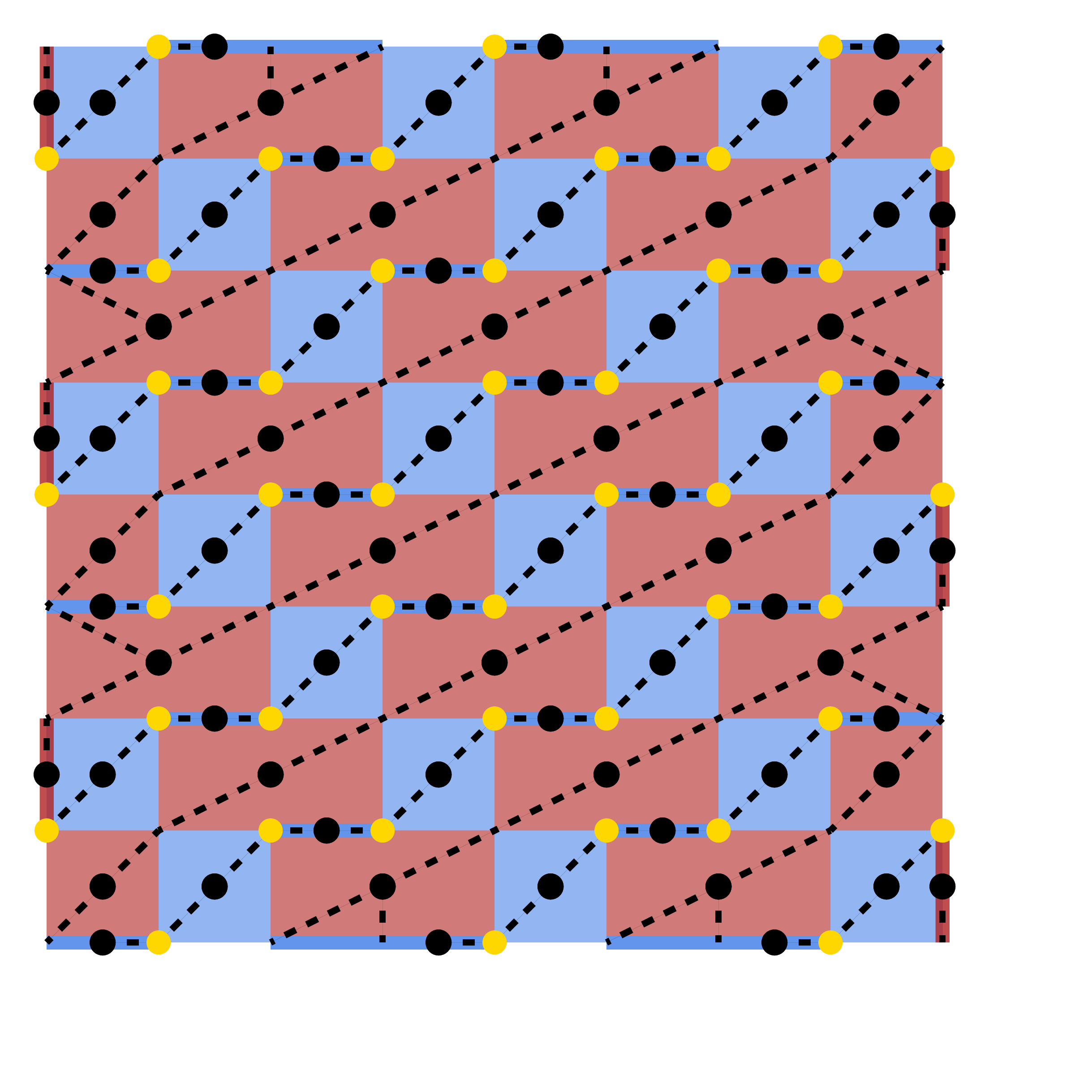}
        \label{fig:l3_highweight_XZZX}
    \end{subfigure}
\hfill
    \begin{subfigure}{0.32\textwidth}
    \caption{}
        \includegraphics[width = 0.9\linewidth]{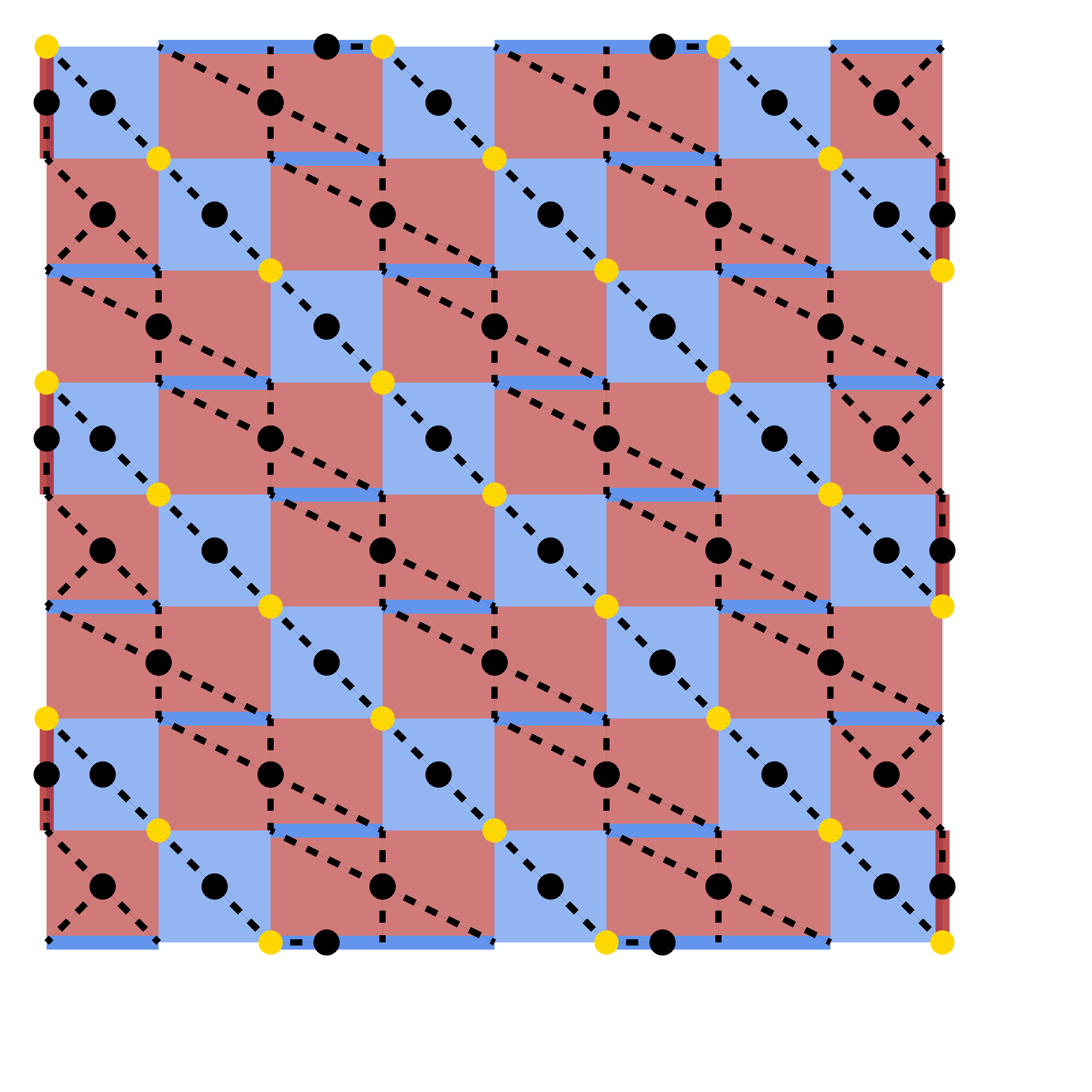}
        \label{fig:l3_highweight_ZXXZ}
    \end{subfigure}
    \caption{We list the low/high-weight graphs for the CSS (\textbf{a}/\textbf{d}), XZZX$\square$-deformed (\textbf{b}/\textbf{e}) and ZXXZ$\square$ (\textbf{c}/\textbf{f})- deformed compass code with $\ell = 3$. }
    \label{fig:graphs_l3}
\end{figure}

\begin{figure*}[tp!]
    \centering
    \begin{subfigure}{0.32\columnwidth}
    \caption{}
        \includegraphics[width = 0.9\linewidth]{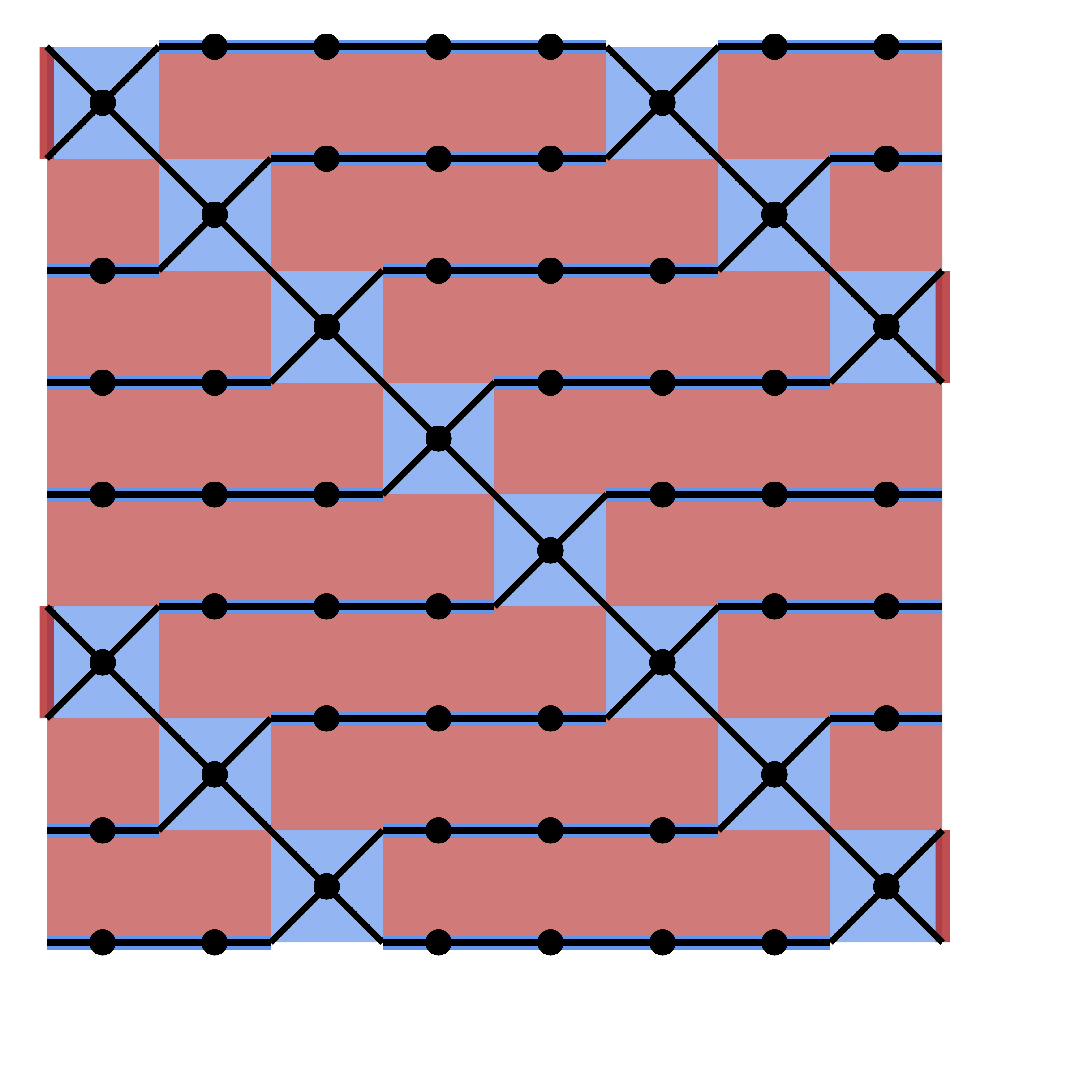}
        \label{fig:l5_lowweight_CSS}
    \end{subfigure}
    \hfill
    \begin{subfigure}{0.32\columnwidth}
    \caption{}
        \includegraphics[width = 0.9\linewidth]{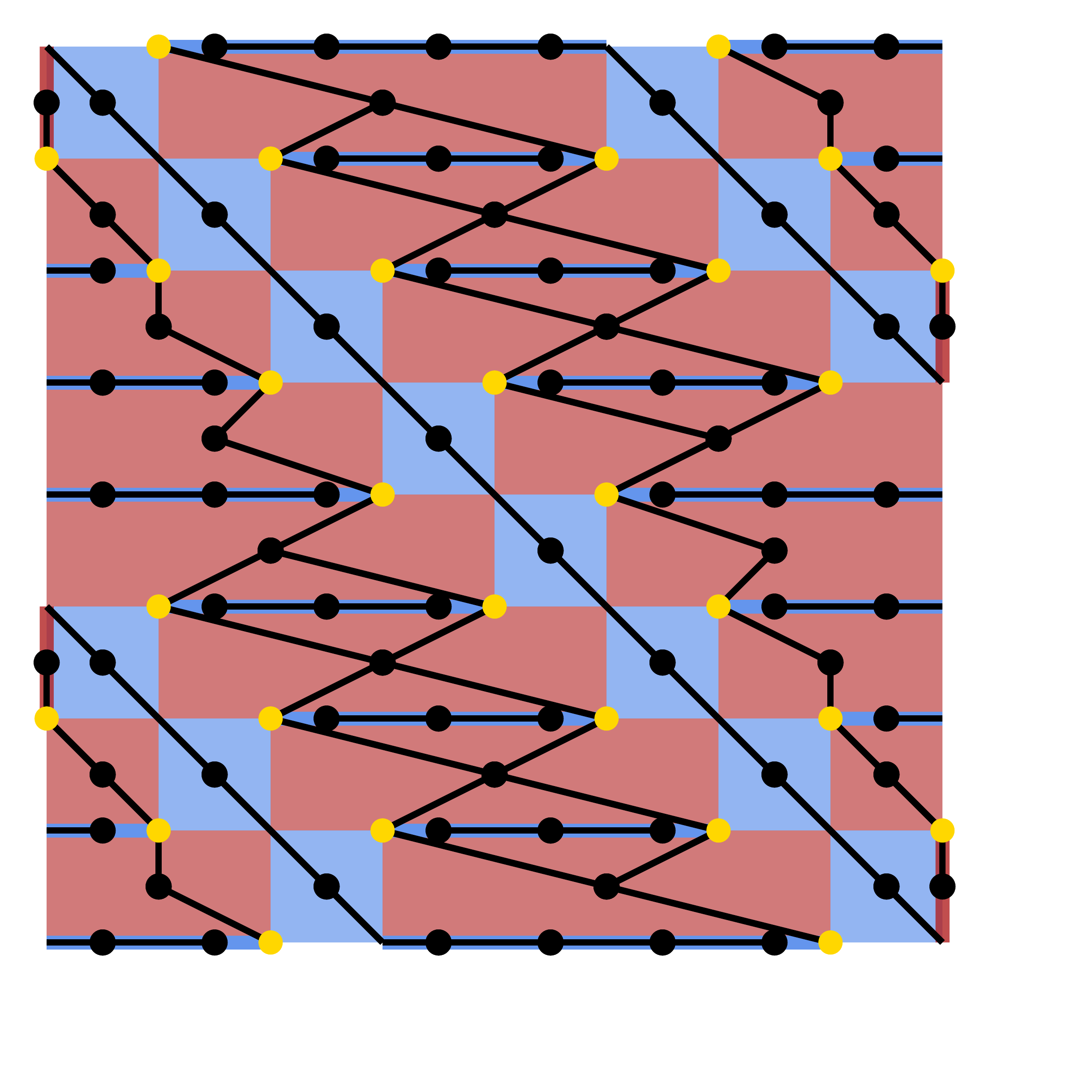}
        \label{fig:l5_lowweight_XZZX}
    \end{subfigure}
    \hfill
    \begin{subfigure}{0.32\columnwidth}
    \caption{}
        \includegraphics[width = 0.9\linewidth]{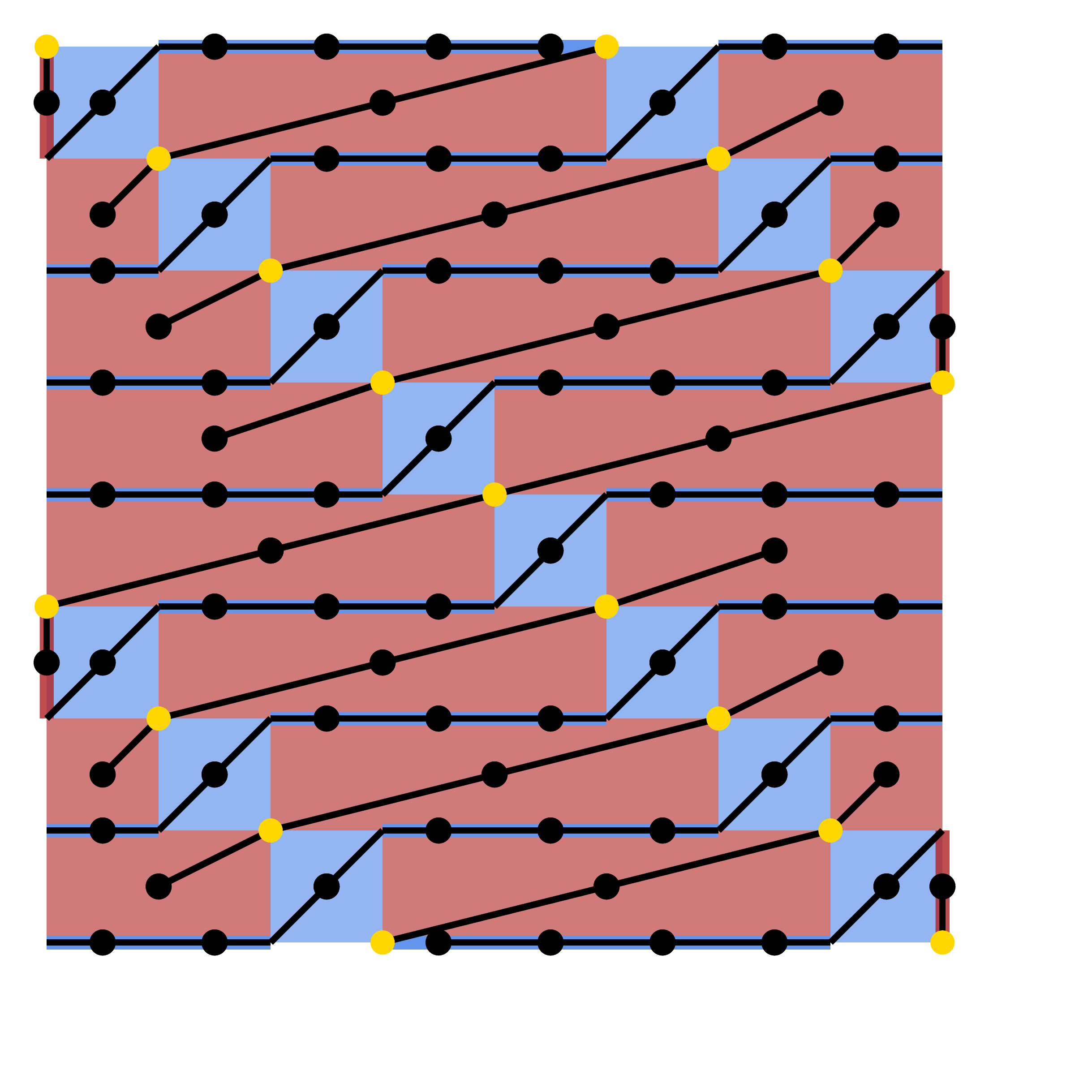}
        \label{fig:l5_lowweight_ZXXZ}
    \end{subfigure}
    \\
\begin{subfigure}{0.32\columnwidth}
    \caption{}
        \includegraphics[width = 0.9\linewidth]{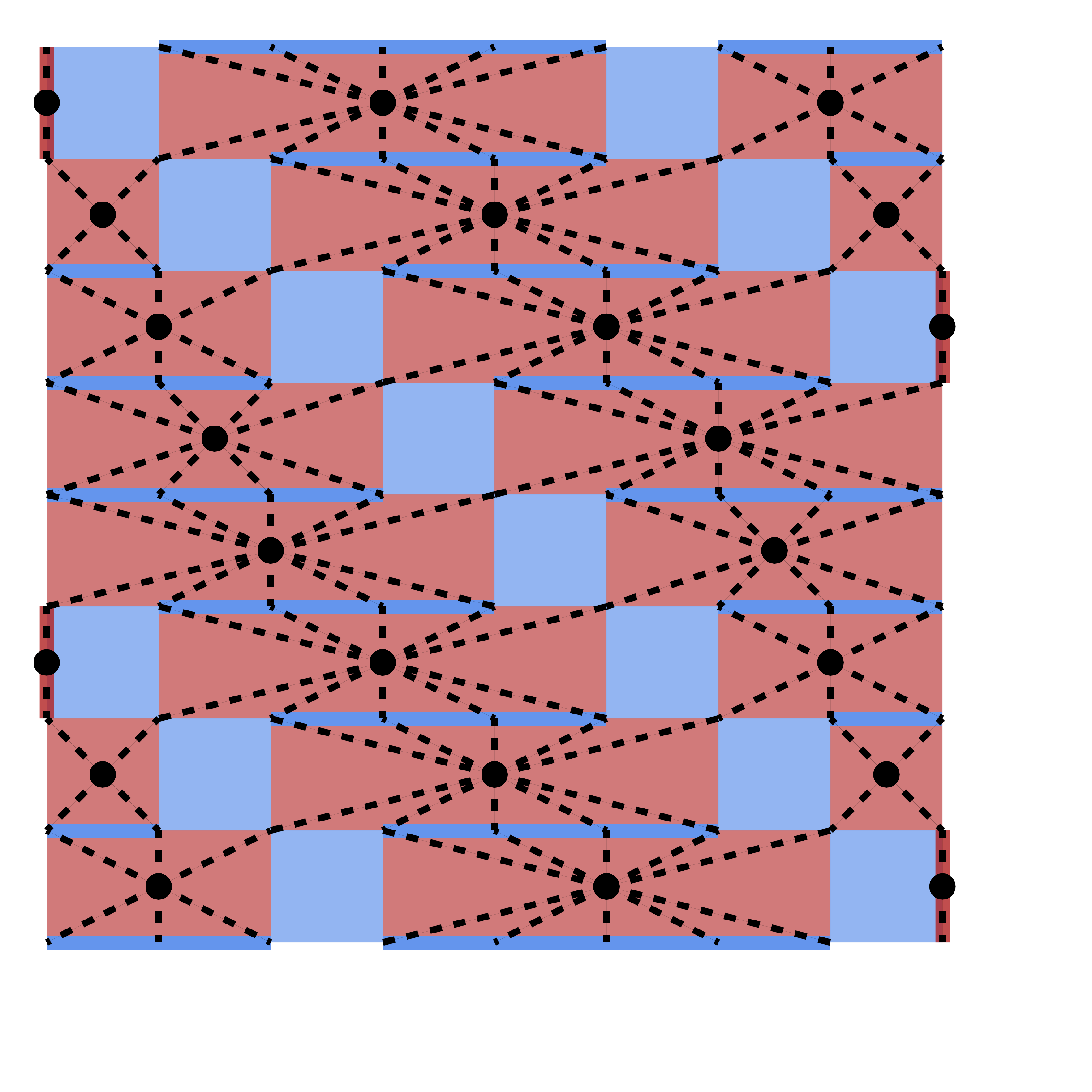}
        \label{fig:l5_highweight_CSS}
    \end{subfigure}
    \hfill
    \begin{subfigure}{0.32\columnwidth}
    \caption{}
        \includegraphics[width = 0.9\linewidth]{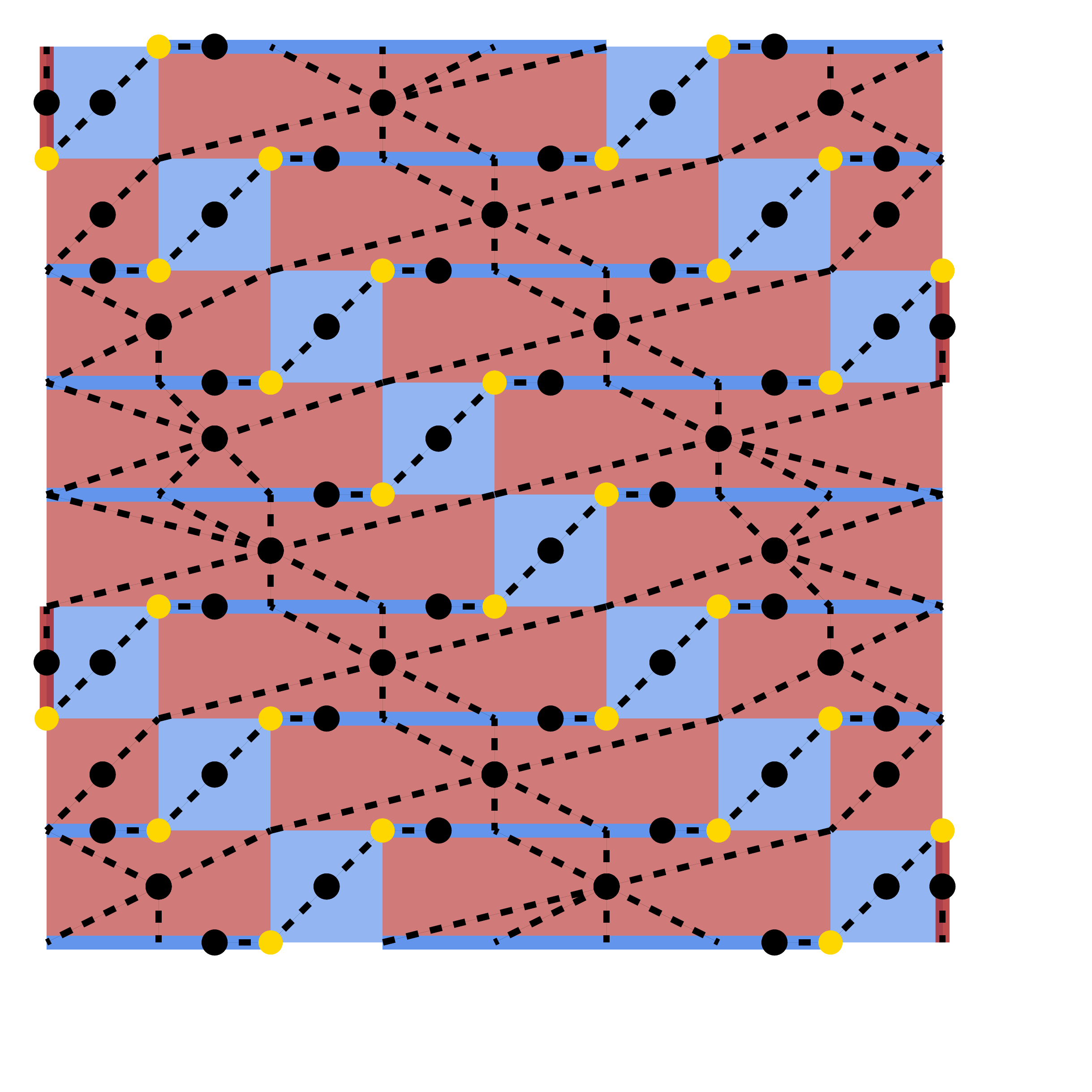}
        \label{fig:l5_highweight_XZZX}
    \end{subfigure}
    \hfill
    \begin{subfigure}{0.32\columnwidth}
    \caption{}
        \includegraphics[width = 0.9\linewidth]{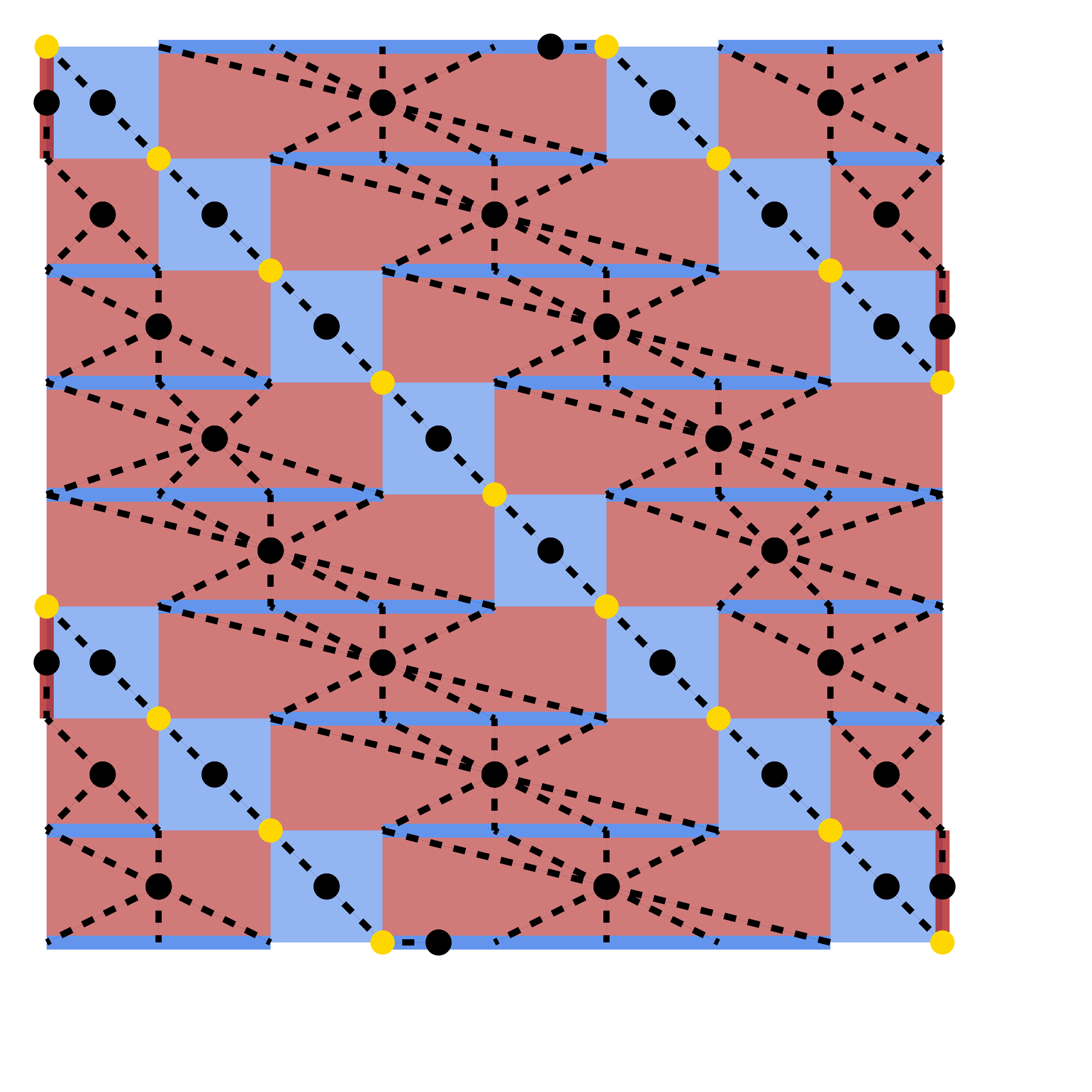}
        \label{fig:l5_highweight_ZXXZ}
    \end{subfigure}
    \caption{We list the low/high-weight graphs for the CSS (\textbf{a}/\textbf{d}), XZZX$\square$-deformed (\textbf{b}/\textbf{e}) and ZXXZ$\square$-deformed (\textbf{c}/\textbf{f}) compass code with $\ell = 5$. }
    \label{fig:graphs_l5}
\end{figure*}

The low(high)-weight decoder graphs (Figures \ref{fig:l3_lowweight_CSS}-\ref{fig:l3_highweight_CSS} and \ref{fig:l5_lowweight_CSS}-\ref{fig:l5_highweight_CSS}) for the CSS compass codes correspond to the $X$($Z$) decoder graphs since all qubits experience the noise model in Equation \ref{eq:NoiseModel}. We see that both the low and high-weight decoder graphs are highly connected. 

The XZZX$\square$-deformed codes have low-weight decoder graphs that are partitioned by the diagonals created by the weight-4 stabilizers which take the form XZZX (Figures \ref{fig:l3_lowweight_XZZX} and \ref{fig:l5_lowweight_XZZX}). Additionally, we see that the regions between these partitions do not increase in complexity regardless of the elongation parameter. The maximum degree of the vertices on these graphs is four. The high-weight graphs are highly connected, however, increasing decoding complexity as the elongation parameter grows (Figures \ref{fig:l3_highweight_XZZX} and \ref{fig:l5_highweight_XZZX}). 

Low-weight graphs of the ZXXZ$\square$-deformed compass codes are composed of disjoint segments or repetition codes (Figures \ref{fig:l3_lowweight_ZXXZ} and \ref{fig:l5_lowweight_ZXXZ}). The repetition codes are not of the same length, which could lead to more significant finite-size effects (Figure \ref{fig:Full_l5_10_ZXXZ}). Additionally, the high-weight graphs (Figures \ref{fig:l3_highweight_ZXXZ} and \ref{fig:l5_highweight_ZXXZ}) are partitioned in a fashion similar to the low-weight graph of the XZZX$\square$-deformed codes. This allows the ZXXZ$\square$-deformed codes to suppress low rate errors more efficiently than the XZZX$\square$-deformed codes. 

\FloatBarrier
\newpage
\section{Threshold Plots}\label{appendix:Thresholds}
It is conventional to use the finite-size scaling hypothesis to determine the thresholds of codes whose stabilizers can be mapped to the generalized random-bond Ising model or the $\mathbb{Z}_2$ random plaquette gauge model \cite{wang2003confinement}. However, there are some limitations to applying this method when finite-size effects are significant, which is the case for some of the codes and noise parameters we consider. We attempt to suppress these effects by applying the finite-size scaling fit to data from simulations of codes with high distances. In particular, we use odd distances between 27 and 43 for the deformed codes and distances between 11 and 19 for CSS codes. We also evaluate the extent of the finite-size effects by studying the logical error rate near the estimated threshold at higher distances. We note that, in general, the $X$ and $Z$ thresholds of the codes we consider do not coincide. As a result, the total threshold is determined by the lower one of the two in the thermodynamic limit. In this work, we are interested in evaluating the overall performance of the code at sufficiently large distances.

 The thresholds were estimated using finite-size scaling analysis near an observed crossing point. The numerical fit was a quadratic function of $(p-p_{th})L^{1/\nu}$ where $p$ is the physical error rate, $p_{th}$ is the threshold, $L$ is the distance and $\nu$ is a critical parameter. We observe strong agreement of the fitted curves with the numerical data.

\begin{figure*}[hbtp!]
    \centering
    \begin{subfigure}{0.47\linewidth}
      \caption{}
      \includegraphics[width = \linewidth]{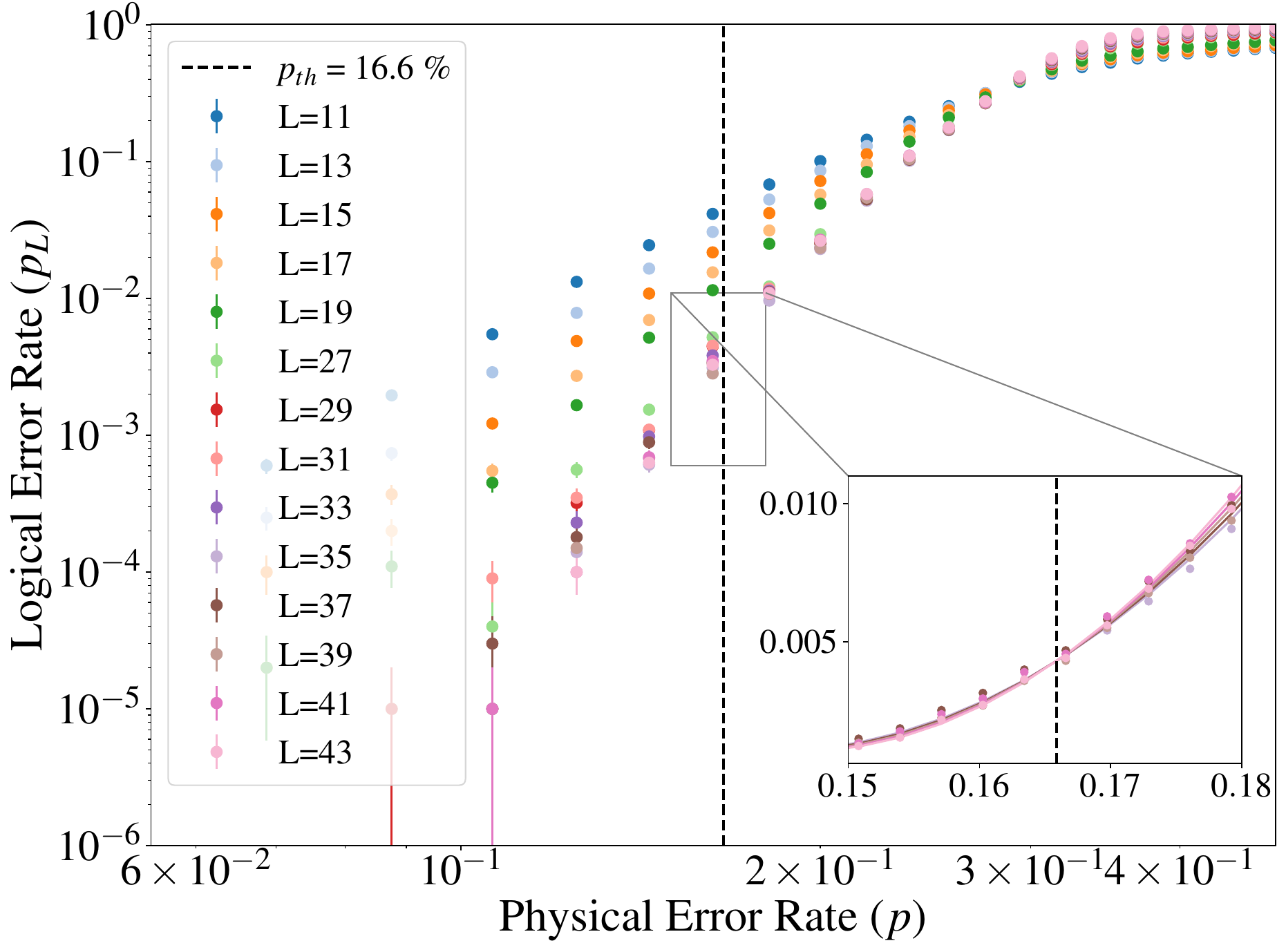}
      \label{fig:Thresh_l5_10_ZXXZ}
    \end{subfigure}
        \hfill 
    \begin{subfigure}{0.47\linewidth}
        \caption{}
        \includegraphics[width = \linewidth]{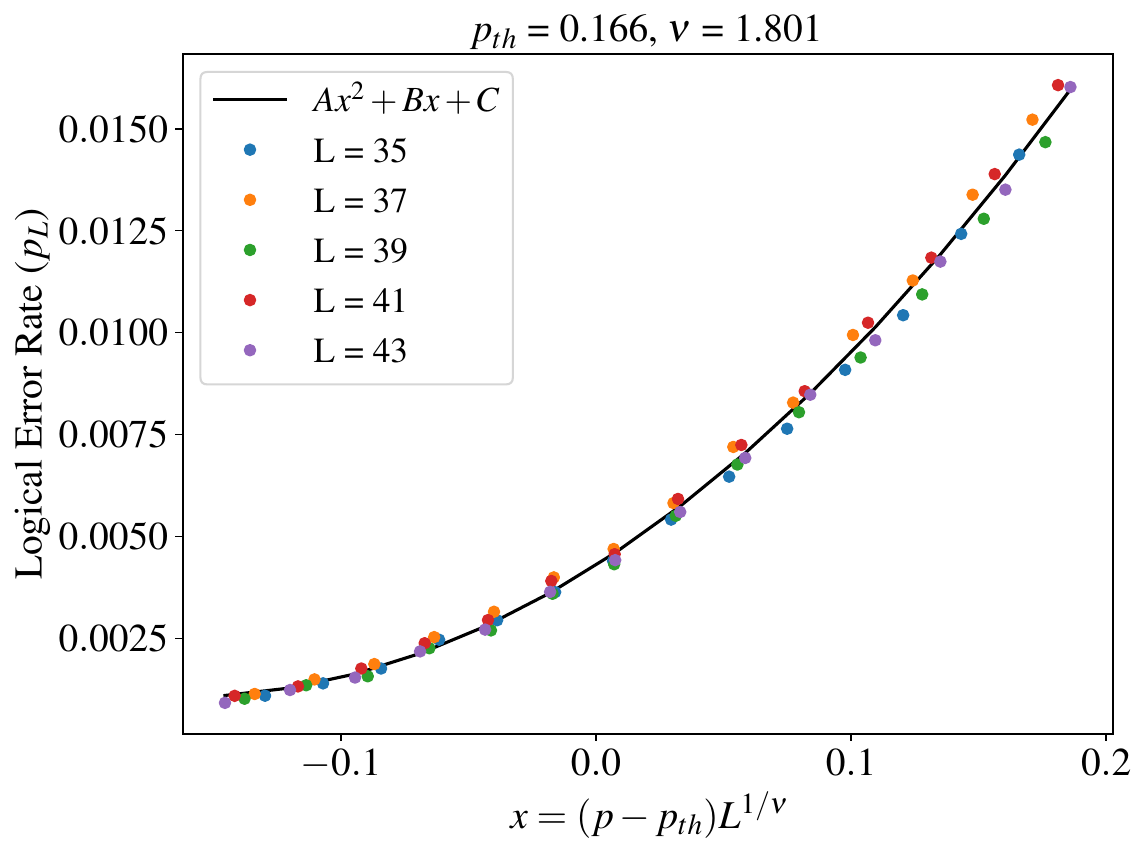}
        \label{fig:FSS_l5_10_ZXXZ}
    \end{subfigure}
    \\
    \begin{subfigure}{0.47\linewidth}
        \caption{}
        \includegraphics[width =\linewidth]{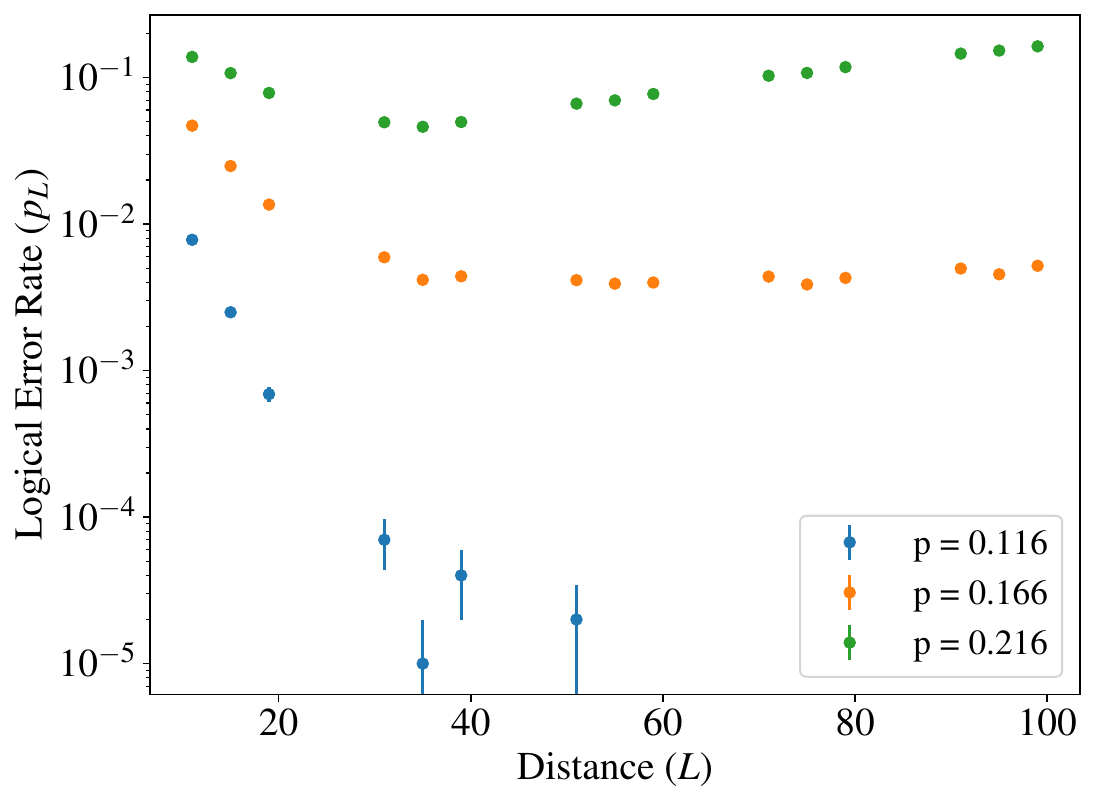}
        \label{fig:ThreshConv_l5_10_ZXXZ}
    \end{subfigure}
    \caption{\textbf{a}) Threshold plot for ZXXZ${\square}$-deformed code with elongation parameter $\ell = 5$ and bias $\eta = 10$. \textbf{b}) Finite-size scaling plot corresponding to the fit shown in the inset of \textbf{a}. \textbf{c}) Plot of logical error rates versus distances for physical error rates below ($p = 0.116$), at ($p = 0.166$), and above ($p = 0.216$) threshold. In \textbf{a}, we see two different physical error rates at which groups of the curves cross. The set of curves that intersect at the larger physical error rate have smaller distances, indicating that they suffer from finite-size effects. From \textbf{c}, we observe how the logical error rates at the threshold value decrease for distances less than $L=35$, which would indicate that our threshold is higher than what we report. However, the logical error rates at the threshold value stabilize after that distance.}
    \label{fig:Full_l5_10_ZXXZ}
\end{figure*}

Finite-size effects arise from many sources in the codes we consider. For example, elongated compass codes are constructed from repeated unit cells that increase in size with elongation. Thus, larger system sizes are required to capture the structure of the code. Finite-size effects are further amplified by the biased noise models and Clifford deformations we consider. The magnitude of these effects for the XZZX and XY surface codes was calculated in \cite{xiao2024exact}. The authors found that reliable threshold estimation using finite-size scaling requires distances comparable to the bias. One can see how these effects impact threshold plots in Figure \ref{fig:Thresh_l5_10_ZXXZ}, where we observe that lower distance codes (11 to 19) seem to have a crossing point at a higher physical error rate than the threshold we report. However, curves from higher distances cross at a lower physical error rate, indicating a lower threshold. We also observe these effects when analyzing logical error rate fluctuations with respect to distance near the estimated threshold. The expectation is that the logical error rate stabilizes at the threshold, decreases for physical error rates below the threshold and increases for those above the threshold.

\begin{figure*}[hbtp!]
    \centering
    \begin{subfigure}{0.47\linewidth}
    \caption{}
    \includegraphics[width=\linewidth]{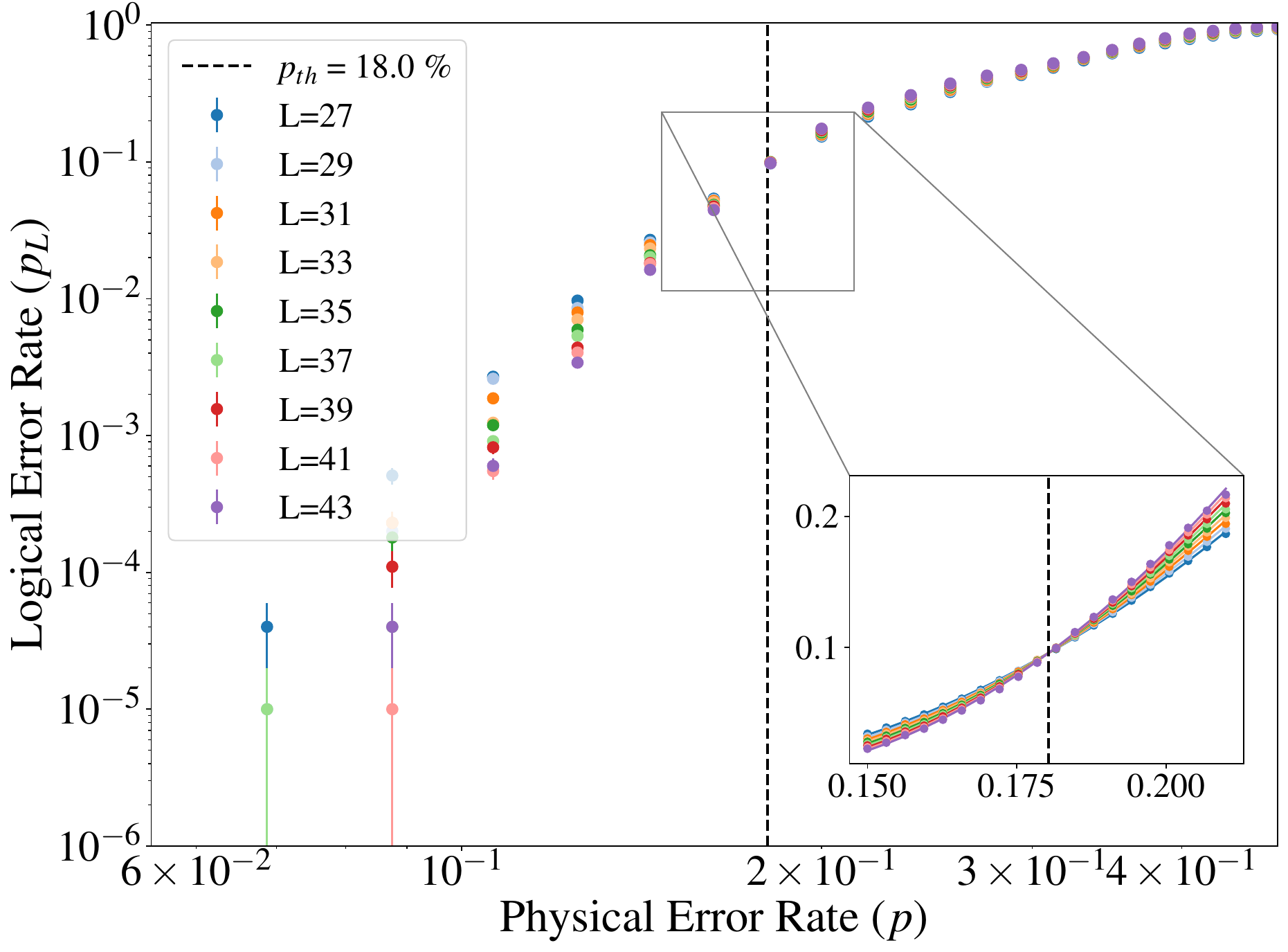} \label{fig:l_3_bias_10_XZZX_threshold_inset}
    \end{subfigure}
    \hfill
    \begin{subfigure}{0.47\linewidth}
    \caption{}
    \includegraphics[width=\linewidth]{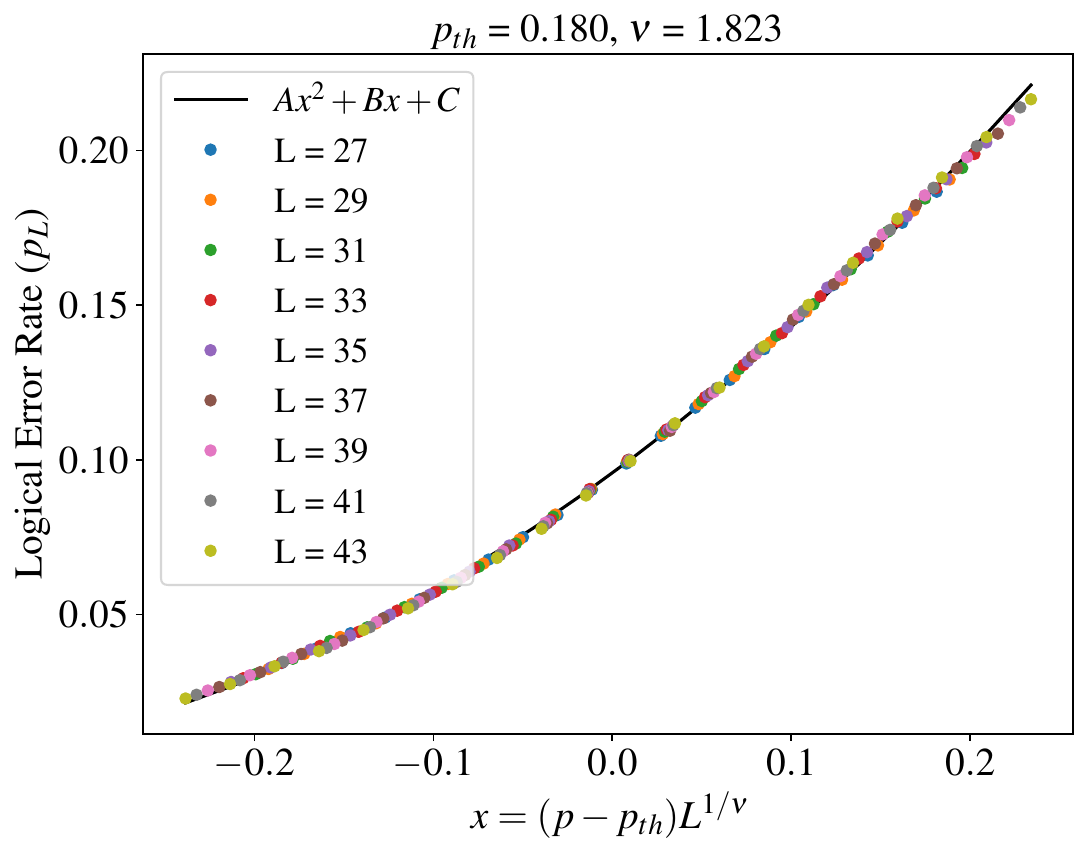}
    \label{fig:l_3_bias_10_XZZX_FSS}
    \end{subfigure}
    \\
    \begin{subfigure}{0.47\linewidth}
        \caption{}
       \includegraphics[width=\linewidth]{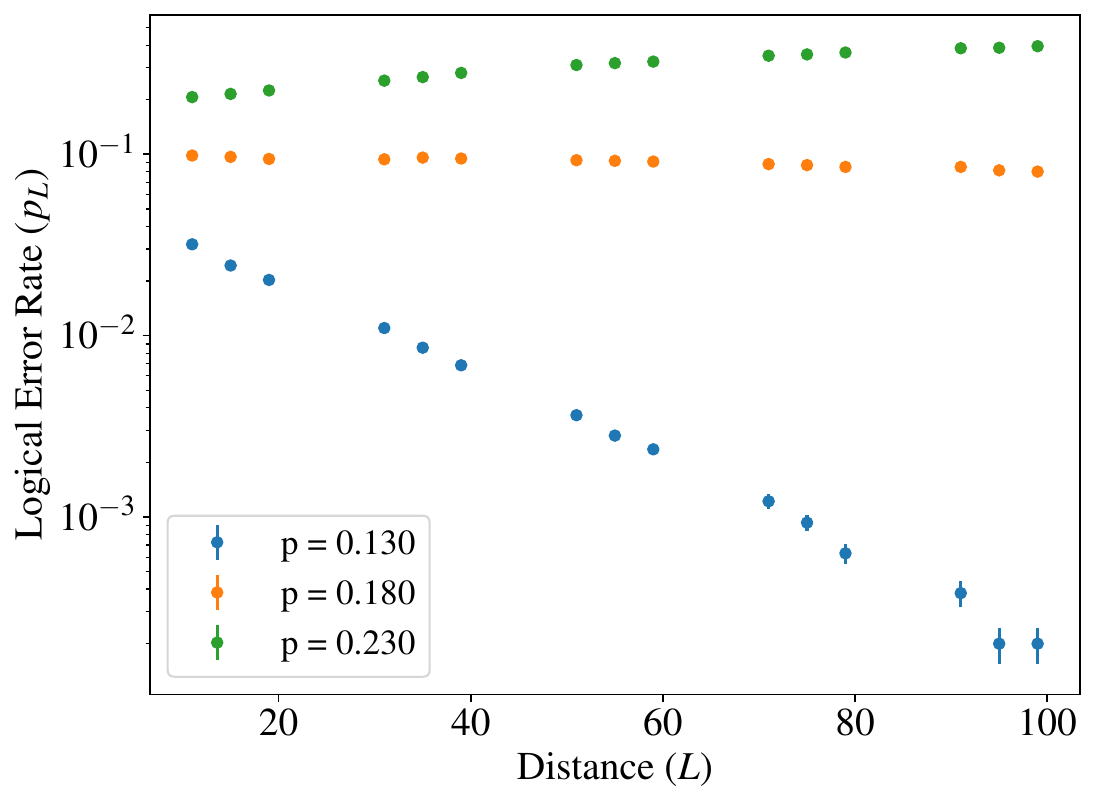} \label{fig:l_3_bias_10_XZZX_ThreshConv}
    \end{subfigure}
\caption{\textbf{a}) Threshold plot for XZZX${\square}$-deformed code with elongation parameter $\ell = 3$ and bias $\eta = 10$. \textbf{b}) Finite-size scaling plot corresponding to the fit shown in the inset of \textbf{a}. \textbf{c}) Plot of logical error rates versus distances for physical error rates below ($p = 0.13$), at ($p = 0.18$), and above ($p = 0.23$) threshold. We observe that the logical error rate remains constant at the threshold while it decreases (increases) for physical error rates below (above) threshold.}
\label{fig:l3_bias_10_XZZX_Full}
\end{figure*}

\begin{figure*}[hbtp!]
    \centering
    \begin{subfigure}[t]{0.47\linewidth}
      \caption{}
      \includegraphics[width = \linewidth]{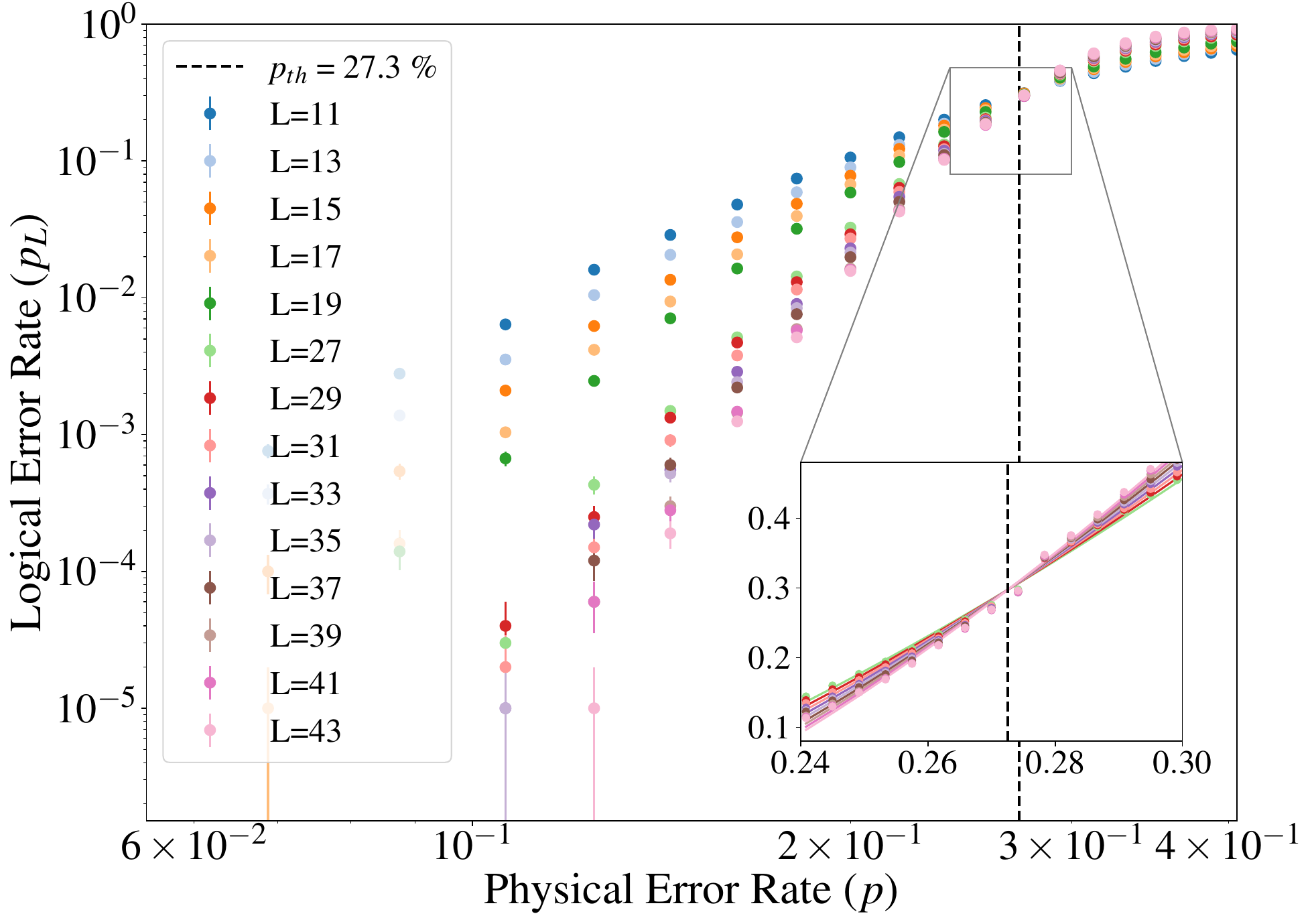}
      \label{fig:Thresh_l3_10_ZXXZ}
    \end{subfigure}
        \hfill 
    \begin{subfigure}[t]{0.47\linewidth}
        \caption{}
        \includegraphics[width = 0.9\linewidth]{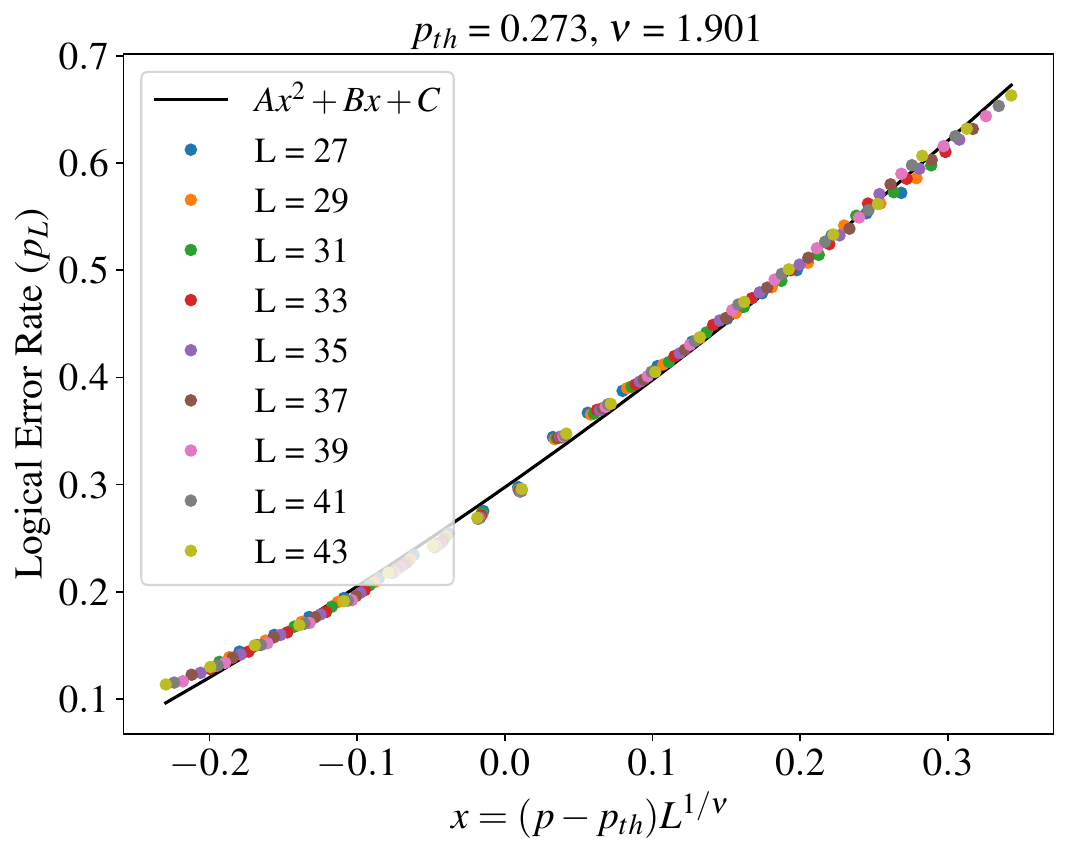}
        \label{fig:FSS_l3_10_ZXXZ}
    \end{subfigure}
    \\
    \begin{subfigure}{0.47\linewidth}
        \caption{}
        \includegraphics[width = \linewidth]{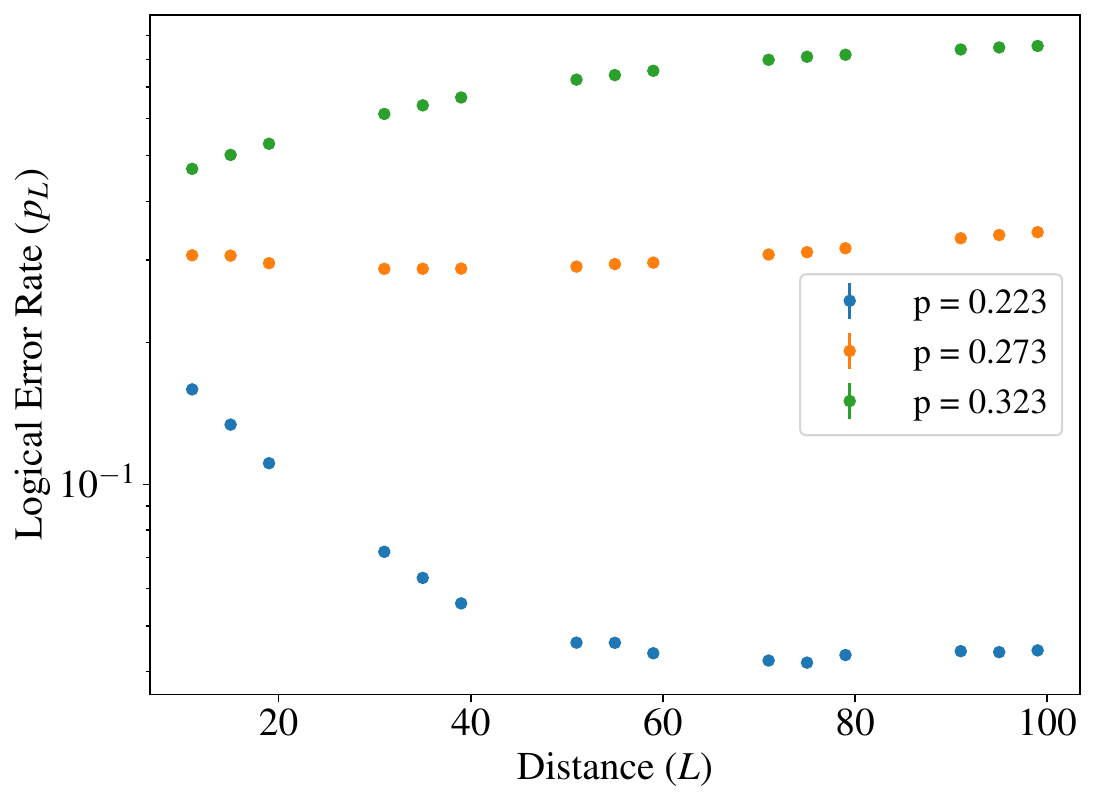}
        \label{fig:ThreshConv_l3_10_ZXXZ}
    \end{subfigure}
    \caption{\textbf{a}) Threshold plot for ZXXZ${\square}$-deformed code with elongation parameter $\ell = 3$ and bias $\eta = 10$. \textbf{b}) Finite-size scaling plot corresponding to the fit shown in the inset of \textbf{a}. The fit is applied to curves with distances 27-43. We include curves of lower distances in \textbf{a} to show that all curves intersect near the threshold value reported. \textbf{c}) Plot of logical error rates versus distance for physical error rates below threshold ($p = 0.223$), at threshold calculated by the finite-size scaling method ($p = 0.273$), and above threshold ($p = 0.323$) threshold. We observe that the logical error rate is nearly constant at the threshold calculated by the finite-size scaling method while it increases for physical error rates above the approximate threshold. However, we also observe that the logical error rates begin to increase after $L = 60$ at $p = 0.223$. This implies that the threshold in the thermodynamic limit is below 22.3\%.}
    \label{fig:Full_l3_10_ZXXZ}
\end{figure*}

\begin{figure*}[hbtp!]
    \centering   
    \begin{subfigure}{0.47\linewidth}
    \caption{}
    \includegraphics[width=\linewidth]{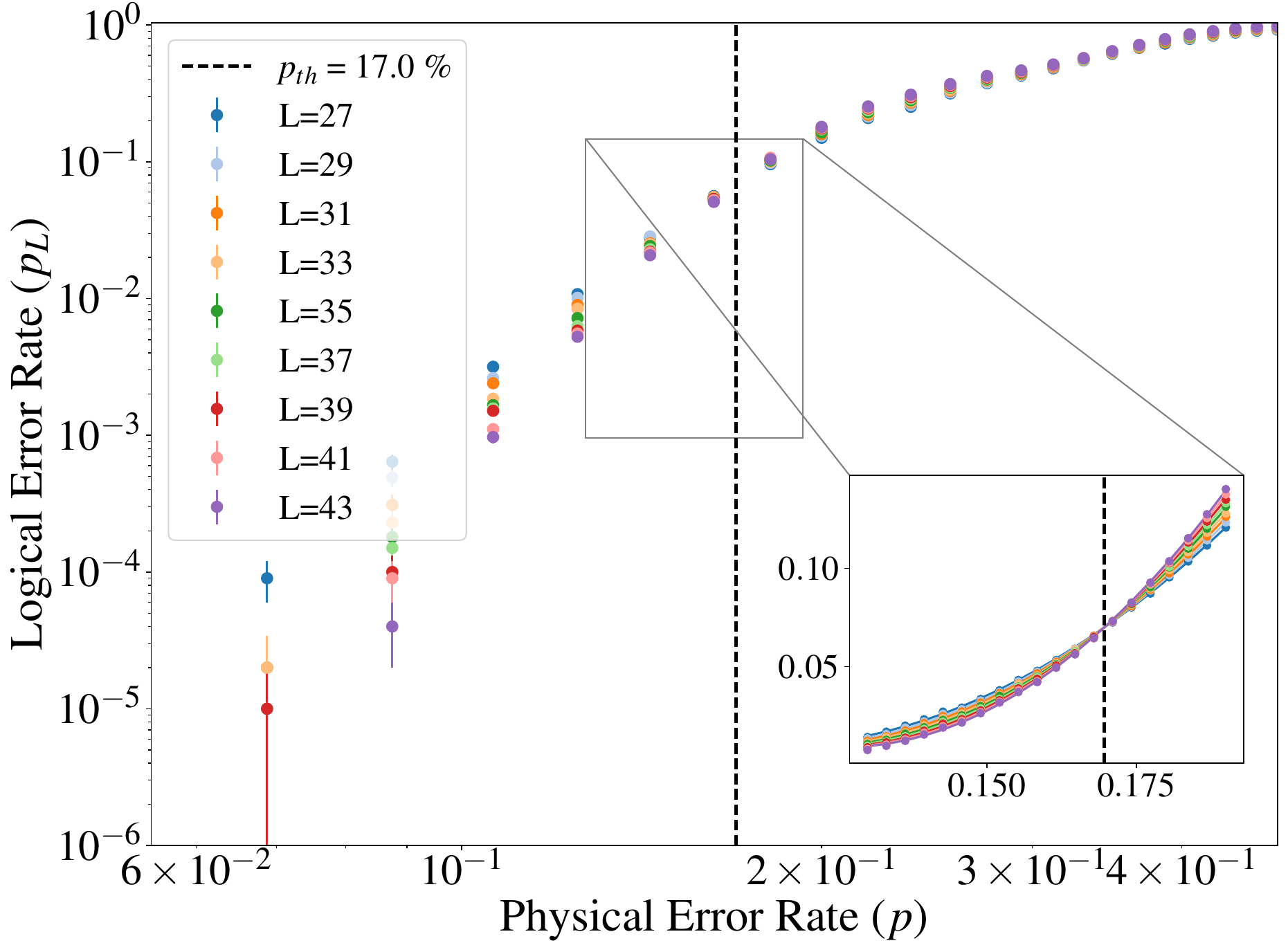}
    \label{fig:Thresh_l5_10_XZZX}
    \end{subfigure}
        \hfill 
        \begin{subfigure}{0.47\linewidth}
        \caption{}
        \includegraphics[width = \linewidth]{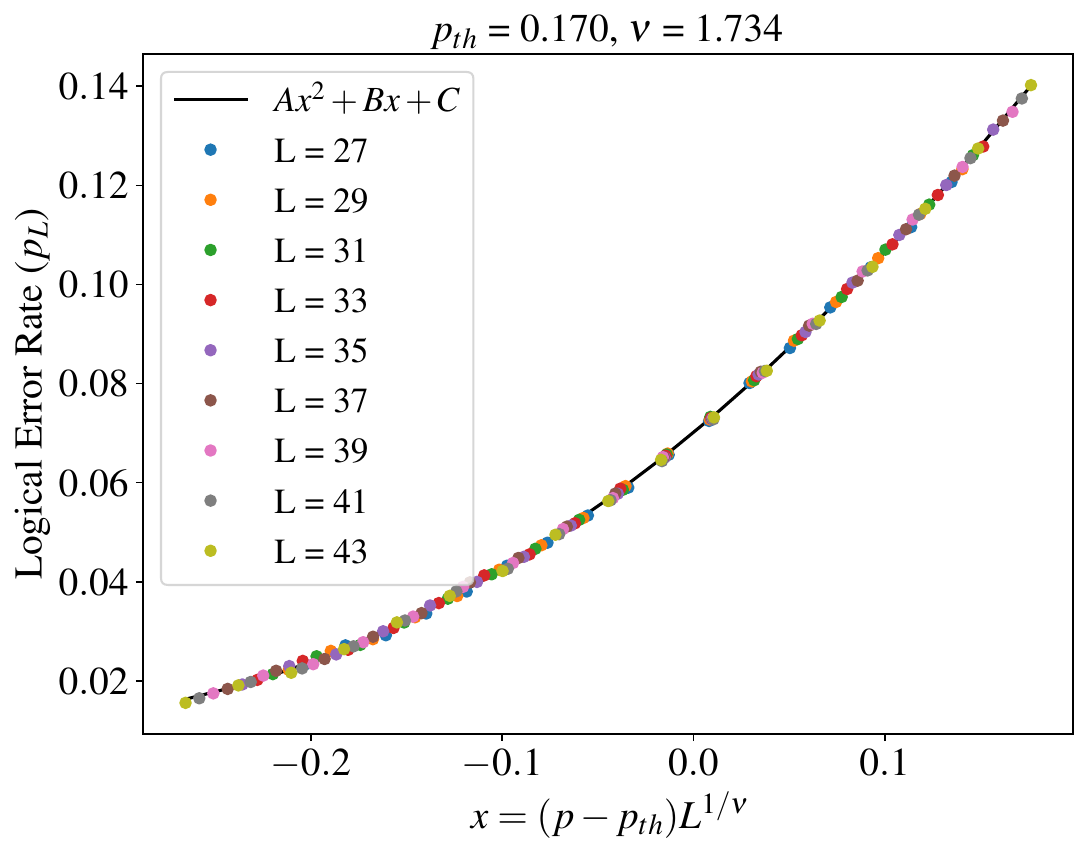}
        \label{fig:FSS_l5_10_XZZX}
    \end{subfigure}
    \\
    \begin{subfigure}{0.47\linewidth}
        \caption{}
        \includegraphics[width = \linewidth]{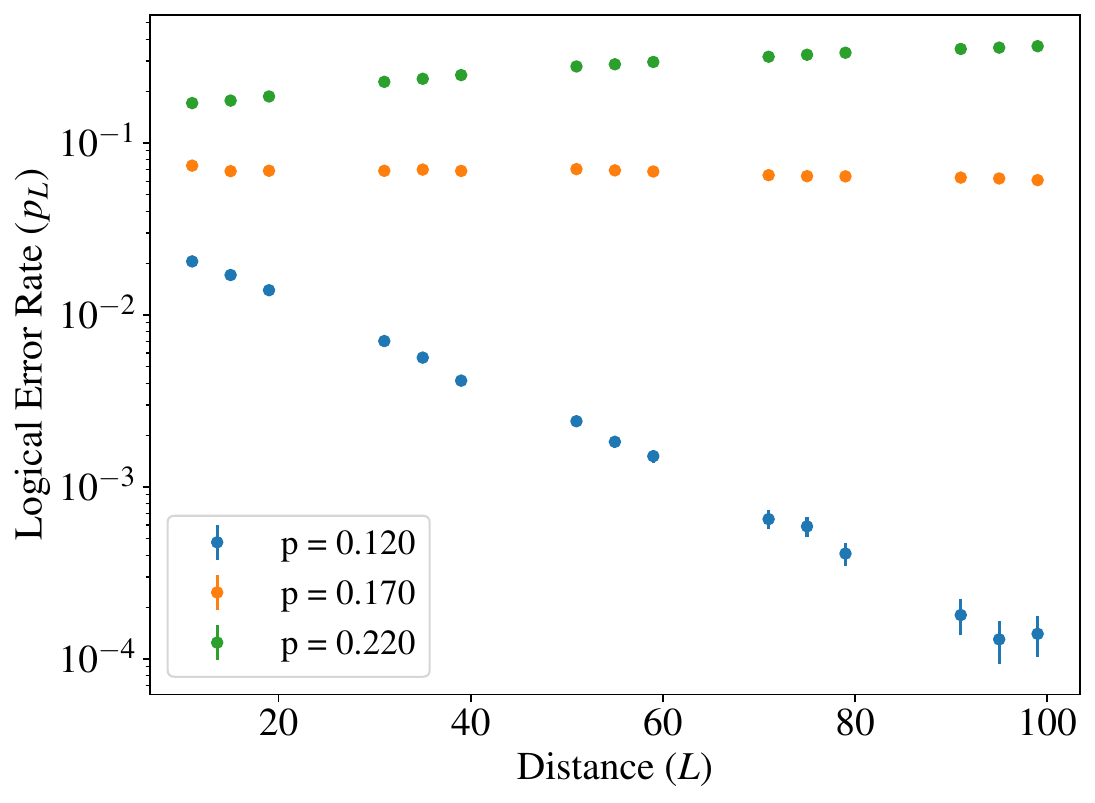}
        \label{fig:ThreshConv_l5_10_XZZX}
    \end{subfigure}
    \caption{\textbf{a}) Threshold plot for XZZX${\square}$-deformed code with elongation parameter $\ell = 5$ and bias $\eta = 10$. \textbf{b}) Finite-size scaling plot corresponding to the fit shown in the inset of \textbf{a}. \textbf{c}) Plot of logical error rates versus distances for physical error rates below ($p = 0.12$), at ($p = 0.17$), and above ($p = 0.22$) threshold. We observe that the logical error rate remains constant at the threshold while it decreases (increases) for physical error rates below (above) threshold.}
    \label{fig:Full_l5_10_XZZX}
    \end{figure*}

Code capacity threshold values reported are physical error rates which yield logical error rates with asymptotic behavior as the distance approaches 100 unless stated otherwise. Additionally, the logical error rate is suppressed as the distance of the code is increased, provided that the physical error rate is below the reported threshold. We show results from finite-size scaling fits and threshold stability simulations in Figures \ref{fig:l3_bias_10_XZZX_Full}-\ref{fig:Full_l5_10_ZXXZ}. We use a similar method to determine phenomenological thresholds by simulating codes with distances up to 40.

\end{document}